\documentclass{aa}
\usepackage{graphicx}
\usepackage{natbib}
\usepackage{hyperref}
\bibpunct{(}{)}{;}{a}{}{,}
\usepackage{txfonts}

\usepackage{subfiles}
\usepackage{tikz}
\usepackage{listings}
\usepackage{widetext}

\newcommand{\MSun}{\mbox{${M}_\odot$}}
\newcommand{\Brutus}{\mbox{Brutus}}
\newcommand{\Hermite}{\mbox{ph4}}
\newcommand{\PhFour}{\mbox{ph4}}
\newcommand{\HermiteGRX}{\mbox{Hermite\_GRX}}
\newcommand{\HermiteGRP}{\mbox{Hermite\_GR1P}}
\newcommand{\Sapporo}{\mbox{Sapporo}}
\newcommand{\AMUSE}{\mbox{AMUSE}}
\newcommand{\avg}[1]{\ensuremath{\langle #1 \rangle}}
\newcommand{\vect}[1]{\ensuremath{\boldsymbol{#1}}}
\newcommand{\unit}[1]{\ensuremath{\, \mathrm{#1}}}
\newcommand{\BigO}[1]{\ensuremath{\mathcal{O}\left(#1\right)}}
\newcommand{\deriv}[2]{\ensuremath{\frac{\mathrm{d}#1}{\mathrm{d}#2}}}
\newcommand{\DeltaTau}{\ensuremath{\kappa}}
\newcommand{\Cscaling}{\ensuremath{v/c}}

\let\apgt\ga
\let\aplt\la
\usepackage[normalem]{ulem}
\usepackage{amstext}

\newcommand{\includefig}[1]{\IfFileExists{Figs/#1.pdf_tex}{\subimport{Figs/}{#1.pdf_tex}}{
\begingroup%
  \makeatletter%
  \providecommand\color[2][]{%
    \errmessage{(Inkscape) Color is used for the text in Inkscape, but the package 'color.sty' is not loaded}%
    \renewcommand\color[2][]{}%
  }%
  \providecommand\transparent[1]{%
    \errmessage{(Inkscape) Transparency is used (non-zero) for the text in Inkscape, but the package 'transparent.sty' is not loaded}%
    \renewcommand\transparent[1]{}%
  }%
  \providecommand\rotatebox[2]{#2}%
  \ifx\svgwidth\undefined%
    \setlength{\unitlength}{158.875bp}%
    \ifx\svgscale\undefined%
      \relax%
    \else%
      \setlength{\unitlength}{\unitlength * \real{\svgscale}}%
    \fi%
  \else%
    \setlength{\unitlength}{\svgwidth}%
  \fi%
  \global\let\svgwidth\undefined%
  \global\let\svgscale\undefined%
  \makeatother%
  \begin{picture}(1,1.00015736)%
    \put(0,0){\includegraphics[width=\unitlength]{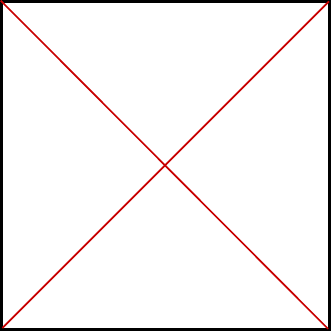}}%
    \put(0.30098955,0.86360472){\color[rgb]{0,0,0}\makebox(0,0)[lb]{\smash{Placeholder}}}%
    \put(0.30109181,0.0828228){\color[rgb]{0,0,0}\makebox(0,0)[lb]{\smash{Placeholder}}}%
  \end{picture}%
\endgroup%
}}

\begin{document}

\title{Chaos in self-gravitating many-body systems}

\subtitle{Lyapunov time dependence of $N$ and the influence of general relativity}

\author{
S.~F.~Portegies~Zwart \inst{1}
\and
T. C. N. Boekholt \inst{2}
\and
E. H. Por \inst{3} 
\and
A. S. Hamers \inst{4}
\and
S.L.W. McMillan \inst{5}
}

\institute{
Leiden Observatory, Leiden University, PO Box 9513, 2300 RA, Leiden, The Netherlands \\
\email{spz@strw.leidenuniv.nl}
\and
Rudolf Peierls Centre for Theoretical Physics, Clarendon Laboratory, Parks Road, Oxford, OX1 3PU, UK \\
\email{tjarda.boekholt@physics.ox.ac.uk}
\and
Space Telescope Science Institute, 3700 San
Martin Drive, Baltimore, MD 21218, USA
\email{epor@stsci.edu}
\and
Max-Planck-Institut f\"ur Astrophysik, Karl-Schwarzschild-Str. 1, 85741 Garching, Germany
\and
Department of Physics, Drexel University, Philadelphia, PA 19104, USA
}
\date{Received March 29, 2021; accepted March St. Juttemis}

\abstract{
In self-gravitating $N$-body systems, small perturbations introduced
at the start, or infinitesimal errors that are produced by the
numerical integrator or are due to limited precision in the computer,
grow exponentially with time. For Newton's gravity, we confirm earlier
results that for relatively homogeneous systems, this rate of growth
per crossing time increases with $N$ up to $N \sim 30$, but that for
larger systems, the growth rate has a weaker scaling with $N$.  For
concentrated systems, however, the rate of exponential growth
continues to scale with $N$. In relativistic self-gravitating systems,
the rate of growth is almost independent of $N$. This effect, however,
is only noticeable when the system's mean velocity approaches the
speed of light to within three orders of magnitude. The chaotic
behavior of systems with more than a dozen bodies for the usually
adopted approximation of only solving the pairwise interactions in the
Einstein-Infeld-Hoffmann equation of motion is qualitatively different
than when the interaction terms (or cross terms) are taken into
account. This result provides a strong motivation for follow-up
studies on the microscopic effect of general relativity on orbital
chaos, and on the influence of higher-order cross-terms in the
Taylor-series expansion of the Einstein-Infeld-Hoffmann equations of
motion.
}

\keywords{stars: kinematics and dynamics -- 
methods: numerical
}

\maketitle

\section{Introduction}
\label{sec:introduction}

Soon after the first gravitational $N$-body problems were computed
\citep{1960ZA.....50..184V,1964ApNr....9..313A,1968BAN....19..479V},
\cite{1964ApJ...140..250M} questioned the validity of such
simulations. The nature of his concern was based on the intrinsically
chaotic behavior of Newton's law of gravity. Errors during integration
are introduced by the limited precision of the computer, together with
the limited accuracy of the numerical integration scheme. The
exponential growth of both sources of errors then contributes to the
lack of reproducibility in $N$-body simulations
\citep{2020MNRAS.493.3932B}.

In an attempt to acquire a converged solution to the 25-body problem,
\cite{1970A&A.....7..249H} integrated Newton's equations of motion
using various precisions on a 60~bit CDC~6400, a 64~bit IBM~360/365,
and an 80~bit Honeywell-Bull CII~90--80. He found that although identical initial realizations were used, he acquired a different answer
for the final positions and velocities of the 25 objects in his
calculations, even with the same algorithm and time step. He argued
that the difference in precision of the floating-point unit was
responsible for the lack of reproducibility of the results. Because the
IEEE standard for floating-point arithmetic (IEEE 754) was only
introduced in 1985, the discrepancy identified by
\cite{1970A&A.....7..249H} could also have originated from differences
in round-off in the least significant digit of the various machines.
We still encounter this problem in today's Graphical Processing Units
(GPU), which have limited precision \citep{6169579}. These sources of
errors, time step, and round-off make individual solutions to the
$N$-body problem notoriously unreliable, although statistically, they
may still have the correct phase-space distribution characteristics
\citep{2015ComAC...2....2B,2020MNRAS.493.1913H}. The common agreement
on round-off among computer manufacturers, compiler designers, and
operating systems hides this problem by generating identical output
when the same initial realization is run using different hardware and
operating systems, so long as the same source code is
run on a single core, and compilers are not too different.

It is a common assumption among $N$-body practitioners that the
microscopic unpredictability and the consequential loss of
reproducibility is irrelevant so long as the global phase-space
characteristics are still representative of true physics. It
remains an article of faith that such a statistical validity
holds for the Newtonian N-body problem \citep{1991pscn.proc...47H}.
Chaos, however, leads to unpredictability due to temporal
discretization, round-off, and uncertainty in the initial realization
\citep{1964ApJ...140..250M}. Unpredictability due to chaotic behavior
and the consequential loss of reproducibility may then lead to incorrect
physical results.

After the pioneering work of \cite{1970A&A.....7..249H}, the problem
of characterizing chaos in self-gravitating $N$-body systems received
little attention until the late 1980s \citep[except possibly in][in
which the focus was on the computer's precision]{1979CeMec..20..209Z},
when computers became sufficiently powerful to address the problem for
larger $N$
\citep{1988ltdb.conf..329H,1991ApJ...374..255K,1992ApJ...386..635K,1992ApJ...399..627K,
1993ApJ...415..715G,1994JPhA...27.2879G,
1994ApJ...428..458K,1994ApJ...436L.111F}.
\cite{1993ApJ...415..715G} and \cite{1994ApJ...436L.111F} provided
excellent overviews of the underlying arguments for this chaotic
behavior in few- and many-body systems, respectively.

Even today, the chaotic $N$-body problem is hard to address
adequately using digital computers, and there is still no analytic
solution. We address three aspects of the problem here: 1) the
veracity of a solution for $N=4$ to $1024$, 2) the scaling of the
growth of errors to large $N \apgt 10^5$, and 3) the effect of general
relativity on the chaotic behavior of $N$-body systems. The terminology used in this text is explained in the glossary in \cite{2018CNSNS..61..160P}. Loosely
speaking, a reprehensive solution is a solution in which the errors
introduced during integration exceed the system size. For a
veracious solution, this is not the case. For $N=3$,
\cite{2015ComAC...2....2B} demonstrated that veracious $N$-body
solutions give statistically indistinguishable results as an ensemble
of converged solutions. They called this behavior {\em Nagh Hoch}, to
signify the importance of consistent statistical ensemble average
behavior of veracious solutions to the chaotic self-gravitating
$N$-body problem. It is not clear if this concept also holds for
larger $N$.

\cite{2020MNRAS.493.1913H} studied {\em Nagh Hoch} by running
ensembles of reprehensible numerical solutions for planetary systems.
They integrated a single $3\cdot 10^{-5}$ mass planet in a circular
orbit at a distance of 0.29 \cite[in dimensionless $N$-body
  units,][]{1971Ap&SS..13..284H} from the star, together with a test
particle in the same plane at apocenter and with an eccentricity of
0.19 and semimajor axis 0.21. The calculations were conducted for 200
$e$-foling timescales using four different algorithms for solving the
equations of motion. They found that for a sufficiently large sample
of initial realizations and a tolerably small time step, the results
of the various integrators are statistically indistinguishable. We
conclude from their simulation results that their adopted reprehensive
N-body algorithms for a system of two planets complies to {\em Nagh
  Hoch}. They argue that a relative integration error $\aplt 0.05$ is
sufficient to preserve the quality, consistent with
\cite{2015ComAC...2....2B}, who argued that a time step smaller than
$2^{-5}$ is sufficient to preserve ergodicity in the outcome space.

Ideally, one would like to perform large $N$-body simulations to a
converged solution, but this is unrealistic, even on modern digital
computers. We can achieve converged solutions for $N$ up to about $1$k
particles for several crossing times, leading to a series of veracious
solutions. At the moment, however, a longer evolution or a larger number of
particles are too costly. However, we demonstrate that the chaotic
behavior of these systems is reprehensive and confirm that they show
statistically indistinguishable chaotic behavior. We subsequently
perform reprehensible $N$-body simulations for up to $128$k particles.

The interest in scaling to $N\gg3$ is in part motivated by
understanding the chaotic nature of galaxies. Dense stellar systems,
such as globular clusters [$N = {\cal O}(10^6)$], are highly chaotic
\citep{1924BuAst...5...72P,1999CeMDA..73..159C}. Galaxies ($N
\rightarrow \infty$) are considered collisionless because their
relaxation time exceeds the Hubble time
\citep{2008gady.book.....B}. For sufficiently large $N$ the background
potential becomes smooth, and the collisionless assumption becomes
increasingly applicable
\citep{2004CeMDA..88..379M,2009CeMDA.105..379M}. However, due to the
point-particle granularity of the potential, the microscopic
exponential instability remains present in the system
\citep{2000chun.proc..229V}. Even large $N$-systems are therefore
affected by the chaos in small-$N$ subsystems. It is nonetheless
unclear how the microscopic exponential instability propagates to the
macroscopic structure of the stellar system as a whole. At which value
of $N$ does the system exhibit the transition from chaotic small-$N$
to smooth large-$N$ systems?

\cite{1988ltdb.conf..329H} argued that the dynamics $N$-body systems
under Newton's equations of motion are dominated by encounters at an
impact parameter of about $r/N^{1/2}$. Since chaos is driven by
encounters, the Lyapunov timescale then has a similar
scaling\footnote{Throughout this manuscript, we use the terms Lyapunov
  exponent and Lyapunov timescale for brevity, where we should write
  the largest positive global Lyapunov exponent (or timescale, for
  that matter).}. In a pioneering study, \cite{1993ApJ...415..715G}
discussed this scaling and found a transition in the chaotic behavior
around $N\simeq 32$. They argued that the Lyapunov timescale is
proportional to the dynamical crossing time $t_\lambda \propto \gamma
t_{cr}/\ln(\ln(N))$ over all values of $N$. For large $N$ ($ \apgt
32$), the constant $\gamma$ is smaller, leading to a weaker dependence
on the Lyapunov timescale \citep[see
  also][]{1992ApJ...386..635K,1998NYASA.867..320K}.
\cite{2002ApJ...580..606H} found a similar transition, but argued in
favor of scaling $t_\lambda \propto 1/\ln(N)$ over the entire range of
$N$. In this latter study, however, the perturbed particle is evolved
in a static background \citep{2002ApJ...580..606H}, and it is not
clear if their different scaling resulted from this particular
assumption or from the slightly different potential. We therefore extend
the range for which the Lyapunov time was determined to $N=128$k using
reprehensive $N$-body solutions, allowing us to test the scaling of
the Lyapunov timescale to large $N$ in an actual self-gravitating
system. In addition, we chose various initial density profiles: a smooth profile
(as in \cite{1993ApJ...415..715G}), a Plummer profile \cite[1911, as
  in][]{2002ApJ...580..606H} , and a King profile \citep[$W_o = 12$,][ just for
  fun]{1966AJ.....71...64K}.

The change in slope near $N\simeq 32$ pinpoints the transition
from chaotic behavior driven by local few-body relaxation (for $N\aplt
32$) to far-field many-body relaxation. This transition can globally
be understood by comparing the dynamical crossing time $t_{\rm cross}$
with the two-body relaxation time $t_{\rm rlx}$\, , which can be
approximated by
\citep{1971swng.conf..443S,1971ApJ...166..483S,1971ApJ...164..399S,1987degc.book.....S}
\begin{equation}
t_{\rm rlx} \simeq \left( {N \over 8 \ln(N)} \right) t_{\rm cross}.
\end{equation}
This relation indicates that for $N \apgt 32$, the relaxation
timescale exceeds the crossing time. It remains unclear, however,
which underlying process determines the slope.

\cite{2002ApJ...580..606H} argued in favor of the same scaling for all
$N$ and suggested that a galaxy with $10^{12}$ stars would have a
Lyapunov timescale of $t_\lambda \approx t_{cr}/30$ or $t_\lambda
\sim 8$\,Myr, whereas \cite{1993ApJ...415..715G} argued for a Lyapunov
timescale that is roughly an order of magnitude larger. Regardless
of either of the two scaling relations, the Galaxy would be subject to
chaotic motion on a fraction of the crossing timescale, and would therefore
not be representable by the collisionless Boltzmann equation
\citep{1872Bolzmann}.

For the Solar System, there seems to be a difference as well, this time, in terms of
chaotic behavior. In the Newtonian case, resonances between Jupiter
and Mercury greatly enhance the eccentricity of the orbit of the
latter planet. This may eventually lead to a collision between Mercury
and the Sun \citep{1992Icar...95..148L,1992Natur.357..569M}. However,
the inclusion of relativistic effect tends to stabilize the system,
resulting in the much-reduced probability that Mercury collides with the
Sun \citep{2009Natur.459..817L}. Based on our results, we tend to
agree with the increased stability of the Solar System when
considering relativistic mechanics. Interestingly, the global
Lyapunov timescale for the Solar System is not dissimilar from the
galactic result, being $\sim 5$\,Myr
\citep{1992Icar...95..148L} to $\apgt 6.8$\,Myr
\citep{1986AJ.....92..176A} or even slightly longer
\citep{1993ARA&A..31..265D,2015ApJ...799..120B}.

We argue that ensembles of reprehensive $N$-body solutions, from $N=4$
to $1024$ ($1k$), give statistically the same chaotic behavior as the
converged solutions. This trend holds at least up to $N=1k$, for
which we acquire converged solutions for $10$ H\'enon time units
(equivalent to $\text{four}$ crossing times). It becomes rather unpractical to
continue with converged solutions for $N \geq 1k$. 

Regardless of the large-$N$ behavior of the chaotic self-gravitating
systems under Newton's forces, we also study the relativistic case.
Rather than solving Einstein's field equations directly, we address
relativistic dynamics through an expansion to the gravitational field in
terms of $\Cscaling$, which expresses the speed of light ($c$) in
dimensionless $N$-body units in terms of the velocity of the particles
($v$). The zeroth order in this expansion represents Newton's equations
of motion, which does not depend on the velocity. The first-order
post-Newtonian term (the 1-PN) is proportional to $v^2/c^2$, describes
the motion of $N$ Schwarzschild black holes, and is known as the
Einstein-Infeld-Hoffmann (EIH) equation \citep{1938AMath.65..100E}.
\cite{LorentzDroste_KNAW1917} worked out a first generalization for
these post-Newtonian $N$-body equations of motion, but the final
formulation was realized by \cite{1938AMath.65..100E}.

Our study is motivated by the generally adopted view that the
radiation-induced dissipation in general relativity quenches the
chaotic behavior of the $N$-body system.  Spyrou \citep[1975, but see
  also,][]{2002PhDT.........8W,2003PhRvD..68b4004C,2011PhRvD..84j4038G,2014APS..APRM15008N}
\nocite{1975ApJ...197..725S}
demonstrated that the pairwise post-Newtonian expansion to 2.5th order
is chaotic in democratic three-body systems. They also studied the
gravitational-wave signal for such systems
\citep{2006ApJ...640..156G}. Chaotic behavior was not demonstrated for
relativistic systems with more than three particles.

We adopt the EIH equation and compare the degree of chaos measured in
systems from $N=3$ to $N=1k$ with the Newtonian solutions ($v/c
\rightarrow 0$). Our approach, however, is limited to first-order
post-Newtonian correction terms for the EIH equations of motion. The
1-PN terms are not dissipative and therefore are not expected to
result in less chaotic behavior when compared to Newton's equations of
motion.

The motivation for performing our a study is somewhat academic
because it is currently unclear to which degree large-$N$ relativistic
systems appear in nature or how frequently they emit gravitational
waves in an observable wavelength. Galactic nuclei are probably the
most promising places to find multiple supermassive black
holes orbited by intermediate-mass black holes or other compact
objects. \cite{2006ApJ...641..319P} argued that the Galactic center
would be populated by a steady population of a dozen intermediate-mass
black holes within a few milliparsec of the supermassive black hole.
In addition, there could be many stellar-mass black holes and a rich
population of neutron stars in the Galactic center
\citep{2004ApJ...613..326M,2015MNRAS.447.3096R}. These black holes
may merge, producing observable gravitational-wave signals
\citep{2005PhRvL..95l1101P,2017ApJ...848L..12A} and high-velocity
recoiling black holes \citep{2007PhRvL..98w1102C}. Similar, but
less extreme, situations may be present in the cores of some globular
clusters \citep{2010MNRAS.402..371B,2021MNRAS.503.3371B}.
\cite{2011ApJ...741L..12B} even speculated that clusters composed only of dark objects might exist, which could be copious sources of
gravitational radiation. Recently, \cite{2021NatAs.tmp..116G} argued
that the globular cluster Palomar 5 may host such a central collection
of compact objects and that in the coming 100\,Myr, its entire core may be
composed of black holes. With the current rather large virial
radius, the dynamics in this cluster is not expected to be subject to
strong relativistic effects, however.

It would be interesting to determine the
gravitational-wave signature of such chaotic systems and maybe even
search for them in the data collected by gravitational-wave
observatories. An analysis like this was done for hierarchical triples
\citep{2011PhRvD..83h4013G,2017ApJ...834..200M,
2018PhRvD..98f4012R,2020PhRvD.102f4033L, 2021PhRvD.103f3003W}, but
not for $N>3$.

\section{Methods}


In the gravitational many-body problem, $N$ objects move under the
attractive influence and space-time distortions of each other. We use
three independently developed implementations of the force-law and
integration method. One of these is the code \PhFour\, \citep[see
  Appendix\,\ref{sec:implementation},][]{2014DDA....4530301M}, which
is designed for the problem of $N$ point masses under a Newtonian
force law with regular hardware and compiler-supported precision (see
sec.\,\ref{Sect:Hermite}) For chaotic systems, one may desire more
control over the precision and accuracy of the integrator because
reprehensible $N$-body calculations provide insufficient trust due to
round-off and integration errors that grow exponentially with time. On
the other hand, veracious calculations are also reprehensible, but are
statistically indistinguishable from the converged solutions
\citep{2015ComAC...2....2B}. We therefore also perform calculations
using \Brutus\, \citep{2014ApJ...785L...3P}, which allows us to
control these parameters with a tolerance $\epsilon$ and word-length
$L_w$ (see\,\ref{Sect:Brutus}).

The relativistic calculations are performed up to the first post-Newton
order using the pairwise approximation and the full EIH
equations of motion. The relativistic $N$-body code is called
\HermiteGRX,\, and it is described in sect\,\ref{Sect:Hermite_GR}. The
equations of motion in \PhFour\, and \HermiteGRX\, are integrated
using the Hermite algorithm \citep{1991ApJ...369..200M} or an
adaptation of it to accommodate the velocity dependence of the
acceleration. In \Brutus,\, the equations of motion are solved with a
second-order \cite{PhysRev.159.98} scheme.

We finally use these three methods to study chaos in the large-$N$ limit. Large here means $\sim 1$k for the post-Newton case and
converged Newton solutions, and up to $N=128$k for reprehensible
Newtonian solutions. Each of the codes is interfaced as a community
code to the Astronomical Multipurpose Software Environment
\citep{2018araa.book.....P}. We describe these implementations in
Appendix\,\ref{sec:AMUSE}.

\subsection{Regular $N$-body calculations}\label{Sect:Hermite}

According to Newton's laws of motion, the acceleration $\vec{a}_i$ on
particle $i$ is given by the sum over all other particles
\citep{Newton:1687}:
\begin{equation}
\vec{a}_i = -\sum_{\substack{j=1 \\ j \ne i}}^{n} {Gm_j \over r_{ij}^3} \vec{r}_{ij}
\label{Eq:Newton}
.\end{equation}
Here $m_i$ is the mass of particle $i$, and $\vec{r}_{ij}$ is the relative
position vector from particle $i$ to $j$: $\vec{r}_{ij} \equiv
\vec{r}_i-\vec{r}_j$. Newton's constant is $G= {\cal
  O}(10^{-10}$\,N/kg$^2$/m$^2$), but in our calculations, $G \equiv 1$
\citep{1971Ap&SS..13..284H,1986LNP...267..233H}.

Integration was performed in 64-bits using IEEE 754 Standard for
floating-point arithmetic under the Linux operating system with kernel
version 5.8.0-48-generic. CPU calculations were performed on an 192
core Intel Xeon E7-8890 v4 workstation running at 2.20\,GHz.
Parallelization was realized using the Message Passing Interface
(version Open MPI 4.0.3) \citep{2008NewA...13..285P}. For simulations
with $N>1$k we adopted the GPU version of \PhFour, which uses the
\Sapporo\, GPU library \citep{2007NewA...12..641P,2009NewA...14..630G}
running on Xeon E-2176M CPU with Quadro P2000 Max-Q GPU. The GPU has
$4$\,GB GDDR5 RAM, but is not equipped with error correction, potentially leading to non-IEEE compliant errors in the calculations.

\subsection{Arbitrarily precise $N$-body calculations}\label{Sect:Brutus}

The Hermite algorithm is a fourth-order scheme that reaches a relative
energy error $dE/E \simeq 10^{-15}$ with a time-step parameter $\eta
\simeq 10^{-4}$ for a distribution of equal-mass objects in a
virialized homogeneous distribution in space.  All our calculations
were performed with $\eta = 0.01$.  Round-off and time-step errors
become important for smaller time steps when integrating for more than
$\sim 10$ crossing times, resulting in a systematic growth of the
energy error. This growth scales $\propto \eta^2$. Although small,
typically about $1/10^{16}$ (in relative coordinates), these
errors drive the eventual
irreproducibility of the simulations  through exponential growth. This irreproducibility is
undesirable when one is interested in studying chaotic motion. We
therefore also performed calculations with \Brutus\,
\citep{2015ComAC...2....2B}, an $N$-body code that allows us to
integrate any $N$-body system to arbitrary precision. In \Brutus\, we
control the different sources of error by adopting the
Gragg-Bulirsch--Stoer algorithm
\citep{springerlink:10.1007/BF01386092, 1965SJNA....2..384G}. In this
algorithm, one performs a single step using a Verlet integrator
\citep{PhysRev.159.98}, then repeats that same step using half the
step size. Now the relative error between the two solutions can be
determined by taking the absolute value of the differences in each
coordinate. If this error is smaller than some predetermined
tolerance, the result is accepted, and the next step is
calculated. Otherwise, the same step is repeated with a quarter time
step. This procedure is repeated until the relative error between two
subsequent solutions is smaller than the tolerance
\cite[see][]{1992nrca.book.....P}.

We controlled the discretization error with arbitrary-precision
arithmetic, using the {\tt GMP}
\citep{Granlund12,Granlund:2015:GMM:2911024} and {\tt MPFR} libraries
\citep{10.1145/1236463.1236468} instead of conventional double
precision. This allowed us to control the round-off error by changing
the number of digits, which we express in a word-length $L_w$. A
word-length $L_w = 64$ bits then corresponds approximately to the
usual 16-decimal place precision in standard IEEE~754 floating-point
operations on current regular microprocessors.

In practice, we only specified the tolerance and calculated the
word length $L_w \in \mathbb{Z}$ with \citep{2015ComAC...2....2B},
\begin{equation}
L_w = \textrm{int}(32 - 4 \log_{10}(\epsilon)).
\end{equation}
A converged solution to $n$ decimal places is achieved by iteratively
repeating a calculation that started with one selected realization of
the initial conditions with lower tolerance for each subsequent
calculation. This process is repeated until the first $n$ decimal
places of the final phase-space coordinates of two subsequent
iterations lead to identical values for the first $n$ digits in the
positions and velocities of all particles. When the numerical
solution has achieved this state of convergence, it is deemed to be
definitive \citep{2018CNSNS..61..160P}.

Calculation typically started with tolerance $\epsilon = 10^{-5}$
($L_w = 52$) for $N\leq64,$ reducing the tolerance by a factor $10^5$
upon subsequent calculations. For high values of $N,$ we started with
$\epsilon=10^{-20}$ ($L_w = 112$), reducing the tolerance by a factor
$10^{10}$ upon subsequent calculations. With $\epsilon = 10^{-40}$
($L_w = 192$), all solutions up to $N=1$k have converged to $n=3$
decimal places (the adopted convergence limit).

Calculation time with \Brutus\, scales $\propto N^2$, but due to the
expense of the large mantissa calculations, the offset in computer
time is long. In addition, several recalculations may be needed
before a converged solution is achieved. In
figure\,\ref{fig:scaling_with_N} we present the scaling of the
calculations with \Brutus\, (ochre symbols show the mean of the
computing time for that particular $N$). The lines to higher values
indicate the upper limit for a single most expensive calculation for
each value of $N$. These still scale roughly proportional to $N^2$,
but are often $\sim 100$ times more costly than the average.

\subsection{Einstein-Infeld-Hoffmann solver}\label{Sect:Hermite_GR}

Between 1907 and 1915, Einstein developed general relativity
\citep[see][for an interesting read on the history of this
  development]{2012arXiv1202.2791W,2014grav.book.....P} and viewed
gravity as the result of the curvature of space-time
\citep{1914ZMP....63..215E,1915SPAW.......778E,1915SPAW.......799E,Hilbert:1915tx}. The
Einstein field equations dictate the gravitational interaction between
particles, but these equations are nonlinear and notoriously hard to
solve. \cite{1916AbhKP1916..189S} found the first nontrivial solution
to the Einstein field equations: the Schwarzschild metric describes a
point-like particle. A rotating black hole was first described
analytically as a solution to the field equations by
\cite{1963PhRvL..11..237K}. Today, there are rather standard software
implementations to solve for general relativistic dynamical systems
\citep{2018PhRvD..97h4059M,maria_babiuc_hamilton_2019_3522086}, even
including magnetic fields \citep{2020PhRvD.101j4007M}. However,
simulating multiple black holes in a relativistic context is somewhat
expensive in terms of computer time.

A numerically cheaper solution is the Einstein-Infeld-Hoffmann
equations
\citep{2014PhRvD..89d4043W,2014grav.book.....P,2021PhRvD.103f3003W},
in which the acceleration of body $i$, $\vec{a}_{i}$ is given by
\begin{eqnarray}
  \nonumber \vect{a}_i &= &-\sum_{j\not=i} \frac{G m_j}{r_{ij}^3}
  \vect{r}_{ij}
\nonumber \\
&&
+ \frac{1}{c^2} \sum_{j\not=i} \frac{G
    m_j}{r_{ij}^3} \vect{r}_{ij} \left[ 4\frac{G m_j}{r_{ij}}+5\frac{G
      m_i}{r_{ij}} + \sum_{k\not=i,j} \frac{G m_k}{r_{jk}} \right.
\nonumber \\
&& \qquad \qquad \qquad 
\left.+ 4 \sum_{k\not=i,j} \frac{G m_k}{r_{ik}} - \frac12 \sum_{k\not=i,j} \frac{G m_k}{r_{jk}^3}(\vect{r}_{ij} \cdot \vect{r}_{jk})
\right.
\nonumber \\
&& \qquad \qquad \qquad 
\left.
- v_i^2 + 4\vect{v}_i \cdot \vect{v}_j - 2 v_j^2 + \frac32 (\vect{v}_j \cdot \vect{n}_{ij})^2 \right]
\nonumber \\
&&
- \frac{7}{2c^2} \sum_{j\not=i} \frac{G m_j}{r_{ij}} \sum_{k\not=i,j} \frac{G m_k}{r_{jk}^3} \vect{r}_{jk}
\nonumber \\
&& 
+ \frac{1}{c^2} \sum_{j\not=i} \frac{G m_j}{r_{ij}^3} \vect{r}_{ij} \cdot (4\vect{v}_i - 3 \vect{v}_j) \vect{v}_{ij}.
\label{eq:EIHequationsofmotion}
\end{eqnarray}

Here $\vect{n}_{ij} = \vec{r}_{ij}/r_{ij}$,
$\vec{v}_{ij} \equiv \vec{v}_i - \vec{v}_j$, and
$\vec{\hat{r}}_{ij} = \vec{r}_{ij}/|{\vec{r}_{ij}}|$ is the unit vector
along $\vec{r}_{ij}$.
To conserve energy, an addition term has to be introduced that depends on
the 1\,PN approximation, 
\begin{eqnarray}
E &=& \frac{1}{2} \sum_{i} m_i \left ( v_i^2 - \sum_{j \ne i} \frac{Gm_j}{r_{ij}} \right )
\nonumber \\
&& 
+ \frac{1}{c^2} \sum_{i} m_i \biggl [ \frac{3}{8} v_i^4 + \frac{3}{2} v_i^2 \sum_{j \ne i} \frac{Gm_j}{r_{ij}} 
  \nonumber \\
&& \qquad \qquad 
+ \frac{1}{2} \sum_{j \ne i} \sum_{k \ne i} \frac{G^2 m_j m_k}{r_{ij} r_{ik}} 
\nonumber \\
&& \qquad \qquad 
- \frac{1}{4} \sum_{j \ne i} \frac{Gm_j}{r_{ij}} \left ( 7 \vec{v}_i \cdot \vec{v}_j
+ (\vec{v}_i \cdot \vec{\hat{r}}_{ij})(\vec{v}_j \cdot \vec{\hat{r}}_{ij}) \right ) 
\biggr ]
\,,
\label{eq:energy}
\\
\vec{P} &=& \sum_{i} m_i \vec{v}_i 
+ \frac{1}{2c^2} \sum_{i} m_i \vec{v}_i 
\left ( v_i^2 - \sum_{j \ne i} \frac{Gm_j}{r_{ij}} \right )
\nonumber \\
&& 
- \frac{G}{2c^2} \sum_i \sum_{j \ne i} \frac{m_i m_j}{r_{ij}} (\vec{v}_i \cdot \vec{\hat{r}}_{ij}) \vec{\hat{r}}_{ij} \,.
\label{momentum}
\end{eqnarray}

Eq.\,\ref{eq:EIHequationsofmotion} gives the full EIH equations of
motion. The first term in Eq.\,\ref{eq:EIHequationsofmotion}
(zeroth-order term in the Taylor expansion) is identical to
Eq.\,\ref{Eq:Newton} and represents Newton's acceleration. The other
terms reflect post-Newtonian corrections. Several of them depend on
velocity, and the penultimate term contains the accelerations of
the other particles, making it expensive to compute.

Eq.\,\ref{eq:EIHequationsofmotion} is first order, but higher-order
post-Newtonian corrections exist, although only under the assumption
of pairwise interactions. The three-body Hamiltonian, and therefore
the corresponding equations of motion, are known in closed form to
second post-Newtonian order ${\cal O}(v^4/c^4)$ \citep{SCHAFER1987336,
2008CQGra..25s5019L}, and the two-body equations of motion up to
$3.5$\,PN order, or ${\cal O}(v^7/c^7)$
\citep{2007LRR....10....2F,2009PhRvD..80l4003I}.

\begin{figure}
\centering
\includegraphics[width=\columnwidth]{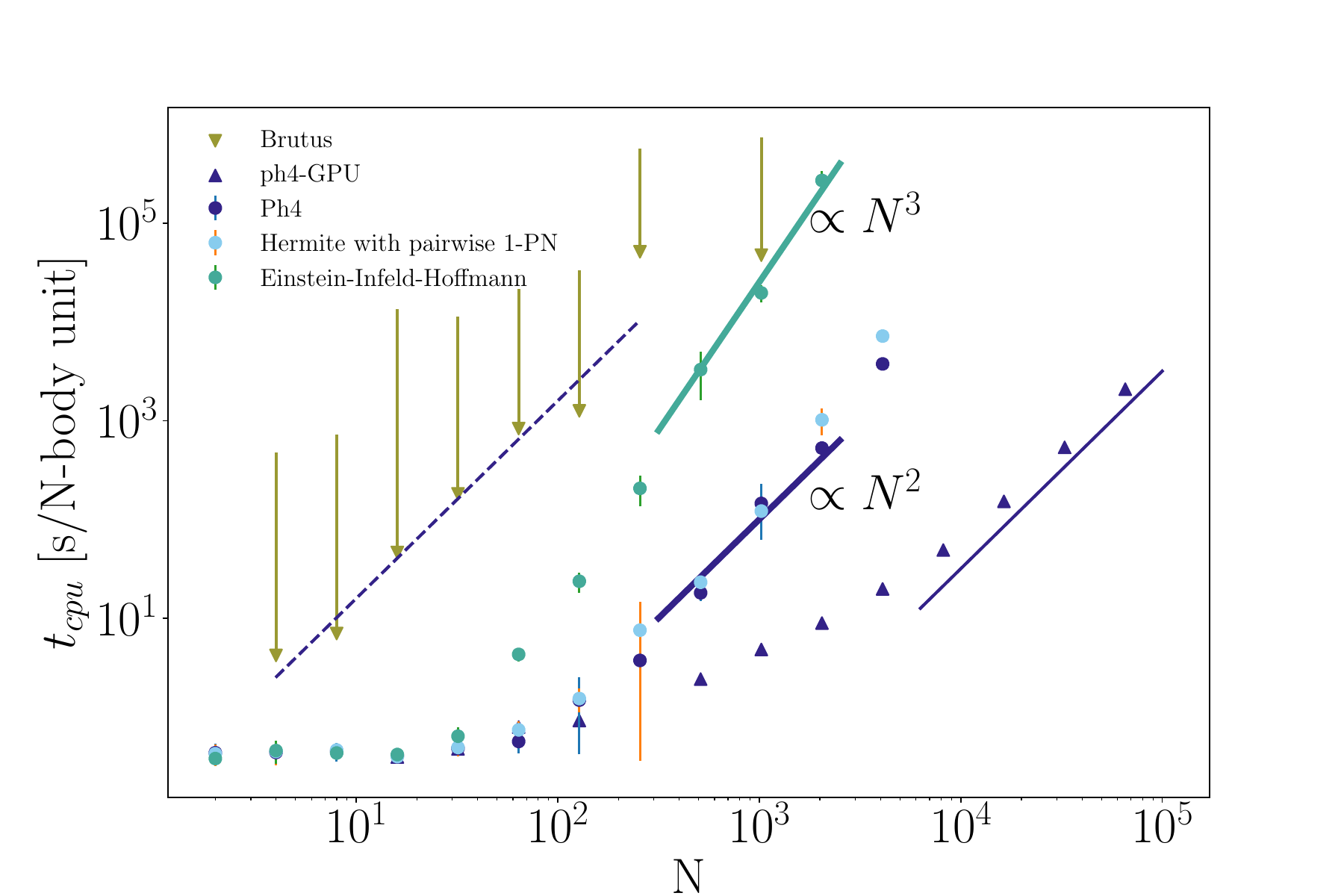}
\caption{Scaling of the various integration methods as a function of
the number of particles (N). The classic Newton integrations scale
as $\propto N^2$, as indicated with the solid and dashed dark blue
lines. The pairwise first expansion scales similarly, but tends to be
slightly slower than the pure Newton expansion. The Einstein-Infeld Hoffmann
equations to first order scale $\propto N^3$, making large
calculations that include the cross-terms unpractical. The scaling
presented for Brutus is based on the calculations presented here.
The bullet points indicate the mean timescale for acquiring a
converged solution, and the line pointed upward ends at the
single most expensive calculation in our sample of simulations for
that particular $N$. For \PhFour,\, we included the regular
implementation as well as the GPU-enabled version (to the right),
running on an Intel Xeon CPU E5620 operating at 2.40GHz and NVIDIA
G96 (Quadro FX580), running on a generic 64-bit Ubuntu Linux
kernel 2.6.35-32. }
\label{fig:scaling_with_N}
\end{figure}

Due to the summations over pairs of particles in
Eq.\,\ref{eq:EIHequationsofmotion}, the motion of one particle due to
a second particle depends on the other particles in the system. As a
consequence, the EIH equations of motion scale as ${\cal O}(N^3)$,
rather than the usual scaling to ${\cal O}(N^2)$ for Newton's
case. This scaling is confirmed in figure\,\ref{fig:scaling_with_N}.

\begin{figure}
\centering
\includegraphics[width=\columnwidth]{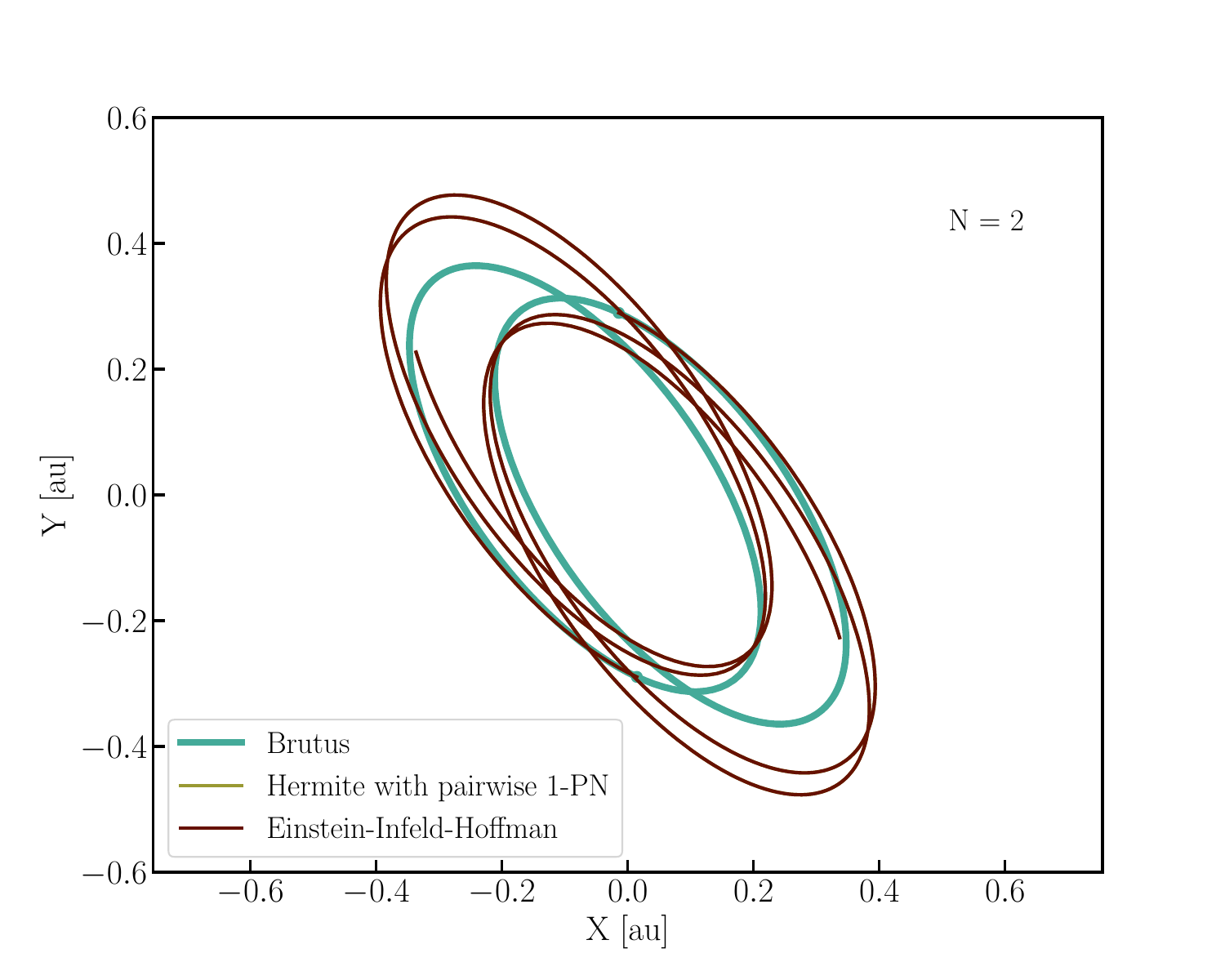}
\caption{Orbital evolution of two black holes of masses
$10^6$\,\MSun\, with an initial separation of $0.819$\,au integrated
for half a day for three integrations, pure Newton expansion, Newton with
expansions to first order, and the full first-order
Einstein-Infeld-Hoffmann equations. The first two are precisely on
top of each other, and we plotted the first-order pairwise solution
last. The last two solutions are identical because the cross-terms
do not lead to deviations from the first-order expansions. The
post-Newton orbits are not closed, as in the Newtonian case
(green). No separate scaling of $v/c$ is applied here because the
system is initialized in physical units.}
\label{fig:N2_BlackHole_orbits}
\end{figure}

\begin{figure}
\centering
\includegraphics[width=\columnwidth]{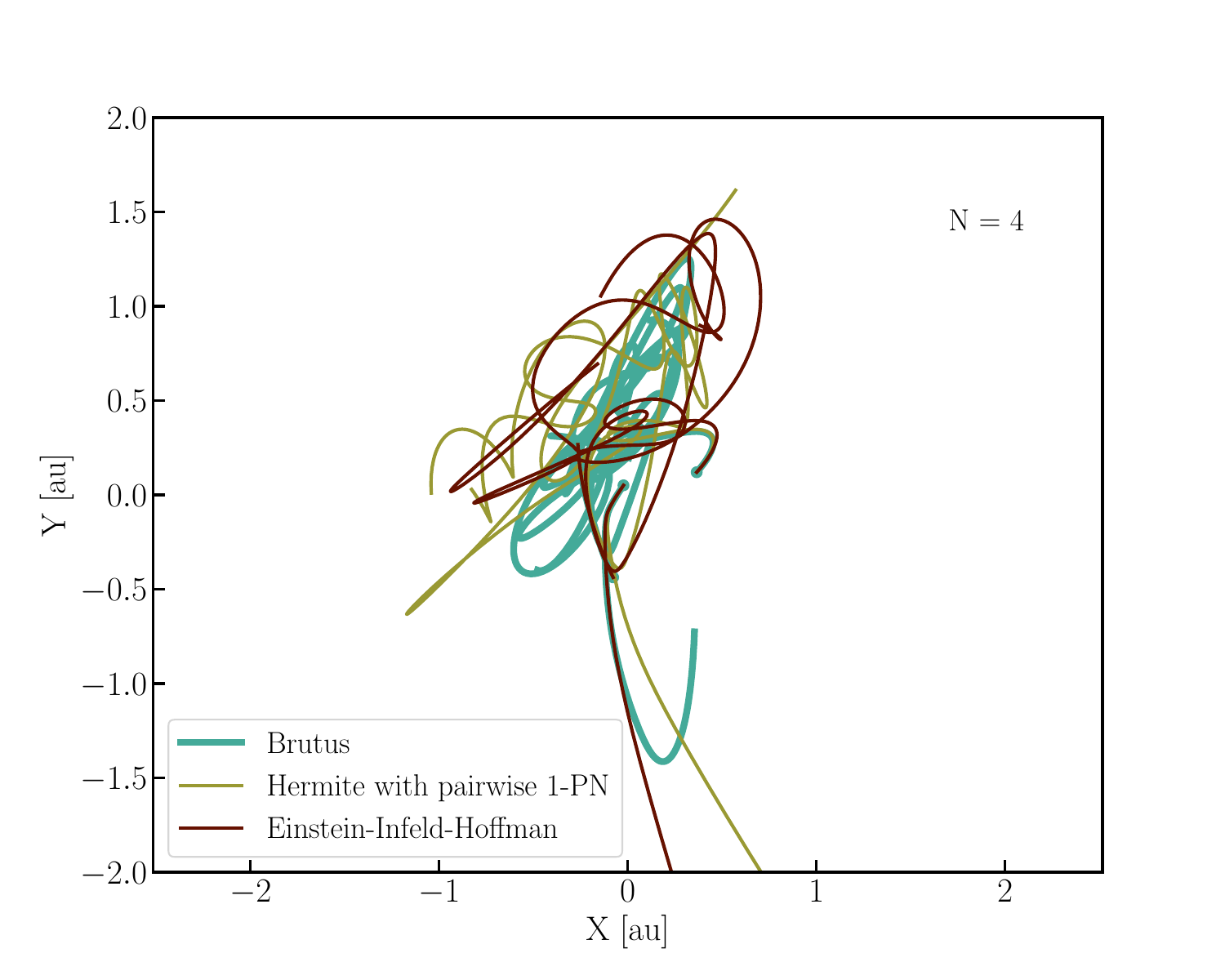}
\caption{Orbital evolution of four of $10^6$\,\MSun\, black holes with
the same integrators as in fig.\,\ref{fig:N2_BlackHole_orbits}. }
\label{fig:N4_BlackHole_orbits}
\end{figure}

\begin{figure}
\centering
\includegraphics[width=\columnwidth]{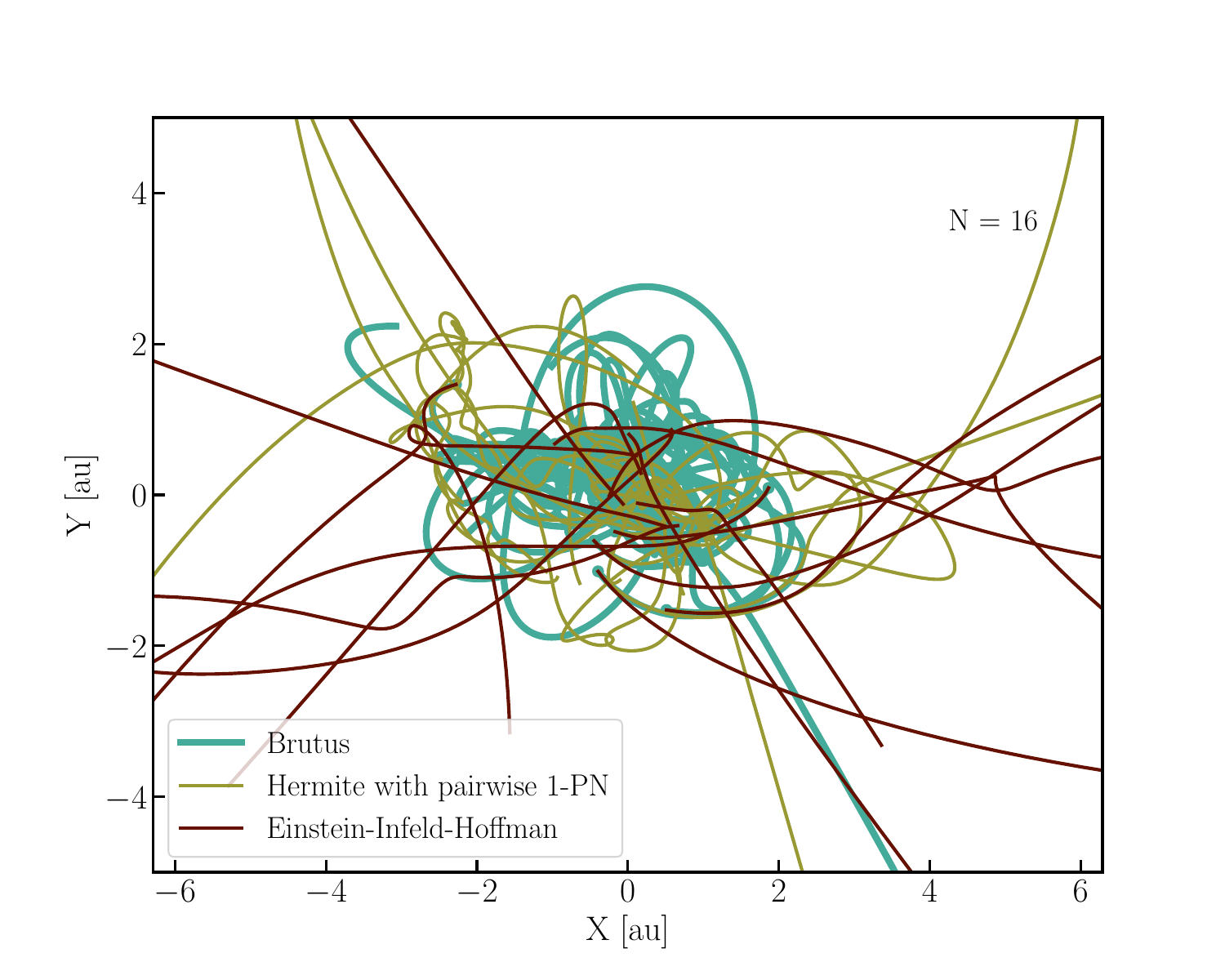}
\caption{Orbital evolution of 16 of $10^6$\,\MSun\, black holes with
the same integrators as in fig.\,\ref{fig:N2_BlackHole_orbits}. }
\label{fig:N16_BlackHole_orbits}
\end{figure}

We implemented the pairwise and the full EIH equations of motion to
1-PN order using a fourth-order Hermite predictor-corrector scheme (see
sect.\,\ref{Sect:Hermite}). We refer to \HermiteGRP\, as the pairwise
equations of motion, and to Hermite-GRX for the EIH solution to 1-PN
order. To illustrate the working of the various implementations, we
present figs.\,\ref{fig:N2_BlackHole_orbits},
\ref{fig:N4_BlackHole_orbits}, and \ref{fig:N16_BlackHole_orbits} for
orbits of the $N=2$, $N=4,$ and $N=16$ Newtonian case, 1-PN pairwise
equations of motion, and for the full EIH equations of motion to first
order. For these simulations, we adopted black hole masses of
$10^6$\,\MSun, in a $1$\,pc cube. Equivalent to specifying the mass
of the system, we can also use the relative speed of light
$\Cscaling$. In H\'enon units, in which $G=M=1$, Newton's kinetic
energy of a system of $N$ bodies with total mass $M$ is $E_{\rm kin} =
0.5M\sigma^2 = 1/4$, with a velocity dispersion $\sigma^2 = 1/2$
\citep{1986LNP...267..233H}.

All the initial conditions in this study are virialized according to
Newton's equations of motion (see
sect.\,\ref{Sect:initialconditions}), and therefore in H\'enon units,
the scaled velocity $v = 1/\sqrt{2}$, which sets the scaling of our
$N$-body simulations. We specify the relative scaling with respect to
mass or size by changing the speed of light, or $v/c$. In the
numerical implementation, this parameter is specified through the
parameter $\zeta$, which is the reciprocal of $v/c$ (see
sect.\,\ref{sect:speedoflight}). For clarity, in the main paper, we
only use $v/c$ as free parameter. 

\subsection{Initial conditions}\label{Sect:initialconditions}


Initial conditions were generated in standard IEEE double precision
($L_w=64$). This introduces a discrepancy with the low-$N$ $(\leq 64)$
experiments, for which the initial iteration was performed with $L_w =
52$, but if convergence is already achieved for $\epsilon=10^{-10}$
($L_w=72$), round-off at the 13th mantissa (the limiting precision for
$L_w = 52$) cannot have propagated to the first 3 decimal places.

We adopted the initial conditions from \cite{1993ApJ...415..715G}.
All objects then have the same mass and are distributed in a unit cube
in phase space (position and velocity) using dimensionless $N$-body
units. After generating random positions and velocities, the system
was moved to the center-of-mass frame and scaled to virial equilibrium
for the Newtonian solution.  The simulation runs with a
finite speed of light also used the identical initial realizations
as for the Newtonian case.  The slight deviations from virial
equilibrium in the relativistic initial conditions have no effect on
our results because we started measuring the phase-space distance at
$t=1$, and by that time, the system was well virialized.  We validated
and confirmed this statement by recalculating all simulations for
$N=16$ with $v/c = 0.01$, for which we scaled the initial
conditions to virial equilibrium for $v/c=0.01$ by adapting masses and
velocities according to \cite{1964ApJ...140.1512B}, however. The difference
between the Newtonian and the relativistically virialized initial
conditions was negligible.

The number of runs performed varied for each code and the number of
particles, as listed in Table\,\ref{tab:number_of_runs}.  In the
second column, we list the number of runs performed with \PhFour. For
the other runs, with \HermiteGRX\, and \Brutus,\, the same initial
realizations were adopted, but sometimes this was a subset.  These
calculations were performed up to $N=1$k using the same number of runs
with the same initial conditions as in \cite{1993ApJ...415..715G}.

\begin{table}
\begin{tabular}{lrrr}
\hline
$N$ & \multicolumn{3}{|l|}{$N_{\rm runs}$} \\
& \PhFour & \HermiteGRX & \Brutus \\
\hline
3 & 100 & --- & --- \\
4 & 200 &200 & 200 \\
8--64 & 100 &100 & 100 \\
128 & 20 & 10 & 10 \\
256 & 10 & 10 & 3 \\
512 & 10 & 2 & 2 \\
1024 & 10 & -- & 2 \\
2048--16384 & 10 & -- & -- \\
65536 & 6 & -- & -- \\
131072 & 2 & -- & -- \\
\hline
\end{tabular}
\caption{Number of simulations performed per implementation and number
of particles. The simulations for different codes use exactly the
same initial realizations: The 200 initial realizations for $N=4$
are identical for the \PhFour, \Brutus,  and \HermiteGRX.}
\label{tab:number_of_runs}
\end{table} 

We performed an additional series of simulations using \PhFour\, and
\HermiteGRX, but with realizations generated using a
\cite{1911MNRAS..71..460P} sphere and a King model \citep[$W_o =
  12$,][]{1966AJ.....71...64K}.

The main reason not to perform larger simulations including the
EIH equations is their unfavorite scaling of the computer time with
$N$, which we depict in fig.\,\ref{fig:scaling_with_N}. We estimate
approximately $\text{one}$ year of integration for $N\sim 10^4$ with the
current CPU implementation.

\subsection{Measuring the Lyapunov timescale}\label{Sect:Ly_timescale}

Measuring the Lyapunov timescale for the gravitational $N$-body
system is not trivial. Several methods for deriving this quality have
been proposed. One method uses the geodesic-deviation vector-technique
\citep{1972gcpa.book.....W,2003PhLA..312..175N} for two nearby orbits
with projection operations and with time as an independent variable
\citep{2003PhLA..313...77W}, and the two-nearby realizations without
projection operations and with time as an independent variable. We
adopted the last, which may be more expensive to calculate, but
is considerably simpler for large $N$, and it is least affected by
underlying assumptions. This same technique was adopted in
\cite{1993ApJ...415..715G}, which means that our analysis at least starts from
the same assumptions. Strictly speaking, the Lyapunov timescale is
defined properly from some starting point until the system dissolves
\citep{2009MNRAS.392.1051U,2013ARep...57..429M}. Because this
definition is rather unpractical, particularly for large $N$, we stopped
the calculations at $10 \text{ } N$-body time units \citep[equivalent
to][]{1993ApJ...415..715G}.

The degree of chaos in the simulation was measured using the evolution
of the phase-space distance between two almost identical initial
realizations (see \S\,\ref{Sect:initialconditions}). The second
realization was constructed by increasing the Cartesian $x$ coordinate
of a randomly selected particle with a value of $10^{-7}$ (in
dimensionless $N$-body units). Just to emphasize, this initial
displacement is 10 million times shorter than the size of the initial
extent of the $N$-body system. The perturbed realization is therefore
not in strict equilibrium, but deviates from Newton's equilibrium
potential energy by ${\cal O}(10^{-7}/N^2)$.

We integrated both initial realizations to $10\text{ } N$-body time units
while saving a snapshot every $0.1 N$-body time units,
resulting in 100 snapshots per run. The phase-space distance was
determined by taking the difference in position and velocity between
the same particles in each snapshot, and summing them,
\begin{equation}
\ln (\delta) = {1\over2} \ln \left[ \sum (\vec{r}_{b}-\vec{r}_{a})^2 + (\vec{v}_{b}-\vec{v}_{a})^2 \right].
\end{equation}
This leads to a phase-space distance as a function of time. To
calculate the Lyapunov exponent, we only used the data from $t=1$ to
a maximum of either $t=10$, or the first moment in which the phase-space
distance exceeded $0.1$.  The choice of starting the Lyapunov
timescale measurements at $t=1$ guarantees that the system is in
virial equilibrium even in the most relativistic cases.
We subsequently performed a least-squares fit to
the phase-space distance evolution. The fitted slope (in $\log$ space)
to this phase-space distance evolution gives the Lyapunov exponent.
The Lyapunov timescale $t_\lambda$ is the reciprocal of the Lyapunov
exponent.

This procedure is slightly different than what was used in
\cite{1993ApJ...415..715G}, who adopted $t_\lambda = 9
/(\ln(\delta_{t=10}) - \ln(\delta_{t=1}))$, but results in a better
estimate of the global Lyapunov timescale. We stopped our measurement
when $\delta \geq 0.1$ because due to conservation of the phase-space
characteristics, the system then grows on a relaxation timescale,
rather than on a Lyapunov timescale, and $\delta$ saturates when it
becomes on the order of unity \citep{2002JSP...109.1017H}. For very chaotic
systems, the procedure adopted by \cite{1993ApJ...415..715G} leads to
an underestimate of the Lyapunov exponent and therefore to an
overestimate of $t_\lambda$, as is the case for $N\apgt 1$k King
models (see, e.g., in
fig.\,\ref{fig:KingW12_Lyapunov_timescale}).


\section{Results}

\subsection{Chaos in large-$N$ Newtonian systems}

\begin{figure}
\centering
\includegraphics[width=1.0\columnwidth]{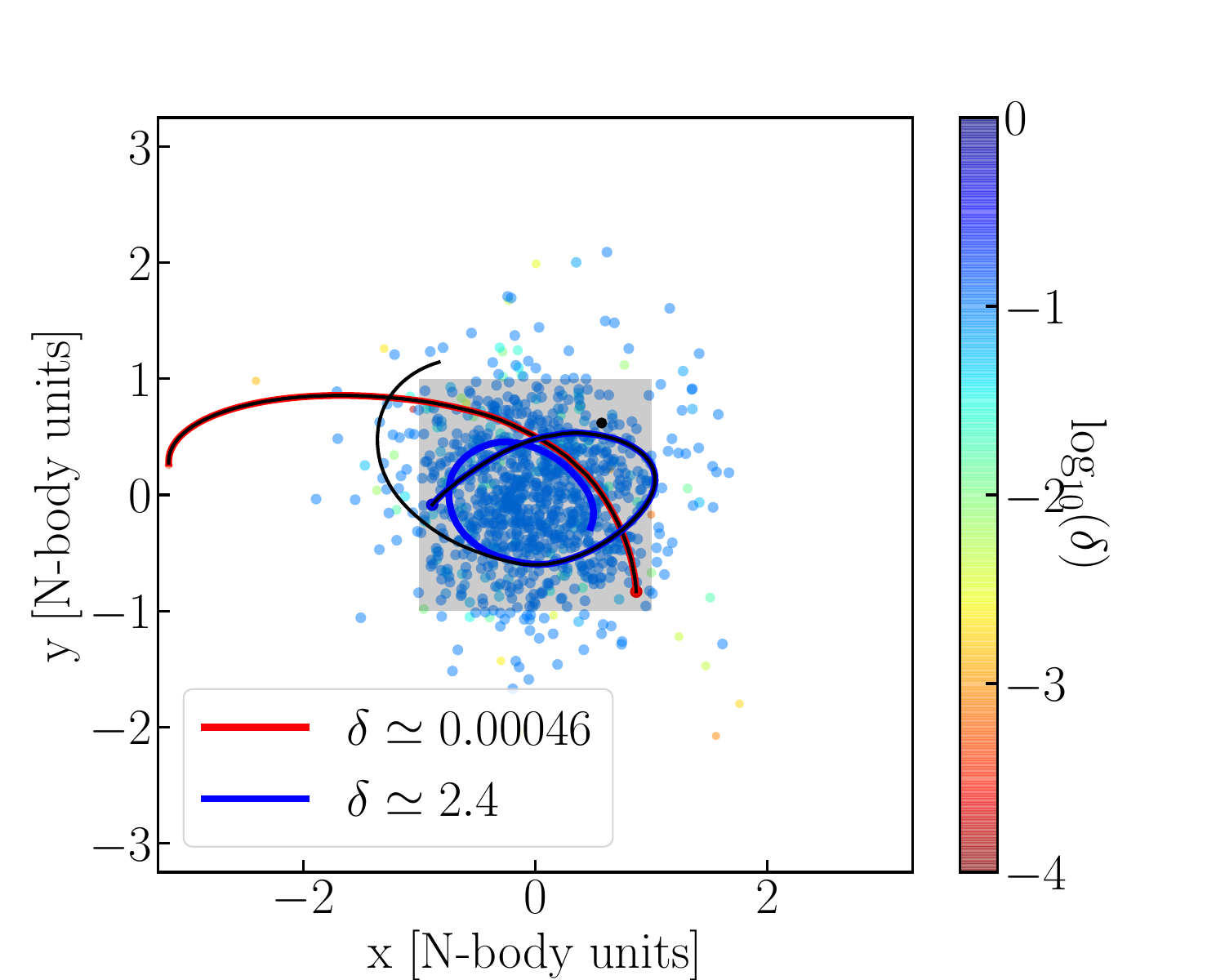}
\caption{Distribution of phase-space distances in
a cluster with 1024 particles. Units are dimensionless N-body units
\citep{1971Ap&SS..13..284H}. The gray shaded region indicates the
initial conditions in a virialized unit cube. One particle,
indicated with the black bullet point, is displaced by $10^{-7}$
along the Cartesian $x$ coordinate. The final conditions (at $t=10$)
of the unperturbed particles are represented with the bullet
points. The color and size of the points represents the phase-space
distance measured over the duration of the simulation ($10 \text{ }N$-body
time units), and ranges over $ \text{about four}$ orders of magnitude. The
majority of objects experience considerable change in their orbits,
but some are hardly perturbed. Calculations were performed using
\Brutus\, until convergence to 3 decimal places, which requires a
tolerance of $\tau = 10^{-40}$.}
\label{fig:Cluster_example}
\end{figure}

In fig.\,\ref{fig:Cluster_example} we show an example for an $1$k-body
system, starting with the initial conditions of
\cite{1993ApJ...415..715G}. The gray square in the middle represents these initial conditions; particles, according to
\cite{1993ApJ...415..715G}, are initialized in a unit cube. One
calculation (bullet points) gives the result of the unperturbed
solution. The perturbed solution is not shown, but the colors of the
particles give the phase-space distance between the final perturbed
and unperturbed solutions. The black bullet point toward the
top right corner of the gray area identifies the (randomly selected)
particle for which the initial $x$-coordinate was increased by
$10^{-7}$. The least (red) and most (blue) chaotic particles are
represented as lines. The overplotted thin black curves show the
orbit of the perturbed solution.

\begin{figure}
\centering
\includegraphics[width=\columnwidth]{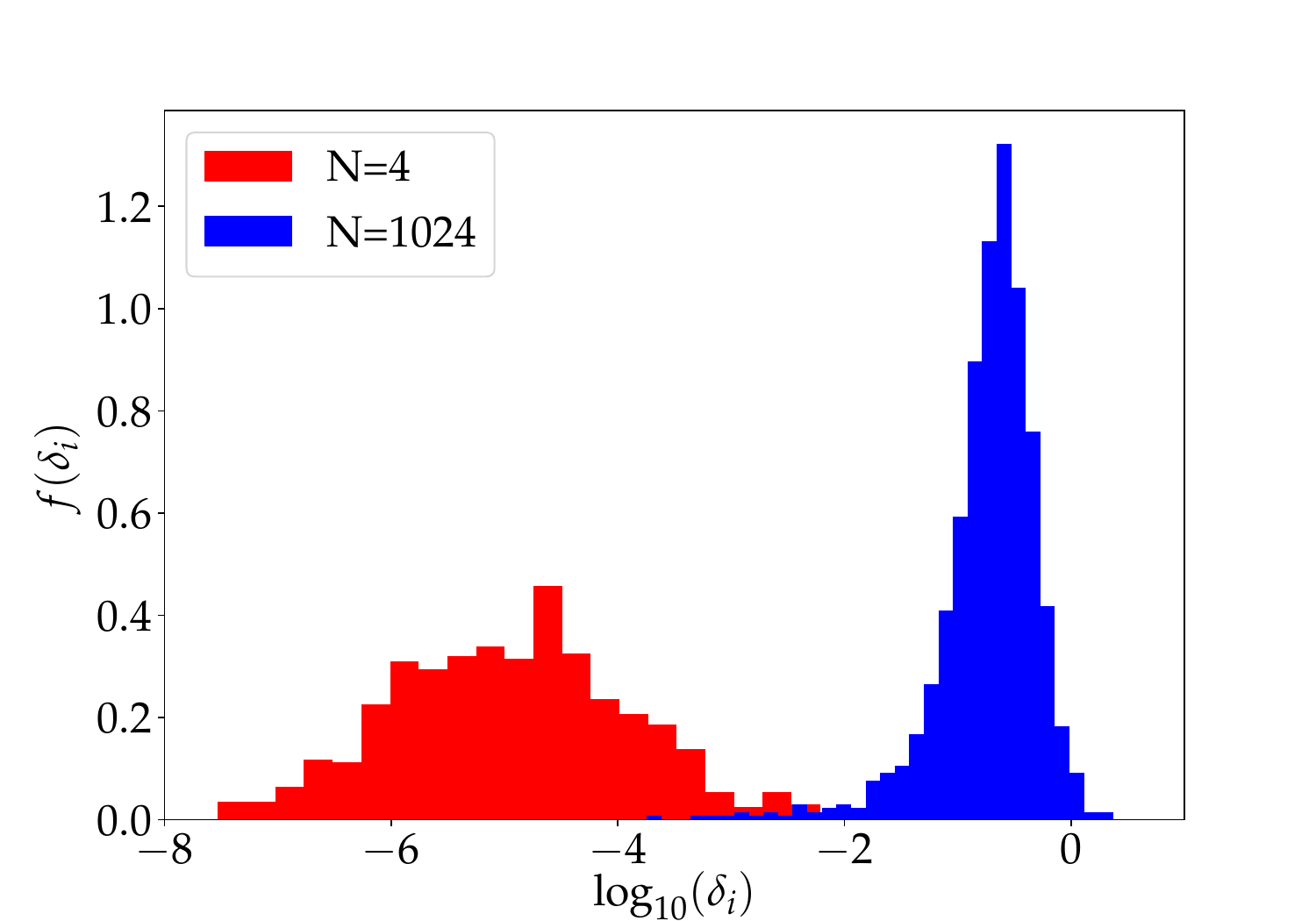}
\caption{Distribution of phase-space distances for individual
  particles $\delta_i$ in the simulations with $N=4$ (red) and those
  with $N=1024$ (blue) after integrating for $t=10$~$N$-body time
  units.  The data for $N=4$ are the result of 200 runs. For $N=1024,$
  we adopted the run used in
  fig\,\ref{fig:Cluster_example}. Calculations were performed using
  \Brutus\, until the solution was converged. }
\label{fig:hist_delta_N1024t10_Brutus}
\end{figure}

Figure\,\ref{fig:Cluster_example} illustrates Miller's (1964)
\nocite{1964ApJ...140..250M} point that a small perturbation in a
single object leads to large variations in the final phase-space
distribution. In fact, most objects experience a strong variation,
whereas only a minority of objects are hardly affected. In
fig.\,\ref{fig:hist_delta_N1024t10_Brutus} we plot the distribution
of phase-space distances ($\log_{10}(\delta)$) for the calculation of
fig.\,\ref{fig:Cluster_example}.

The degree to which particles are affected by a small initial
perturbation depends on the number of particles in the system. This is
illustrated in fig.\,\ref{fig:hist_delta_N1024t10_Brutus}, where we
show the distribution of phase-space distance between a perturbed and
an unperturbed solution for the same simulation as in
fig.\,\ref{fig:Cluster_example} using $1$k particles, and compare this
distribution with $200$ simulations of $N=4$. The small-$N$ systems
(red histogram) exhibits a much weaker response to a perturbing
particle than the large-$N$ systems (blue); few-body systems are less
chaotic than large-$N$ systems (at least for this selection of initial
conditions, and under Newton's forces).

\begin{figure}
\centering
\includegraphics[width=1.0\columnwidth]{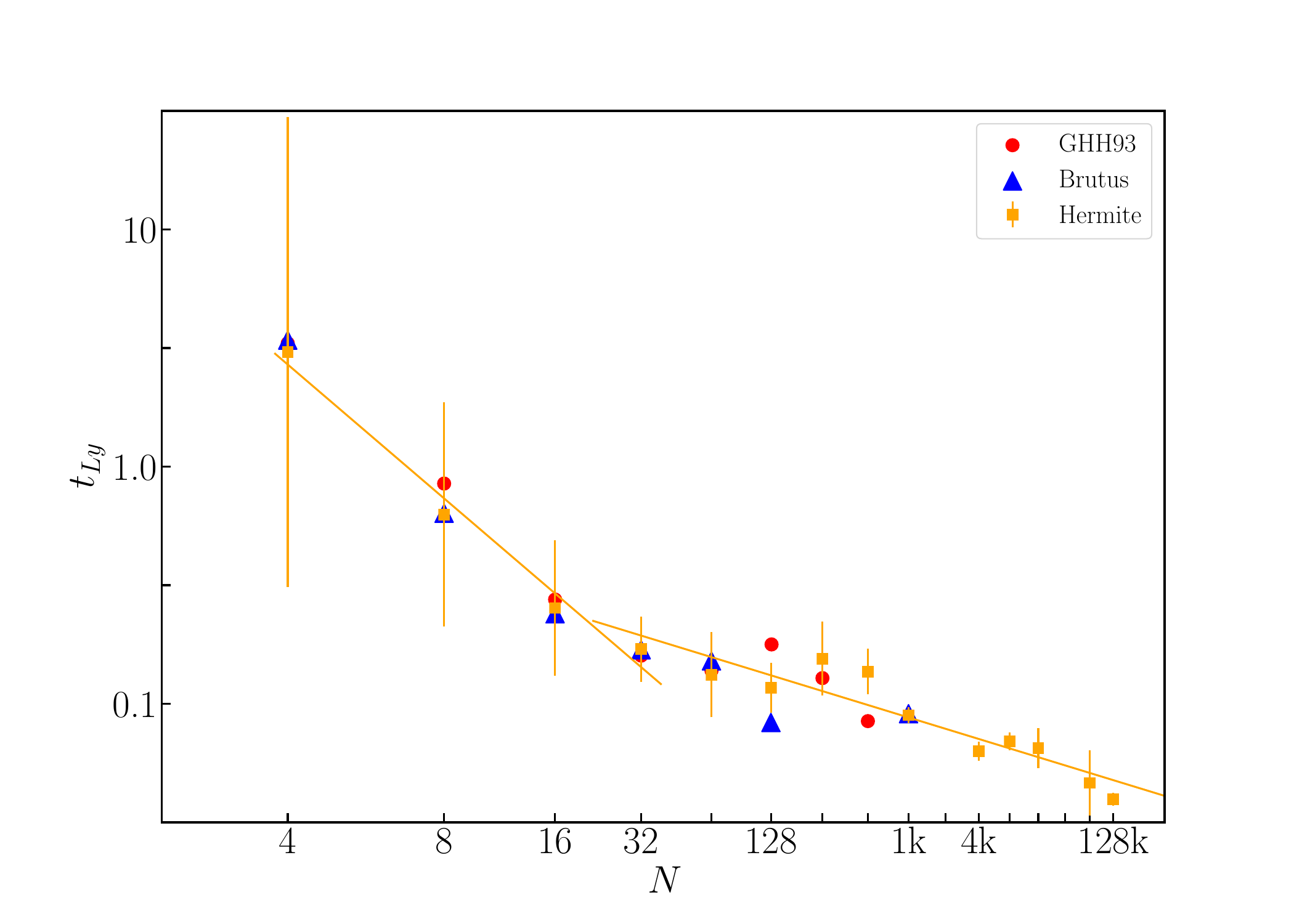}
\caption{Estimate of the Lyapunov timescale as a function of the
  number of particles. Here the horizontal axis is not linear, but in
  $\ln(\ln(N))$ to illustrate the scaling proposed in
  \cite{1993ApJ...415..715G}. The different symbols and colors
  represent different calculations (see legend). The vertical bars,
  plotted for Newton's Hermite only, show the root-mean-square of the
  dispersion in the series of solutions. The error bars in the results
  obtained with {\Brutus} are statistically indistinguishable from the
  presented bars. }
\label{fig:Newton's_Lyapunov_timescale}
\end{figure}

The different behavior for small-$N$ systems compared to large-$N$
systems motivated \cite{1993ApJ...415..715G} and
\cite{2002ApJ...580..606H} to conduct their analysis and study the
source of chaos in small versus large $N$-body systems. In
fig.\,\ref{fig:Newton's_Lyapunov_timescale} we show the results of
\cite{1993ApJ...415..715G} and compare them with converged solutions
using \Brutus\, up to $N=1$k and reprehensible solutions using
\Hermite\, for up to $N=128$k. The consistency between the results
obtained by \cite{1993ApJ...415..715G} (red), \Brutus\, (blue), and
\Hermite\, (ochre) gives us confidence in the validity of the
nonconverged (reprehensible) $N$-body solutions by
\cite{1993ApJ...415..715G} and using the regular Hermite algorithm
implemented in \PhFour\, without going through the elaborate process
of reaching a converged solution for $N>1$k.

The scaling we observe in fig.\,\ref{fig:Newton's_Lyapunov_timescale}
is consistent with that found by \cite{1993ApJ...415..715G} over the
entire range they explored, from $N=4$ to $N=512$. We therefore
conclude that 1) reprehensible simulations are adequate for studying
short-timescale Lyapunov exponent measurements for relatively
homogeneous systems, and 2) the Lyapunov timescale $t_\lambda \propto
\gamma t_{\rm cr}/\ln(\ln(N))$, with $\gamma = -1.39$ for $N\aplt 32$
and shallower with $\gamma = -0.498$ for $N\apgt 32$.

\subsection{Chaos in large-$N$ relativistic systems}

To study the degree of chaos in the relativistic regime, we
used the EIH equations of motion with initial realizations (masses,
positions, and velocities) identical to those used in the Newtonian
simulations. Therefore the latter initial conditions are in virial
equilibrium for the Newtonian case, but not for the highly
relativistic cases. Whereas the Newtonian $N$-body initial
realizations and calculations were scale free, we have to relax this
assumption when introducing the speed of light. Since our
calculations scale with mass $M$, size $R$, or velocity $v$, we have
the option to qualify the scaling by just changing $\Cscaling$. Here
$\Cscaling \rightarrow 0$ corresponds to the Newtonian case (because
in that case, $c \rightarrow \infty$); higher values of $v/c$ indicate a
more relativistic regime.

We started by confirming that for $\Cscaling \rightarrow 0,$ we reproduce
the results from fig\,\ref{fig:Newton's_Lyapunov_timescale}. As long
as $\Cscaling \aplt 10^{-4}$, the results of the relativistic
integration can hardly be distinguished from the Newtonian case (see
also figure\,\ref{fig:HGRX_c_dependence}).

\begin{figure}
\centering
\includegraphics[width=\columnwidth]{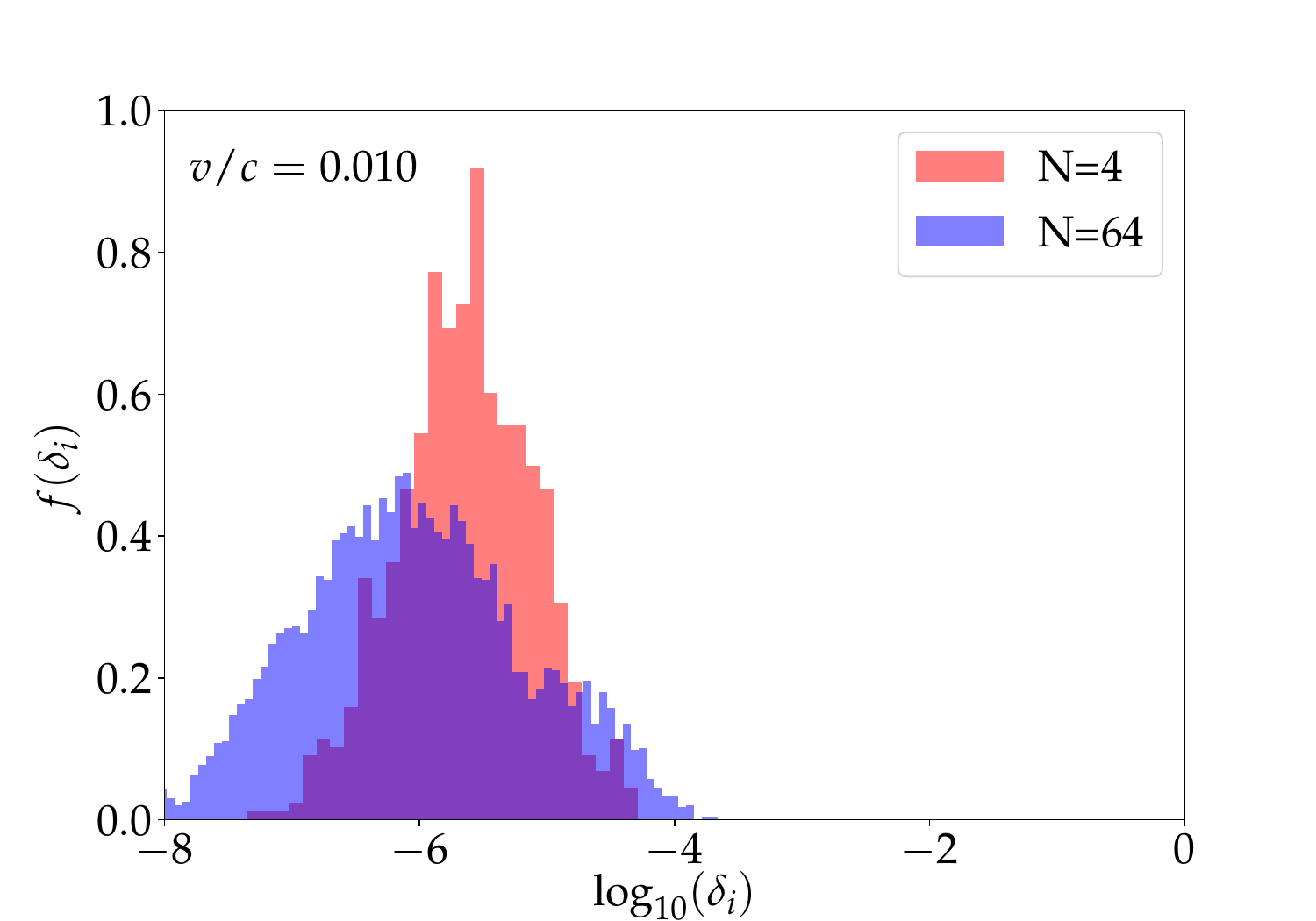}
\includegraphics[width=\columnwidth]{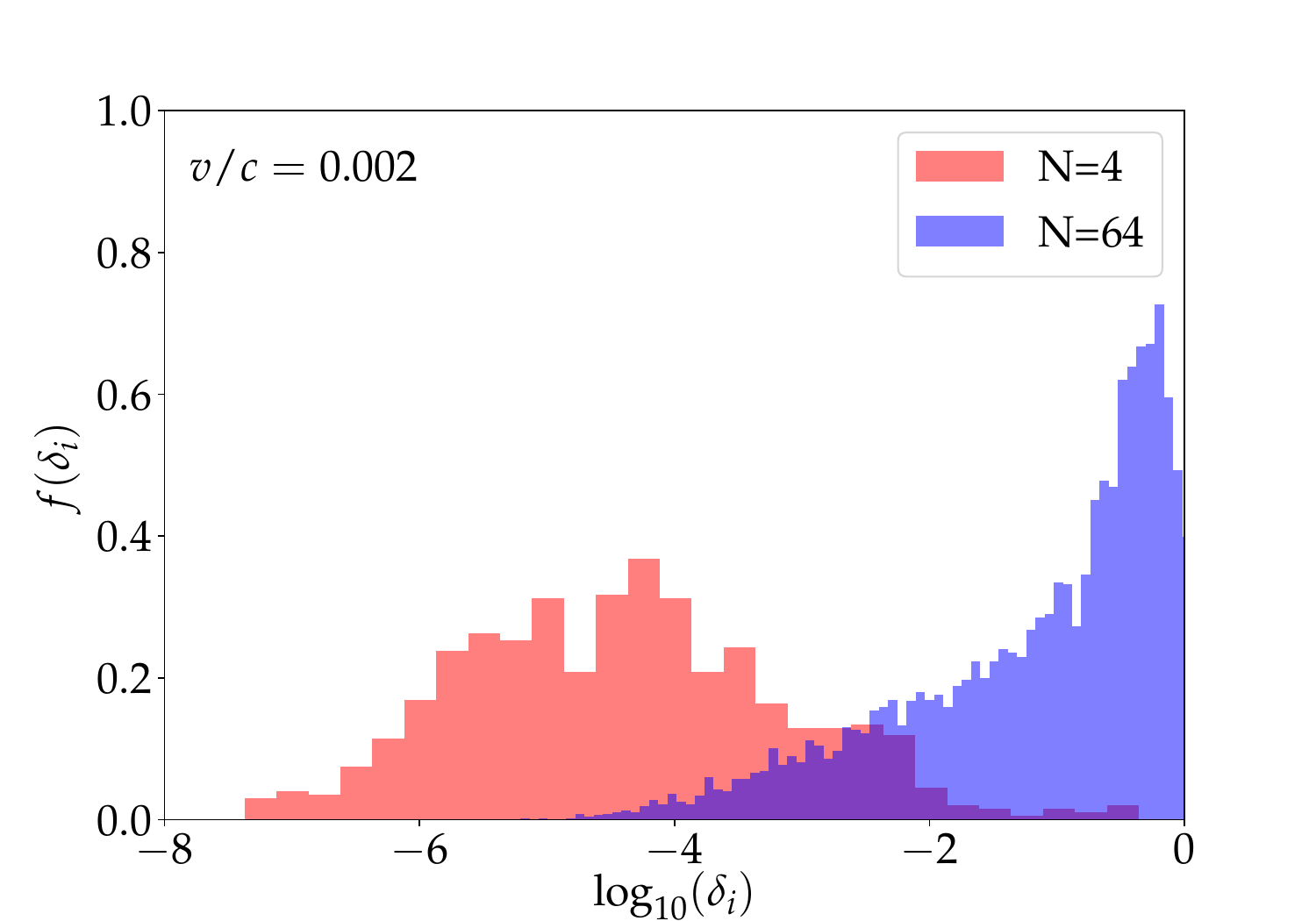}
\caption{Distribution of phase-space distances for individual
  particles in the simulations with $N=4$ (red) and for $N=64$ (blue)
  after integrating for $t=10$~$N$-body time units.  The data for 200
  runs were used for each histogram. Calculations were performed
  using \HermiteGRX\, using $\Cscaling = 0.01$ (top panel) and for
  $\Cscaling=0.002$ (bottom panel). }
\label{fig:hist_delta_N1024t10_HGRX}
\end{figure}

\begin{figure}
\centering
\includegraphics[width=\columnwidth]{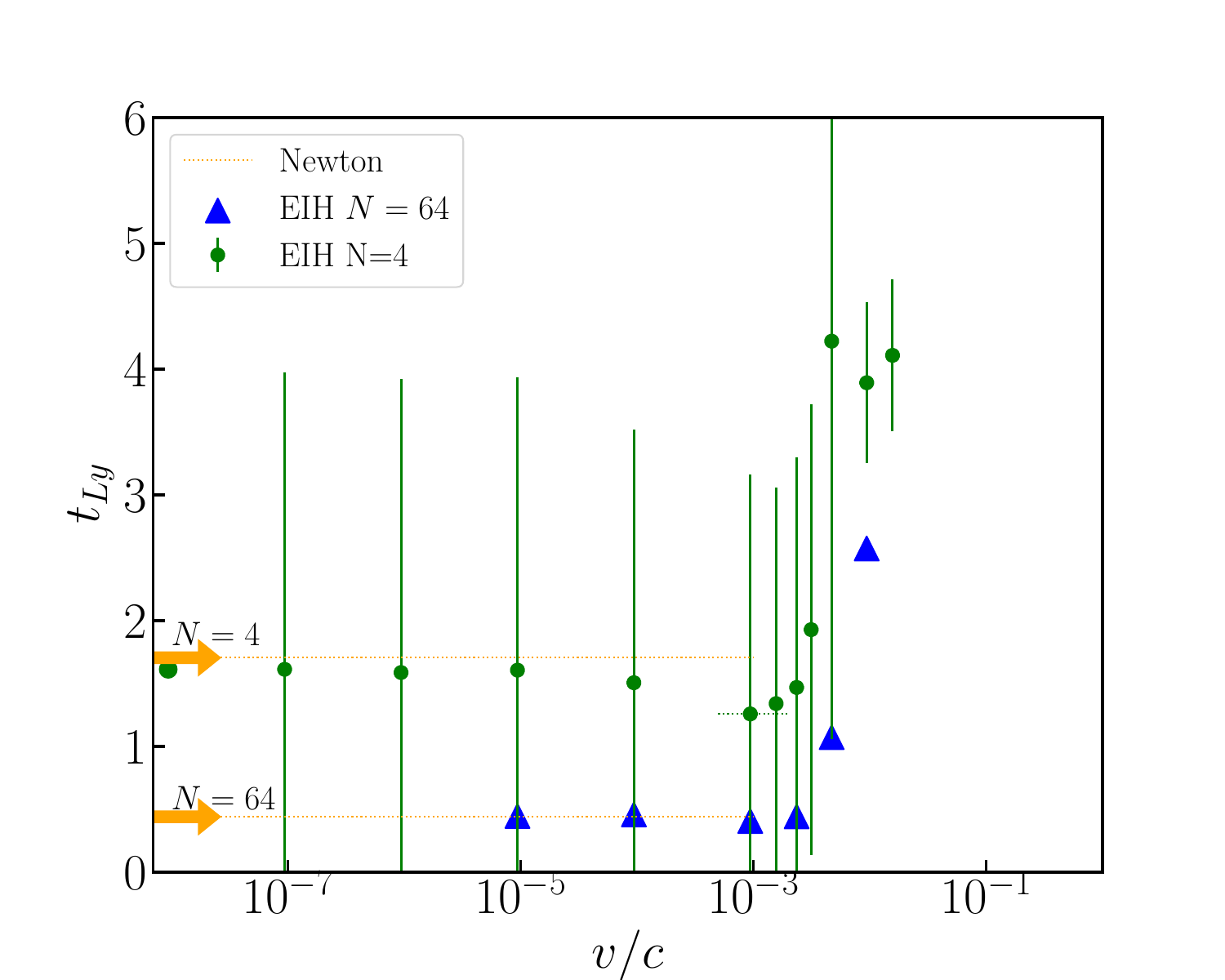}
\caption{Lyapunov timescale as a function of $\Cscaling$ for $N=4$
(green) and $N=64$ (blue). The Newtonian case (run with \PhFour) is
presented as arrows in orange. The vertical bars, only for the
green points, indicate the dispersion in the simulation results.
The short horizontal dotted green line indicates the lowest value
for the Lyapunov timescale reached for $v/c = 10^{-3}$ for $N=4$.
}
\label{fig:HGRX_c_dependence}
\end{figure}

This is further illustrated in
figure\,\ref{fig:hist_delta_N1024t10_HGRX}, where we present two
histograms for $N=4$ (red) and $N=64$ (blue) for $v/c=0.010$ (top
panel) and $v/c = 0.002$ (bottom). In the top panel, the distribution
in phase-space distance for both $N=4$ and $N=64$ are
comparable, although the dispersion for $N=64$ is somewhat larger.
When we reduce the speed of light (expressed as the parameter $v/c$),
both distributions move toward higher values of $\delta$. For
$\Cscaling = 0.002$ (bottom panel), the systems, though still somewhat
relativistic, have mean and median values that already approach the
Newtonian values. We recall that with this adopted scaling,
the system would correspond to a cluster with a total mass of $\sim
1$\,\MSun\, at a size scale of $\sim 436$\,au. Such a cluster of
stars appears insufficiently relativistic for the degree of chaos in
the equations of motion to be affected noticeably.

However, for $\Cscaling = 0.0005$, which corresponds to a size scale four times
larger, or equivalently, to a system four times more massive, the
measurements in the Lyapunov timescale start to deviate from the
Newtonian case. When we further decrease $v/c \aplt 10^{-4}$, the
mean and median values of both distributions are statistically
indistinguishable from the Newtonian case. The dispersion,
particularly noticeable for $N=64$, remains skewed to low values of
$\delta$.

In fig.\,\ref{fig:HGRX_c_dependence} we present the median Lyapunov
timescale for $N=4$ and $N=64$ as functions of $\Cscaling$. For the
asymptotic Newtonian case, $\Cscaling \rightarrow 0$, the Lyapunov
timescale converges to the median for the Newtonian case. The
dispersion in the distribution in the relativistic case remains
somewhat larger, however, even for $\Cscaling \rightarrow 0$, for which
$\langle t_{\rm Ly} \rangle = 1.61 \pm 2.36$, compared to the
Newtonian case, for which $\langle t_{\rm Ly} \rangle = 1.71 \pm 1.68$.

We suspect that these small systematic effects (which are not
statistically significant) could result from a few encounters
sufficiently close to be affected by general relativity.  For the
extreme relativistic case, $\Cscaling \apgt 0.001$, the Lyapunov timescale rises quickly to $\langle t_{\rm Ly} \rangle = 3.89 \pm
0.64$. We present the results for $\Cscaling \apgt 0.02$ (two green
points to the right), even though these are beyond the regime where
the 1-PN Taylor-series expansion to the EIH equations of motion is
valid (see sect.\,\ref{Sect:validity}). We therefore limit further
analysis to $\Cscaling \aplt 0.010$.

For $N=4,$ we observed a minimum in the Lyapunov timescale for $v/c
\sim 10^{-3}$ (signified by the horizontal dotted green line in
fig.\,\ref{fig:HGRX_c_dependence}). The change in
behavior for less and more relativistic systems might be interpreted as a signature that
the adopted Taylor expansion starts to break down, but for $v/c \apgt
10^{-3}$ , the post-Newtonian terms should still be valid. We expect
the low-$N$ configurations to break down earlier when they are evolved
with time because they are more relaxation dominated than the
large-$N$ systems.

With a typical distribution in velocities matching a truncated
Maxwellian, a small fraction ($\sim 2.9$\,\%) of the systems has a
velocity that exceeds the mean dispersion by factor of 3. Even for a
value of $v/c \apgt 0.05,$ the fraction of stars with a velocity $v \apgt
0.3c$ is smaller than $1/10^7$, and it is unlikely that when integrating
over only 10 H\'enon units, the post-Newtonian
Taylor-series expansion breaks down.

\begin{figure}
\centering
\includegraphics[width=\columnwidth]{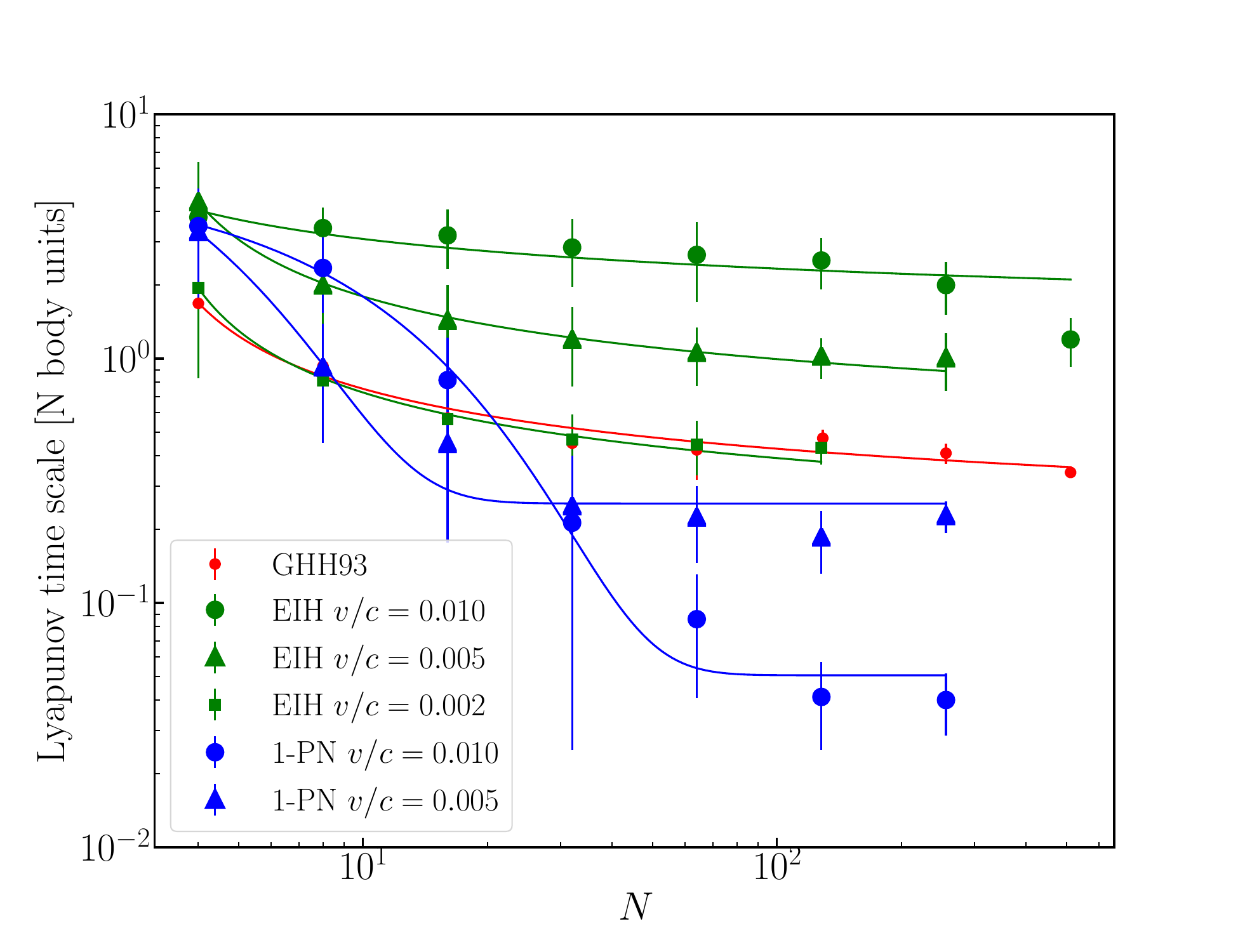}
\caption{Lyapunov timescale as a function of $N$ for
\cite{1993ApJ...415..715G} (red bullets) compared to the various
relativistic solutions. In blue we present the solutions using 1-PN
pair-wise terms with a scaling to the speed of light of
$\Cscaling = 0.010$ and twice this value. For the EIH equations of
motion we show the case for $\Cscaling = 0.010$ in green, twice
and four times this value. The top blue curve fits
$t_\lambda \simeq 0.255 + 13.43 e^{-0.371N}$ for
$\Cscaling = 0.005$, and
$t_\lambda \simeq 0.051 + 5.50 e^{-0.115N}$ for
$\Cscaling = 0.010$. }
\label{fig:Relativistic_Lyapunov_timescale}
\end{figure}

In fig.\,\ref{fig:Relativistic_Lyapunov_timescale} we present
measurements for the Lyapunov timescale for the post-Newtonian
equations of motion. For $N \aplt 4$, the EIH equations of motion as
well as the pairwise 1-PN terms show similar chaotic behavior in the
sense that the relativistic system is less sensitive to initial
perturbations than the Newtonian case. For $N\apgt 20$, the pairwise
1-PN terms result in smaller Lyapunov timescales
compared to Newton's equations of motion, whereas the EIH equations of
motion continue to result in a rather large Lyapunov timescale
compared to the Newtonian case. The Lyapunov timescales for the EIH
equations of motion are roughly twice as large as for the Newtonian
case for $\Cscaling = 0.005$ and roughly four times as large for
$\Cscaling = 0.010$.


\begin{figure}
\centering
\includegraphics[width=\columnwidth]{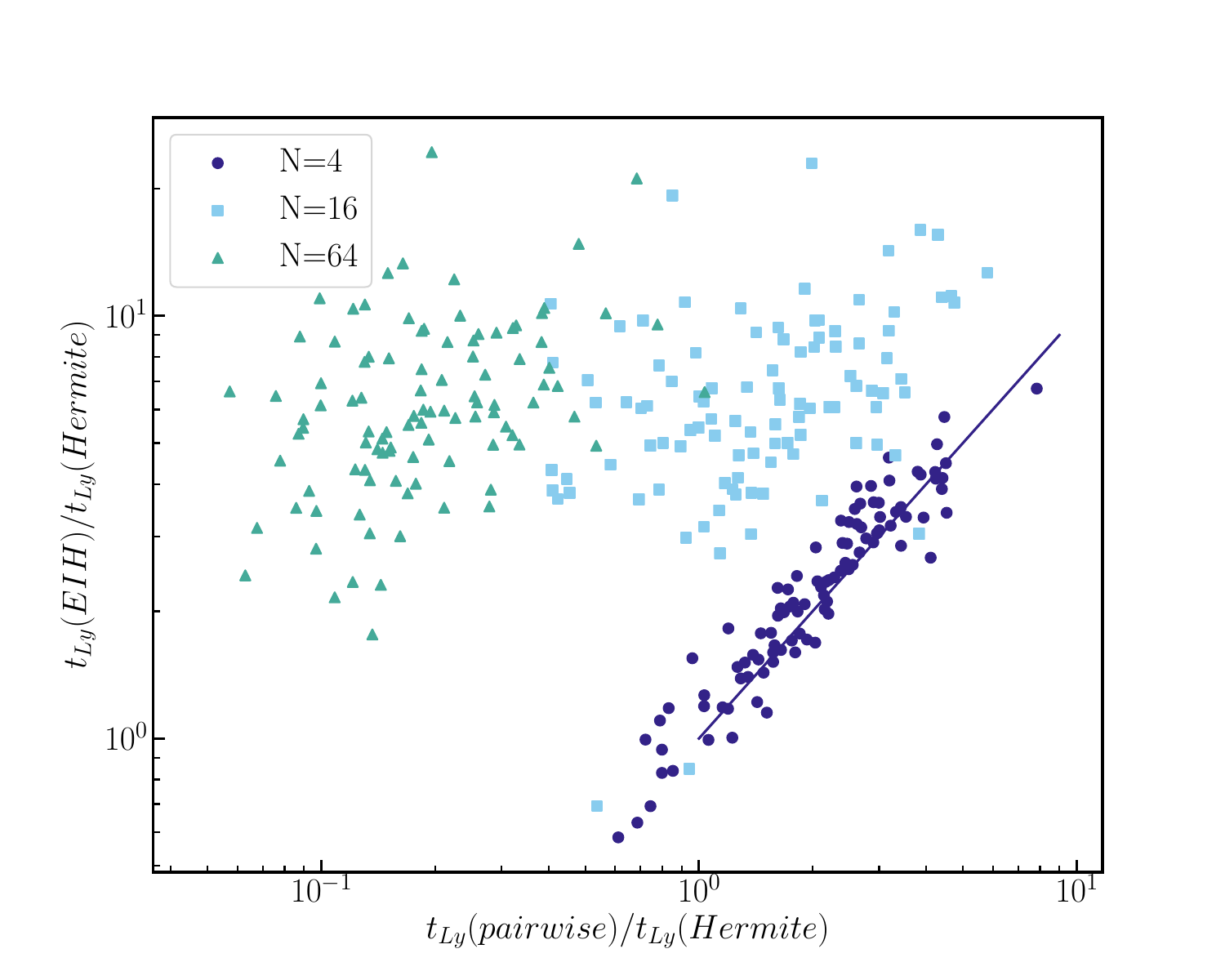}
\caption{Estimate for the Lyapunov timescale for the pairwise 1-PN
  terms as a function of the full EIH equations of motion for three
  choices of $N$, 4 (dark blue), 16 (aquamarine), and 64
  (green). Both, the pairwise 1-PN calculation and the one including
  the cross terms, are represented as fraction of the Newtonian
  solution. All relativistic simulations adopt $\Cscaling = 0.010$. }
\label{fig:Lyapunov_timescale_comparison}
\end{figure}

The effect of $N$ on the degree of chaos in the equations of motion is
further illustrated in
figure\,\ref{fig:Lyapunov_timescale_comparison}. Here we show for
$N=4$, $N=16,$ and for $N=64$ the ratio of the pairwise 1-PN solution
as a function of the full EIH equations of motion. The relativistic
calculations were performed with $\Cscaling = 0.010$.

For $N=4$, the Newtonian case shows a smaller average Lyapunov timescale than the two relativistic cases.  This was also confirmed in the
three-body simulations by \cite{2021arXiv210907013B}, who found no
significant difference in the chaotic behavior of Newtonian
versus relativistic systems.  When $N$ increases, the distribution of
the Lyapunov timescale for the EIH equations of motion continues to
be large compared to the Newtonian case, but the value for the
pairwise 1-PN terms tends to drop to less than 1/10th of the Newtonian
solution. We conclude that if one is interested in the dynamical
behavior of $N \apgt 4$ black holes, the pairwise 1-PN terms do not
reliably represent the chaotic behavior expected for such a
relativistic system. The pairwise 1-PN terms address pairs of compact
objects, whereas the EIH equations of motion should give a more
reliable representation for relativistic systems of $N \geq 3$. We
expect this difference to persist for the higher-order Taylor
expansion terms.

\section{Discussion}

\subsection{Consequences of the 1-PN terms}

Phase-space volume is preserved in a solution to a conservative
Hamiltonian system, but shrinks in a dissipative system
\citep{Shivamoggi2014}, as is the case in general relativity. The
contraction of the phase-space volume gives rise to an attractor, and
as a consequence, can have bounded trajectories
\citep{Berge1987-BEROWC}. It is not a priori clear, however, how
dissipation in an otherwise conservative system affects chaotic motion
\citep{Lakshmanan2003}. We find that in the conservative 1-PN regime,
the chaotic behavior of N-body systems is already affected. In
particular, for $N \apgt 10$, simulations that only include the 1-PN
pairwise terms behave differently than when the 1-PN cross-terms for
the EIH equations of motion are incorporated into the simulations.
The behavior of relativistic systems with $v/c \apgt 0.005$ and for $N
\apgt 10$ is considerably less chaotic than their less-relativistic
$v/c\aplt 10^{-3}$ and Newtonian counterparts.

\subsection{Validity of the post-Newtonian terms}\label{Sect:validity}

In this study, we rely on the post-Newtonian expansion of the EIH
equations of motion. Ideally, we would have adopted full general
relativity in our $N$-body calculations, but this is somewhat beyond
the scope of our study and is numerically challenging.

In an attempt to quantify the validity of the post-Newtonian expansion
adopted here, we compared the apsidal motion of the orbit-averaged
evolution for a two-body system with total mass $M$, semimajor axis
$a,$ and eccentricity $e$. We write the relative velocity in a
circular orbit in terms of the gravitational radius, $r_g = GM/c^2$
, and the speed of light as
\begin{equation}
c \simeq \sqrt{GM/(10 r_g)}.
\end{equation}
The Taylor-series expansion then starts to break down for $v \equiv
c/\sqrt{10} \sim 0.3 c$ \citep{2011PNAS..108.5938W}. During our
$N$-body calculations, we kept track of this velocity to ensure that
the Taylor series expansion in our calculations remained reliable.
However, this safety check does not guarantee that our results are
not affected, particularly for high values of $v/c$.

In the regime in which the Taylor-series expansion of the EIH equations of
motion breaks down, the 1-PN terms adopted here are insufficient to
capture the correct physical behavior. In this case, the 2-PN terms
become essential for the correct physical interpretation of the
numerical results. By definition, the 2-PN terms are smaller than the
1-PN terms because the former scale as $v/c$ and the latter as $v^2/c^2$.
On the other hand, both terms approach each other for more
relativistic systems, with $v \rightarrow c$. It is somewhat tricky
to give an absolute measure when the 2-PN terms should be used in
addition to the 1-PN terms. In a general $N$-body problem, stars may
approach each other at a short distance with relatively high
velocities with respect to $c$. When such an encounter is a
one-time event in the nondissipative limit, the lack of precision in
the PN terms is not expected to make a great difference in the
eventual results \citep{2011PNAS..108.5938W}. Reprehensive simulations
are therefore sufficient to derive the largest global positive
Lyapunov exponent for the system.

In an attempt to quantify the relative importance of 1-PN with respect
to the 2-PN terms, we compared the apsidal motion of the orbit-averaged
evolution for a two-body system with a total mass $M$, semimajor axis
$a,$ and eccentricity $e$. We write \citep{2020Univ....6...53I}
\begin{equation}
\dot{\omega}_{2PN} / \dot{\omega}_{1-PN} \simeq (r_g/a) (1/12) (28-e^2)/(1-e^2).
\end{equation}

\begin{figure}
\centering
\includegraphics[width=\columnwidth]{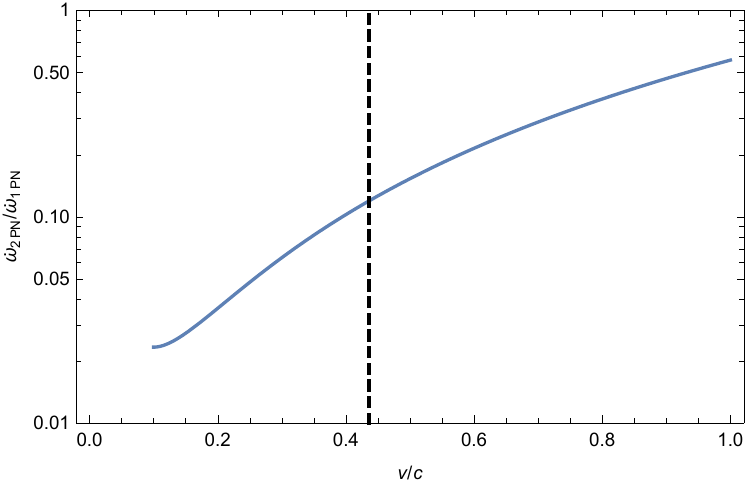}
\caption{Relative importance in the apsidal motion of the 1-PN terms
in comparison to the 2-PN terms. }
\label{fig:Apsidalmotion_1PN}
\end{figure}

In fig.\,\ref{fig:Apsidalmotion_1PN} we show as a function of
$v/c$ the relative drift in the apsidal motion for the 1-PN and 2-PN
terms for two bodies in a circular orbit at $100 r_g$,\begin{eqnarray}
v/c &=& (1/c) \sqrt{GM/a} \sqrt{(1+e)/(1-e)}.
\end{eqnarray}
The boundary at which the post-Newtonian expansion is no longer reliable
is indicated by the dashed vertical line, near $v/c \simeq 0.4$, which
happens when the two objects approach within $10 r_g$. If we compare
this boundary to the range in $v/c$ in
figure\,\ref{fig:HGRX_c_dependence}, we find that all lie below the
boundary, and we therefore argue that the increase in the Lyapunov
timescale toward the relativistic regime is physical and not a
numerical artifact.

\subsection{Other initial density profiles}

The initial conditions adopted in \cite{1993ApJ...415..715G} have a
homogeneous phase-space distribution within the adopted limits, and
they do not represent any observed stellar systems
\citep{2010ARA&A..48..431P}. Clusters of stars are better represented
with a \cite{1911MNRAS..71..460P} distribution or a
\cite{1966AJ.....71...64K} model. For this reason, we also performed a
series of calculations with these distributions. One series of
Newtonian calculations used Plummer models and King models with
dimensionless depth of $W_0=12$ (which is rather concentrated), and
one set of calculations used the EIH equations of motion for the King
model case (with identical initial realizations).

\begin{figure}
\centering
\includegraphics[width=\columnwidth]{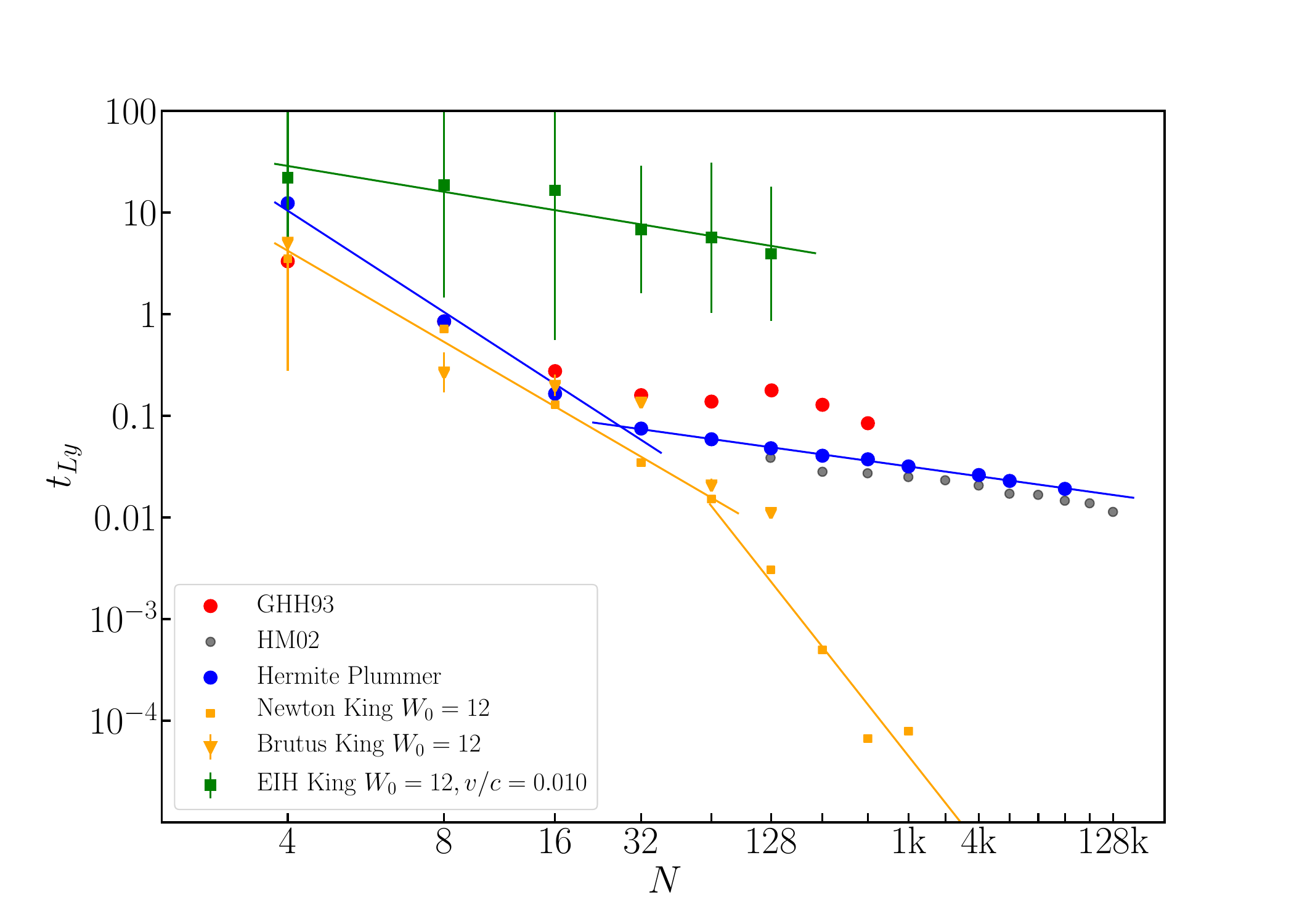}
\caption{Estimated Lyapunov timescale for the Newtonian case with
particles distributed in a Plummer sphere (blue) and a King model
(yellow). In addition, we show results for the King model, but for
the EIH equations of motion with $\Cscaling = 0.010$ (green). Here
the $x$-axis is in $\ln(\ln(N))$. }
\label{fig:KingW12_Lyapunov_timescale}
\end{figure}

In figure\,\ref{fig:KingW12_Lyapunov_timescale} we present the
Lyapunov timescales for these simulations as functions of $N$. The
Newtonian Plummer case shows a slightly smaller Lyapunov timescale
than the homogeneous distribution used in
\cite{1993ApJ...415..715G}. The Plummer distribution is consistent
with the initial conditions adopted by \cite{2002ApJ...580..606H}, and
their results are consistent with our results for the Plummer sphere,
see figure\,\ref{fig:KingW12_Lyapunov_timescale} (black points).

The Newtonian King model with $W_0=12$ is considerably more chaotic
than the homogeneous initial realization adopted in
\cite{1993ApJ...415..715G}, at least given that for $N\apgt 10^3$ we
were unable to measure a reliable Lyapunov timescale because the
phase-space distance grew beyond $\delta = 0.1$ within 1 $N$-body time
unit. King models with a central potential depth of $W_0=12$ turn out
to be considerably more chaotic than Plummer models (for $N\apgt 40$),
while Plummer models are expected to behave more chaotically than a
homogeneous distribution of particles. This is not a complete
surprise because the choice of $W_0=12$ places the model near core
collapse (at least in its density profile). The growth of an initial
phase-space distance between two subsequent calculations with almost
identical initial relations is then dominated by few-body interactions
in the core. One could argue that the entire chaotic behavior of the
star cluster is driven by few-body interactions in the cluster
center. Since some complex three-body interactions are fundamentally
unpredictable \citep{2020MNRAS.493.3932B}, the dynamical evolution of
the entire cluster will be unpredictable.

The extreme relativistic case, with $\Cscaling = 0.01$, shows a
similar characteristic again as the homogeneous initial realizations,
but a rather different scaling when adopting the King model. In the
Newtonian case, King models tend to have considerably smaller Lyapunov
timescales, but when extremely relativistic, they tend to be more
regular than the Newtonian case
Fig\,\ref{fig:Relativistic_Lyapunov_timescale}.

\section{Conclusions}

We have numerically analyzed the rate at which neighboring solutions
of the equations of motion for $N$ self-gravitating bodies diverge in
the Newtonian regime, but also with the 1-post-Newtonian expansion
terms for the pairwise approximation and the Einstein-Infeld-Hoffmann
equations of motion. Our results can be interpreted as the rate of
growth of the error in an $N$-body solution, caused by uncertainties
in the initial conditions or errors produced numerically during
integration.

Our Newtonian simulations were repeated with higher precision and
accuracy until a converged solution was achieved. Due to computer
limitations, this was performed for $N$ up to $1024$ particles. For
large particle numbers (and for the relativistic simulations), we used
reprehensible $N$-body solutions, and we confirmed them to be veracious
for $N$ up to $1$k particles.

The motivation to study the growth of errors stems from our desire to
understand the role of chaos in these systems. The macroscopic
distribution of material in the Galaxy may not be affected by
microscopic chaos. But the Galaxy is built up of small subsystems of
stars, each of which exhibits chaotic behavior, and the range of the
Newtonian force law causes chaos in these microscopic systems to
propagate to the Galaxy at large. Chaos in the Galaxy is then governed
by the chaos in small $N$ subsystems and not by the global slow (on
timescales longer than the dynamical timescale) variations of
the orbits of stars in a smooth potential.

We confirm the earlier result of \cite{1993ApJ...415..715G} and
\cite{2002ApJ...580..606H} that the divergence in $N$-body systems
grows exponentially, with an e-folding timescale on the order of the
crossing time. Our results agree with the $t_\lambda
\propto t_{\rm cross}/\ln(\ln(N))$ scaling of
\cite{1993ApJ...415..715G} and are inconsistent with a $t_{\rm
cross}/\ln(N)$ scaling.

Our conclusions are listed below.
\begin{enumerate}
\item For a homogeneous distribution of equal-mass particles in virial
equilibrium, the e-folding timescale for the growth of an initial
perturbation in an $N$-body system, the so-called Lyapunov timescale, scales for small systems of $N\aplt 32$ as
$t_\lambda/t_{\rm cross} = (0.88\pm0.12) - (1.39\pm0.13)
\ln(\ln(N))$. For larger systems of $N\apgt 32$, the Lyapunov timescale scales as
$t_\lambda/t_{\rm cross} = (-0.094 \pm 0.129) - (0.498\pm0.066)
\ln(\ln(N))$.
\item For an initial Plummer distribution, the e-folding timescale is
smaller than the homogeneous initial realizations by about a factor
of $5$ but preserves the same trend, or for $N\apgt 32,$ it fits
$t_\lambda/t_{\rm cross} = (-0.475 \pm 0.018) - (0.528\pm0.010) \ln(\ln(N))$.
\item For more concentrated models, such as a King model with $W_0=12$, the
scaling is comparable to the slope observed in the homogeneous
unit-cube or the Plummer distribution, with $t_\lambda/t_{\rm cross}
= (1.346\pm0.110) - (2.212\pm0.107) \ln(\ln(N))$, but extending
somewhat farther, to $N\simeq 64$. For larger $N,$ the slope is much
steeper, $t_\lambda/t_{\rm cross} = (4.970\pm2.03) - (4.813\pm1.147)
\ln(\ln(N))$, indicating that these systems are chaotic for large
$N$ on a timescale smaller than a crossing time.
\item If a small perturbation is introduced into a single particle of
a large $N$-body system, all particles
are affected within a few crossing times.
\item For self-gravitating systems with $\Cscaling \aplt 10^{-3}$, the
phase-space mixing of relativistic $N$-body systems is
indistinguishable from the Newtonian case. This limit is already
reached for a total of ten\, black holes of 10\,\MSun\, confined to a
spherical volume of radius $10^{-3}$ pc.
\item For highly relativistic systems, $\Cscaling \apgt 0.002$, the
EIH equations of motion to 1-PN are considerably less chaotic than
their Newtonian counterpart over all values of $N$. The Lyapunov
timescale scales with $t_\lambda/t_{\rm cross} = 6.63\pm1.68 -
\ln\left(3.72\pm2.04 \ln(N)\right)$.
\item For small $N$ ($\aplt 10$), the pairwise 1-PN terms give similar
phase-space mixing to the EIH equations of motion.
\item For $N > 4$, the pairwise 1-PN corrected equations of motion
become considerably more sensitive to perturbations in the initial
conditions compared to the EIH equations of motion. The former show
considerably shorter Lyapunov timescales compared to their
Newtonian counterparts, whereas the latter has even longer Lypaunov
timescales.
\item We conclude that the Galaxy is intrinsically chaotic on
a very short timescale because of the chaotic behavior of
microscopic few-body interactions in the centers of star
clusters. The chaotic behavior of these small-$N$ systems propagates
on a local crossing timescale to the entire star cluster, affecting the orbits of neighboring stars and clusters, and eventually,
the entire Galaxy.
\item The pairwise terms for $N > 3$ give a different dynamical
  behavior for relativistic $N$-body systems compared to the full EIH
  equations of motion.
\end{enumerate}

Considering the effect of the full EIH equations of motion on a
relativistic cluster of compact objects, and the potential
consequences for observations with laser interferometric gravitational
wave observatories, we look forward to implementing and study the
effect of higher-order cross-terms in general relativistic $N$-body
simulations. We do realize, however, that these calculations do not
have the most favorable scaling of the computer time with respect to
$N$.

\section*{Public data}

The source code, input files, simulation data, and data processing
scripts for this manuscript are available at figshare under DOI {\em
10.6084/m9.figshare.xxxx}.

\section*{Software used for this study}

This work would have been impossible without the following public open-source packages and libraries: Python \citep{vanRossum:1995:EEP},
matplotlib \citep{2007CSE.....9...90H}, numpy
\citep{Oliphant2006ANumPy}, MPI \citep{Gropp:1996:HPI,Gropp2002}, and
AMUSE \citep[][available for download at
\url{https://amusecode.org}]{portegies_zwart_simon_2018_1443252}.
\Sapporo\, GPU library
\citep{2007NewA...12..641P,2009NewA...14..630G}, {\tt MPFR} library
\citep{10.1145/1236463.1236468} of the {\tt GMP} library
\citep{Granlund12}.

A (python notebook) tutorial for students is available at
\url{https://github.com/amusecode/Tutorial}. All the $N$-body codes
used in this study are available in the AMUSE repository at
\url{amusecode.org}.

\section*{Acknowledgments}

It is a pleasure to thank Clifford Will, $\ln(a)$ Sellentin, and Arend
Moerman for discussions. This project was supported by funds from the
European Research Council (ERC) under the European Union's Horizon
2020 research and innovation program under grant agreement No 638435
(GalNUC), and by the National Science Foundation in the U.S. under
grant AST-1814772. This work was performed using resources provided
by the Academic Leiden Interdisciplinary Cluster Environment (ALICE),
and using LGM-II (NWO grant \# 621.016.701).

\noindent
{\bf Energy consumption of this calculation}\\
The calculations using \Brutus\ are elaborate and took about $10^7$
CPU seconds. The other two sets of calculations are comparable in
expense, totaling about a year of single CPU usage. Using the tool
\url{http://green-algorithms.org/}, we calculated our energy
consumption to be about 3.32\,MWh. At Dutch electricity rates, this
would produce about 1.8\,kiloton CO$_2$, but since the computers used
are powered by either Dutch wind or Norwegian hydroelectric power
(through certificates) the net CO2 emission should be negligible.

\begin{appendix}
\section{Implementation of the numerical method}
\label{sec:implementation}
\label{sec:methods}


The Newtonian solver and the post-Newtonian terms are implemented in C
and C++ and interfaced with the Astrophysics Multipurpose Software
Environment \citep[AMUSE for short,][]{2018araa.book.....P}. In this
appendix, we discuss the simple Hermite predictor-corrector Newtonian
$N$-body solver called \PhFour\  and the post-Newtonian solver called
\HermiteGRX. The former solves Newton's equations of motion quite
accurately, but is unable to achieve converged solutions. The other
code adopts the post-Newtonian approach in which we address the
pairwise EIH equation as well as the
so-called cross-terms \citep{1938AMath.65..100E}. All equations are
implemented to 1-PN order.

The Newtonian code, \PhFour, is optimized for parallel operations
using the Message Passing Interface
\citep[MPI,][]{Gropp:1996:HPI,Gropp2002}, and for GPU using the
\Sapporo\, library
\citep{2007NewA...12..641P,2009NewA...14..630G}. The post-Newtonian
implementation, \HermiteGRX\, is parallelized using hyperthreading,
but not using MPI, and it does not support GPU operations. In this
code, however, few-body interactions are regularized using
quaternions.

In the following sections, we discuss the various implementations and
optimizations. We also perform some test calculations to demonstrate the
efficiency and accuracy of the various implementations.

\subsection{Fourth-order Hermite integration scheme}
\label{sec:hermiteintegration}

Here we describe the fourth-order Hermite predictor corrector
implementation in \PhFour\, and \HermiteGRX\, briefly. The first
solves Newton's equations of motion; the second also includes various
solvers for addressing the post-Newtonian expansion terms.

\subsubsection{Predict, evaluate, and correct scheme}
\label{sec:PEC}

The Hermite integration scheme is a family of implicit numerical
methods for solving ordinary differential equations. Introduced by
\cite{makino1991optimal}, the fourth-order integration scheme is
written
\begin{eqnarray}
y(t+h) &=& y(t) + \frac{y^{(1)}(t) + y^{(1)}(t+h)}{2} h \\ \nonumber
& & + \frac{y^{(2)}(t) - y^{(2)}(t+h)}{12} h^2 + \BigO{h^5},
\label{eq:fourth_hermite}
\end{eqnarray}
which has a local truncation error of $\BigO{h^5}$, resulting in a
global truncation error of $\BigO{h^4}$. We denote the $i$-th derivative
with respect to $t$ using $(\cdot)^{(i)}$ or with Einstein's
convention. The sixth- and eight-order schemes, derived by
\cite{nitadori2008sixth}, are not implemented here.

Because the scheme is implicit, a fixed-point iteration to solve
eq.~\ref{eq:fourth_hermite} is needed,
\begin{eqnarray}
y_{[i+1]}(t+h) &=& y(t) + \frac{y^{(1)}(t) + y_{[i]}^{(1)}(t+h)}{2} h \\ \nonumber
& & + \frac{y^{(2)}(t) - y_{[i]}^{(2)}(t+h)}{12} h^2.
\end{eqnarray}
Here we used a truncated Taylor expansion around $t$ as the initial
(boundary) condition
\begin{equation}
y_{[0]}(t+h) = y(t) + h y^{(1)}(t) + \frac{h^2}{2} y^{(2)}(t).
\end{equation}
This sequence converges to the limit $y(t+h)$, which ends the current
time step. In practice, a single iteration suffices when the time step
$h$ is small.

Each integration step then consists of
\begin{enumerate}
\item \emph{prediction} of the positions and velocities at the
next time step $t+h$,
\begin{subequations}
\begin{gather}
\tilde{\vect{r}}_{i+1} = \vect{r}_i + \vect{v}_i h +
\tfrac12 \tilde{\vect{a}_i} h^2 + \tfrac16 \vect{j}_i
h^3,\\
\tilde{\vect{v}}_{i+1} = \vect{v}_i + \vect{a}_i h +
\tfrac12 \vect{j}_i h^2.
\end{gather}
\label{eq:hermite_predictor}
\end{subequations}
Here $\vect{r}$, $\vect{v}$, $\vect{a}$, and
$\vect{j}=\deriv{\vect{a}}{t} \equiv \dot{a}$ represent vectors for
the position, velocity, acceleration, and jerk, respectively.
The jerk $\vec{j}$ is dotted, which in this case does not
indicate a time derivative. Predicted values are indicated with the
$\tilde{(\cdot)}$.
\item Acceleration and jerk are calculated using the predicted
positions and velocities (eq.\,\ref{eq:hermite_predictor}).
\item A subsequent correction is applied to the position and velocity
at the next time step using the predicted accelerations and jerks,
\begin{subequations}
\begin{gather}
\vect{v}_{i+1} = \vect{v}_i + \tfrac12 (\vect{a}_i +
\tilde{\vect{a}}_{i+1}) h + \tfrac{1}{12} (\vect{j}_i -
\tilde{\vect{j}}_{i+1}) h^2,\\
\vect{r}_{i+1} = \vect{r}_i + \tfrac12 (\vect{v}_i +
\vect{v}_{i+1}) h + \tfrac{1}{12} (\vect{a}_i -
\tilde{\vect{a}}_{i+1}) h^2.
\label{eq:corrector_pos}
\end{gather}
\label{eq:hermite_corrector2}
\end{subequations}
\end{enumerate}

The corrected velocities increase the order of the method to
$\BigO{h^4}$. In such a predict, evaluate, and correct (PEC) scheme,
the fixed-point iteration can be described as $P(EC)^n$ for $n$
iterations.

\subsubsection{Variable time step}

We use variable but shared time steps. After every step, a new
time-step size is determined based on the minimum interparticle
collision timescale, calculated from unaccelerated linear motion and
the freefall time,
\begin{equation}
h = \eta \min_{i,j\not=i}{\left( \frac{|\vect{r}_{ij}|}{|\vect{v}_{ij}|}, \frac{|\vect{r}_{ij}|}{|(m_i+m_j) \vect{a}_{ij}|} \right)}.
\end{equation}
Here $r_{ij}$, $v_{ij}$, $a_{ij}$ are the relative distance, velocity,
and acceleration between particles $i$ and $j$, and $m_i$ is the mass of
particle $i$. The minimum is taken over each pair of particles $(i,
j)$, and over the two estimates of the collision time. Here the
time-step parameter $\eta$ is introduced to control the time-step size
and therewith the accuracy (and speed) of the integration scheme.
Ler values of $\eta$ generally correspond to smaller errors and a
longer integration wall-clock time. The default value in \AMUSE, $\eta
= 0.03$, generally leads to acceptable accuracy at a reasonable speed:
in many cases, $\eta = 0.1$ probably suffices
\citep{2041-8205-785-1-L3}. For safety, we adopted $\eta = 0.01$ for
our calculations.

The adopted variable time step removes the time-symmetric properties
of the integration. The fundamental idea behind time-symmetrization is
to prevent systematic drift in any conserved quantity. Time
reversibility then introduces the same drift with opposite sign.

A time-symmetric algorithm exhibits the same drift in both directions
of time, resulting in identical absolute drifts when integrating
forward and backward with time. This is a desirable quality of an
integrator because we consider Nature to conserve energy and angular
momentum \citep[see also][ for a discussion on the arrow of time due
to the chaotic behavior of self-gravitating systems and
uncertainties on the smallest scales]{2018CNSNS..61..160P}.

One can reintroduce time-symmetry by selecting a symmetric time step, for
example, by taking the average of some function at either side of the
integration step \citep{hut1995building},
\begin{equation}
h = \tfrac12 (k(t_b) + k(t_e)).
\end{equation}
Here $k(t)$ is a function to determine the step size at the beginning
$t_b$ and at the end $t_e=t_b+h$ of the integration step. This
implicit expression requires fixed-point iteration to evaluate
\begin{subequations}
\begin{equation}
h_{[0]} = k(t_b),
\end{equation}
\end{subequations}
and the eventual time step when the sequence converges becomes
\begin{subequations}
\begin{equation}
h_{[i+1]} = \tfrac12 \left(k(t_b) + k(t_b + h_{[i]}) \right).
\end{equation}
\end{subequations}
Generally, the sequence converges in a single iteration
\citep{hut1995building}.

\subsubsection{Splitting the jerk}

Calculating the jerk is expensive in terms of computer time because it
requires three passes over all particles.
To avoid evaluating the jerk directly, we use a central numerical
derivative,
\begin{equation}
\vect{j}(t) = \frac{\vect{a}(t + h) - \vect{a}(t - h)}{2h} + \BigO{h^2}.
\end{equation}
Here the accelerations are calculated using the Taylor expanded
positions and velocities of the particles,
\begin{subequations}
\begin{gather}
\vect{r}(t \pm h) = \vect{r}(t) \pm h \vect{v}(t) + \tfrac12 h^2 \vect{a}(t) + \BigO{h^3},\\
\vect{v}(t \pm h) = \vect{v}(t) \pm h \vect{a}(t) + \BigO{h^2}.
\end{gather}
\end{subequations}
The numerical calculation of the jerk is equally expensive as the
analytic calculation because it requires two additional acceleration
calculations per jerk. We reduce the computational complexity by the
time step $h$ of the previous integration step and the backward
derivative. This allows us to reuse the previous steps' positions and
velocities for calculating the jerk at the current time step.

Using a first-order derivative instead of the analytical expression
for the jerk leads to a reduced accuracy, but this is corrected for by
splitting the acceleration into two parts: the Newtonian part, and a
perturbing part,
\begin{equation}
\vect{a}(t) = \vect{a}_{\mathrm{Newton}}(t) + \vect{a}_{\mathrm{pert}}(t).
\end{equation}
We note here that we already introduced a perturbation, which in the EIH
equations of motion will be the post-Newtonian terms. The Newtonian
jerk can now be calculated analytically and at negligible cost compared
to calculating the perturbing acceleration. The perturbing jerk is
calculated from the numerical backward derivative,
\begin{equation}
\vect{j}(t) = \vect{j}_{\mathrm{Newton}}(t)
+ \frac{\vect{a}_{\mathrm{pert}}(t) - \vect{a}_{\mathrm{pert}}(t - h)}{h} + \BigO{h^2}.
\end{equation}
This operation increases the memory requirement by storing two
accelerations for each particle.

The algorithm is made to be self-starting by defining the perturbing
acceleration of the previous integration step
$\vect{a}_{i-1, \mathrm{pert}}$, as it depends on the jerk of the first iteration
$\vect{j}_i$. For this, we chose
$\vect{a}_{i-1, \mathrm{pert}}=\vect{a}_{i, \mathrm{pert}}$, so that
$\vect{j}_{i, \mathrm{pert}}=0$. This decreases the local truncation
error of the first step to $\BigO{h^4}$, but its impact on the results
is small because only one step is taken.

\subsection{Regularization}
\label{sec:regularization}

Regularizing the equations of motion for two bodies (or more) in a
close encounter improves computational performance and accuracy. The
main reason to introduce regularization, however, is to prevent a
devision by zero for extremely close approaches between particles
\citep{KS1965,1999MNRAS.310..745M}. Regularizing the post-Newtonian
expressions is harder than the regular Newtonian case because of the
velocity dependence on the acceleration
\citep{2006MNRAS.372..219M,2008AJ....135.2398M}.  Here we derive the
regularized equations of motion in post-Newtonian few-body encounters
using quaternions \citep{waldvogel2006quaternions}, but we start with
a brief overview on quaternions

\subsubsection{Quaternions}

Quaternions are an extension of complex numbers to three complex
base quaternions $\vect{i}$, $\vect{j}$ , and $\vect{k}$
\citep{waldvogel2006quaternions}. A quaternion $\vect{u}$ is
constructed from four real numbers $u_\ell \in \mathbb{R}$ for
$\ell=0, 1, 2, 3$,
\begin{equation}
\vect{u} = u_0 + u_1 \vect{i} + u_2 \vect{j} + u_3 \vect{k},
\end{equation}
with the multiplicative identities
\begin{equation}
\vect{i}\vect{i} = \vect{j} \vect{j} = \vect{k} \vect{k} = \vect{i} \vect{j} \vect{k} = -1,
\end{equation}
from which we derive the other products,
\begin{align}
\begin{cases}
\vect{i}\vect{j} &= -\vect{j}\vect{i} = \vect{k}, \\
\vect{j}\vect{k} &= -\vect{k}\vect{j} = \vect{i}, \\
\vect{k}\vect{i} &= -\vect{i}\vect{k} = \vect{j}.
\end{cases}
\end{align}
From these, the noncummatative property of quaternion multiplication
is evident. We define the conjugate of quaternion $\vect{u}$,
\begin{equation}
\overline{\vect{u}} = u_0 - u_1 \vect{i} - u_2 \vect{j} - u_3 \vect{k},
\end{equation}
which leads to the definition of the norm
\begin{equation}
|\vect{u}|^2 = \vect{u} \overline{\vect{u}} = \overline{\vect{u}} \vect{u} = u_0^2 + u_1^2 + u_2^2 + u_3^2.
\end{equation}
The star conjugate of $\vect{u}$ then is
\begin{equation}
\vect{u}^\star = u_0 + u_1 \vect{i} + u_2 \vect{j} - u_3 \vect{k}.
\end{equation}
We associate the vector $\vect{r} = (r_0, r_1, r_2)\in \mathbb{R}^3$
to quaternion $\vect{u}$ as
\begin{equation}
\vect{u} = r_0 + r_1 \vect{i} + r_2 \vect{j}.
\end{equation}
When describing real-world coordinates, quaternions have a vanishing
component in $\vect{k}$ \citep{waldvogel2006quaternions}.

\subsubsection{Equations of motion}

Regularization is applied to pairs of particles. They, particle $i=1$
and $i=2$, are located at phase-space coordinate $\vect{r}_i$
with velocities $\vect{v}_i$, and masses $m_i$. The equations of
motion are
\begin{equation}
\ddot{\vect{r}}_i = -\frac{G m_j}{r^3} (\vect{r}_j - \vect{r}_i) + \vect{a}_i.
\end{equation}
The acceleration $\vect{a}_i$ for particle $i$ consists of all
perturbing (post-Newtonian) and Newtonian accelerations, excluding the
Newtonian acceleration between particle $i$ and $j$. The center of
mass
\begin{equation}
\vect{r}_{\mathrm{cm}} = \frac{m_1 \vect{r}_1 + m_2 \vect{r}_2}{m_1 + m_2},
\end{equation}
and the relative position
\begin{equation}
\vect{r}_{12} = \vect{r}_2 - \vect{r}_1,
\end{equation}
we rewrite these equations of motion as
\begin{equation}
\ddot{\vect{r}} = -\frac{\mu}{r^3} \vect{r} + \vect{P},
\end{equation}
and
\begin{equation}
\ddot{\vect{r}}_{\mathrm{cm}} = \frac{m_1 \vect{a}_1 + m_2 \vect{a}_2}{m_1 + m_2}.
\end{equation}
Here $\mu=G(m_1 + m_2)$ is the gravitational parameter, and
$\vect{P}=\vect{a}_2 - \vect{a}_1$ the relative perturbing
acceleration.

The above equations of motion have a singular point at $r=0$. We
integrate the center of mass separately from the relative position,
and rewrite the equation of motion in such in a way as to remove this
singular point. This is achieved by remapping the world position to a
regularized position quaternion $\vect{u}$, from which we calculate
the world position
\begin{equation}
\vect{r} = \vect{u} \vect{u}^\star.
\label{eq:regularization_map}
\end{equation}
The quaternion $\vect{r}$ now has vanishing component \vect{k} and can
therefore be transformed into a vector, from which it follows that
\begin{equation}
r = |\vect{r}| = |\vect{u}|^2 = \vect{u}\overline{\vect{u}}.
\end{equation}
A (nonunique) solution to the inverse of
eq.~\ref{eq:regularization_map} is
\begin{equation}
\hat{\vect{u}} = \frac{\vect{r} + |\vect{r}|}{\sqrt{2(|\vect{r}| + r_0)}}.
\end{equation}
The position vector, $\vect{r}$, is almost entirely oriented in the
negative $r$-direction, and the denominator is close to zero. Without loss of generality, we can avoid large numerical errors by
swapping indices $i=1$ and $2$, resulting in the negation of
$\vect{r}$.

The regularization time $\tau$ is
\begin{equation}
\mathrm{d}t = r~\mathrm{d}\tau.
\end{equation}
The equations of motion for the regularized position is written in
regularized time:
\begin{equation}
\vect{u}^{(2)} = -\tfrac{1}{2} b \vect{u} + \tfrac{1}{2} r \vect{P} \overline{\vect{u}}^\star.
\label{eq:regularized_equations_of_motion}
\end{equation}
Here $b$ is the binding energy of the binary,
\begin{equation}
b = \frac{\mu}{r} - \frac12 |\dot{\vect{r}}|^2 \;\;\; = \;\;\; \frac{\mu - 2 |\vect{u}^{(1)}|^2}{|\vect{u}|^2}.
\label{eq:bindingenergy}
\end{equation} For an unperturbed two-body system, $\vect{P}=0$, the
binding energy $b$ is constant, and the equation of motion describes
the harmonic oscilator. For a perturbed two-body system, the expression
resembles a perturbed harmonic oscilator. The greatest advantage of
this approach is that this equation of motion has no singular points,
not even for $r=|\vect{u}|^2=0$. This improves the performance for small
perturbations when numerically integrating the equations of motion of
a highly eccentric binary, and it improves the accuracy for perturbed
binaries.

The binding energy (eq.~\ref{eq:bindingenergy}) is not regularized, and
errors continue to increase in close encounters
\citep{funato1996time}. This problem is mitigated by also integrating
the binding energy numerically. When the binding energy changes
slowly with time as a function of the perturbing accelerations, we find
\begin{equation}
b^{(1)} = -\avg{\vect{r}', \vect{P}}.
\label{eq:bindingenergy_prime}
\end{equation}
Here the $\avg{(\cdot), (\cdot)}$ is the vectorial scalar
product. An initial condition for $\vect{u}^{(1)}$ is
\begin{equation}
\hat{\vect{u}}^{(1)} = \tfrac12 \vect{v} \overline{\vect{u}}^\star.
\label{eq:regularized_vel}
\end{equation}
This expression gives a small correction to the original derivation by
\cite{waldvogel2006quaternions} and \cite{waldvogel2008quaternions}:
it can be verified by substitution in eq.\,\ref{eq:bindingenergy}, to
derive the binding energy in world coordinates. For the reciprocal
eq.\,\ref{eq:regularized_vel}, the right-hand side multiplication with
$\vect{u}^\star$ leads to
\begin{equation}
\vect{v} = \tfrac{2}{r} \dot{\vect{u}} \vect{u}^\star \equiv
\tfrac{2}{r} \vect{u}^{(1)} \vect{u}^\star.
\end{equation}

One can also formulate the Kustaanheimo-Stiefel \citep[hereafter
  KS][]{KS1965} regularized equations of motion in terms of the
perturbing potential \citep{1975lrcm.book.....S}, which can be
advantageous in some cases, although it only applies to cases when the
potential is independent of velocity (not the case for general
relativity). Moreover, Stiefel and Steifele (1975, see pages 30 and
31)\nocite{1975lrcm.book.....S} argued that it is numerically more
efficient to formulate the equations of motion in terms of the total
energy, not just the Kepler energy (see their eq.\,(A.31)). The
equations of motion can also be cast into the form of regular elements
\citep[][their pages 90 and 91]{1975lrcm.book.....S} that are
advantageous for perturbed two-body systems because the regular
elements remain exactly constant for nonperturbed systems (i.e., the
two-body system is integrated analytically).

\subsubsection{Numerical integration}

We solve the equations of motion using the Hermite scheme in the PEC
formulation (see sect.\,\ref{sec:PEC}), with $\Delta\tau \equiv
\DeltaTau$. The resulting intergration scheme is described below.
\begin{enumerate}
\item Predict the regularized position, regularized velocity, and
binding energy at the end of the current integration step,
\begin{subequations}
\begin{gather}
\tilde{\vect{u}}_{i+1} = \vect{u}_i + \vect{u}^{(1)}_i \DeltaTau +
\tfrac12 \vect{u}^{(2)}_i \DeltaTau^2 + \tfrac16 \vect{u}^{(3)}_i
\DeltaTau^3,\\
\tilde{\vect{u}}^{(1)}_{i+1} = \vect{u}^{(1)}_i + \vect{u}^{(2)}_i
\DeltaTau + \tfrac12 \vect{u}^{(3)}_i \DeltaTau^2,\\
\tilde{b}_{i+1} = b_i + b^{(1)}_i \DeltaTau + \tfrac12 b^{(2)}_i
\DeltaTau^2.
\end{gather}
\end{subequations}
Then predict the center-of-mass position and velocity from
eq.\,\ref{eq:hermite_predictor}.
\item Evaluate the regularized acceleration, regularized jerk, and the
derivatives of the binding energy at the end of the current
integration step, according to
eqs.~\ref{eq:regularized_equations_of_motion}
and~\ref{eq:bindingenergy_prime},
\begin{subequations}
\begin{gather}
\vect{u}^{(3)} = \tfrac12 \left( -b^{(1)} \vect{u} - b
\vect{u}^{(1)} + r^{(1)} \vect{P} \overline{\vect{u}}^\star + r
\vect{P}^{(1)} \overline{\vect{u}}^\star + r \vect{P}
(\overline{\vect{u}}^{(1)})^\star \right),\\
b^{(2)} = -\avg{\vect{r}^{(2)}, \vect{P}} - \avg{\vect{r}^{(1)},
\vect{P}^{(1)}}.
\end{gather}
\end{subequations}
The test-particle integrator by \cite{2014MNRAS.443..355H}
adopts a similar integration scheme for the KS coordinates.
\item Correct the regularized position, regularized velocity, and
binding energy at the end of the current integration step,
\begin{subequations}
\begin{gather}
\vect{u}^{(1)}_{i+1} = \vect{u}^{(1)}_i + \tfrac12 (\vect{u}^{(2)}_i
+ \tilde{\vect{u}}^{(2)}_{i+1}) \DeltaTau + \tfrac{1}{12}
(\vect{u}^{(3)}_i - \tilde{\vect{u}}^{(3)}_{i+1}) \DeltaTau^2,\\
\vect{u}_{i+1} = \vect{u}_i + \tfrac12 (\vect{u}^{(1)}_i +
\vect{u}^{(1)}_{i+1}) \DeltaTau + \tfrac{1}{12} (\vect{u}^{(2)}_i -
\tilde{\vect{u}}^{(2)}_{i+1}) \DeltaTau^2,\\
b_{i+1} = b_i + \tfrac12 (b^{(1)}_i + \tilde{b}^{(1)}_{i+1})
\DeltaTau + \tfrac{1}{12} (b^{(2)}_i - \tilde{b}^{(2)}_{i+1})
\DeltaTau^2.
\end{gather}
\end{subequations}
\end{enumerate}
The corrected velocity in the corrector for the position
makes the scheme fourth order \citep{2006MNRAS.372..219M}.

\subsubsection{Regularized time-step considerations}

The regularized time step is determined using
\begin{equation}
s = \eta \min{\left(\sqrt{\frac{|\vect{u}^{(2)}||\vect{u}|}{|\vect{u}^{(3)}||\vect{u}^{(1)}|}}, \frac{|\vect{u}^{(1)}|}{|\vect{u}^{(2)}|}\right)},
\end{equation}
which is a variation on the time steps suggested
by~\cite{funato1996time}. To advance model time, we need to convert
this $\DeltaTau$ into $h$, which, according to
\cite{funato1996time}, is done with
\begin{equation}
h \equiv T(\DeltaTau) = t^{(1)}_{\tfrac12} \DeltaTau + \tfrac{1}{24} t^{(3)}_{\tfrac12} \DeltaTau^3 + \tfrac{1}{1920} t^{(5)}_{\tfrac12} \DeltaTau^5.
\label{eq:regularized_to_model_timestep}
\end{equation}
Here $t^{(i)}_{\tfrac12}$ is the $i$-th derivative of $t$ with respect
to $\tau$ from the begin time $\tau_b$ to
$\tau = \tau_b + \tfrac12 \DeltaTau$ ,  given by
\begin{subequations}
\begin{gather}
t^{(1)}_{\tfrac12} = |\vect{u}|^2,\\
t^{(3)}_{\tfrac12} = \vect{u}^{(2)} \overline{\vect{u}} + 2
|\vect{u}^{(1)}|^2 + \vect{u} \overline{\vect{u}}^{(2)},\\
t^{(5)}_{\tfrac12} = \vect{u}^{(4)} \overline{\vect{u}} + 4
\vect{u}^{(3)} \overline{\vect{u}}^{(1)} + 6 |\vect{u}^{(2)}|^2 + 4
\vect{u}^{(1)} \overline{\vect{u}}^{(3)} + \vect{u}
\overline{\vect{u}}^{(4)}.
\end{gather}
\end{subequations}
Here $\vect{u}$ and all its derivatives are evaluated at half time-steps, using a Taylor expansion at the beginning of the integration
step $\tau_b$, using the approximated regularized snap and crackle,
\begin{subequations}
\begin{gather}
s_b \equiv \vect{u}^{(4)}_b = -6\frac{\vect{u}^{(2)}_b -
\vect{u}^{(2)}_e}{\DeltaTau^2} - \frac{4\vect{u}^{(3)}_b +
2\vect{u}^{(3)}_e}{\DeltaTau},
\label{Eq:snap}\\
c_b \equiv \vect{u}^{(5)}_b = 12\frac{\vect{u}^{(2)}_b -
\vect{u}^{(2)}_e}{\DeltaTau^3} + 6\frac{\vect{u}^{(3)}_b +
\vect{u}^{(3)}_e}{\DeltaTau^2}.
\label{Eq:crackle}
\end{gather}
\end{subequations}
Equation~\ref{eq:regularized_to_model_timestep} has
$\BigO{\DeltaTau^7},$ ensuring an accuracy of $\BigO{\DeltaTau^6}$ in
the final integrated time. The inverse of this transformation cannot
be found analytically. We use Newton-Rapson iteration,
\begin{subequations}
\begin{gather}
\DeltaTau_{[0]} = \frac{h}{|\vect{u}_b|^2},\\
\DeltaTau_{[i+1]} = \DeltaTau_{[i]} - \frac{T(\DeltaTau_{[i]}) -
b}{|\vect{u}_{\tfrac12}|^2}.
\end{gather}
\end{subequations}
Here $\vect{u}_{\tfrac12}$ is given by a third-order Taylor
expansion. In practice, convergence to machine precision is reached in
four iterations.

\subsubsection{Selecting the regularized particle pair}

Selection of which particles are to be regularized is done by calculating a
regularization criterion for each pair of particles. We then sort the
resulting list of pairs, only keeping the $N_{\mathrm{reg}}$ highest-graded pairs. Here $N_{\mathrm{reg}}$ is a free parameter. We then
regularize all particle pairs if they have the smallest norm of the
relative acceleration
\begin{equation}
Q_{\mathrm{reg}} = \frac{G (m_1+m_2)}{r^2}.
\end{equation}

The resulting complete integration steps are listed below.
\begin{enumerate}
\item Select pairs of particles that need to be regularized and
resolve their step.
\item Predict all unregularized particles and all regularized pairs
of particles.
\item Convert coordinates of regularized particles into world
coordinates.
\item Evaluate acceleration and jerks for all pairs of particles,
excluding the Newtonian interaction between regularized pairs.
\item For regularized pairs, calculate derivatives for the
acceleration, jerk, and binding energies.
\item Correct all unregularized particles and all regularized pairs
of particles.
\item Convert coordinates of regularized particles into world
coordinates to synchronize the whole system of particles.
\end{enumerate}

To symmetrize the global time step, we first need to symmetrize the
time steps in their own coordinate system. Regularized time steps need
to be symmetrized in regularized time and then need to be transformed into a world
time step.

\subsection{Implementation}

\subsubsection{\PhFour}

\newcommand{\kira}{\mbox{\texttt{kira}}}
\newcommand{\starlab}{\mbox{\texttt{Starlab}}}
\newcommand{\sapporotwo}{\mbox{\texttt{sapporo2}}}
\newcommand{\multiples}{\mbox{\texttt{multiples}}}


The $N$-body integrator {\PhFour} uses a fourth-order Hermite scheme
similar to that described in \S\ref{sec:PEC}, with some differences in
detail. Its origin lies in the {\kira} integrator, which was part of
the {\starlab} software suite \citep{1998A&A...337..363P}, but most of
the complicating elements in {\kira} , such as treatments of arbitrary
multiples and close encounters, and stellar and binary evolution, have
been removed to be replaced in the {\AMUSE} model by separate
community modules communicating at the Python level. The data
structures in {\PhFour} are deliberately kept simple, making for a
robust module that is easy to manage as a standalone tool, and ph4
facilitates parallelism as well as GPU acceleration, as discussed
below.

Although {\PhFour} uses individual block time steps internally, its
basic mode of operation, like that of most {\AMUSE} modules (see
section\,\ref{sec:AMUSE}), is to take an $N$-body system, typically
synchronized at some initial time $t_0$, and integrate it forward to
some new time $t_1$.  The {\PhFour} module uses individual block time
steps \citep{1986LNP...267..156M}, with essentially the same logic as
described by \cite{1991ApJ...369..200M} and used in {\starlab}. Each
particle $i$ has its own current time $t_i$ and time step $\delta
t_i$. By rounding all steps down to powers of 2, we open the
possibility that many particles can be advanced simultaneously.
Specifically, it is often the case that at any stage of the
calculation, multiple ``$i$-particles'' have the same next time
$t_{next} = t_i+\delta t_i$, allowing a one-time parallel prediction
of all field positions and velocities,
\begin{subequations}
\begin{gather}
\tilde{\vect{r}}_{j} = \vect{r}_j + \vect{v}_j h_j +
\tfrac12 {\vect{a}_j} h_j^2 + \tfrac16 \vect{j}_j
h_j^3
\nonumber\\
\tilde{\vect{v}}_{j} = \vect{v}_j + \vect{a}_j h_j +
\tfrac12 \vect{j}_j h_j^2
\nonumber
\end{gather}
\label{eq:hermite_predictor2}
\end{subequations}
(where $h_j=t_{next}-t_j$) and parallel computation of the predicted
accelerations and jerks of the $i$-particles,
\begin{eqnarray}
\tilde{\vect{a}}_i &=& -\sum_{\substack{j=1 \\ j \ne i}}^{n}
\frac{Gm_j}{r_{ij}^3} \tilde{\vect{r}}_{ij}, \\
\tilde{\vect{j}}_i &=& -\sum_{\substack{j=1 \\ j \ne i}}^{n}
\frac{Gm_j}{r_{ij}^3} \left[\tilde{\vect{v}}_{ij} +
\frac{3(\tilde{\vect{v}}_{ij}\cdot\tilde{\vect{r}}_{ij})\tilde{\vect{r}}_{ij}}{r_{ij}^2}\right],\label{accandjerk}
\end{eqnarray}
where, as before,
$\tilde{\vect{r}}_{ij}=\tilde{\vect{r}}_i-\tilde{\vect{r}}_j$ and
$\tilde{\vect{v}}_{ij}=\tilde{\vect{v}}_j-\tilde{\vect{v}}_j$.

The corrector step in {\PhFour} differs from the more elegant version
described in \S\ref{sec:PEC}, with the formulation chosen to allow the
use of the traditional Aarseth (1985) time step formula. Using the
available derivative information at the beginning and end of the step,
we can estimate the next two derivatives in the Taylor series for the
position and the velocity \citep{1991ApJ...369..200M}, snap and
crackle,
\begin{eqnarray}
\vect{s}_i ~\equiv~ \vect{\ddot{a}}_i
&=& \frac{-6(\vect{a}_i-\tilde{\vect{a}}_i)
- \delta t_i(4\vect{j}_i+2\tilde{\vect{j}}_i)}{\delta t_i^2} \\
\vect{c}_i ~\equiv~ \vect{\dddot{a}}_i,
&=& \frac{-12(\vect{a}_i-\tilde{\vect{a}}_i)
+ 6\delta t_i(\vect{j}_i+\tilde{\vect{j}}_i)}{\delta t_i^3},
\end{eqnarray}
(see also Eqs.\,\ref{Eq:snap} and \ref{Eq:crackle}), leading to the
correction
\begin{eqnarray}
\vect{r}_{i}
&=& \tilde{\vect{r}}_{i}
+ \frac{\delta t_i^4}{24} \vect{s}_i + \frac{\delta t_i^5}{120} \vect{c}_i \\
\vect{v}_{i},
&=& \tilde{\vect{v}}_{i}
+ \frac{\delta t_i^4}{6} \vect{s}_i + \frac{\delta t_i^5}{24} \vect{c}_i .
\end{eqnarray}
The new time step is (Aarseth 1985)
\begin{equation}
\delta t_i = \eta\sqrt{\frac{|\tilde{\vect{a}}_i||\vect{s}_i| + |\tilde{\vect{j}}_i|^2}
{|\tilde{\vect{j}}_i||\vect{c}_i| + |\vect{s}_i|^2}}\end{equation}
(where $\eta$ is an accuracy parameter), rounded down to a power of 2.
We note in passing that {\PhFour} departs from {\kira} in the use of a
novel and more efficient block-scheduling algorithm, which reorders
the block step ($t_{next}$) list and hence determines the next time
step in ${\cal O}(1)$ steps per $i$ particle.

The specialization of {\PhFour} that causes it to perform well, that is, the removal
of most of the complicating physics, in principle also limits its range of
applicability. The code can be coupled to other physics solvers or
other gravity modules on different scales through the AMUSE framework,
see section\,\ref{sec:AMUSE}.

\subsubsection{\HermiteGRX}

We implemented the EIH equations of motion in regularized and
nonregularized forms in standard C++11 \citep{ISO:1998:IIP}. Our
implementation includes the correction terms for EIH equations of
motion, and we include the calculation for the energy and linear
momentum for validation.

The code is parallelized using threads, but not with MPI, and it is not
GPU-enabled. We implemented the post-Newtonian cross terms using two
particle sets: one set of $N$ particles that are affected by general
relativity, and one set of $n$ particles that is purely Newtonian. In
principle, all particles can be considered relativistic, or all can be
Newtonian. In the first case, the code performance scales with $N^3$,
otherwise with the usual $N^2$. In general, the code scales $\propto
N^3 + nN^2 + n^2$. In the case of a galactic nucleus, $N=1$
represents the supermassive black hole and the rest of the particles
$n$ for the other stars.

We implemented 
\begin{itemize}
\item \emph{1PN EIH} for the full EIH equations of
motion, resulting in an $\BigO{N^3}$ time complexity.
\item \emph{1PN Pairwise}, which neglects the acceleration dependence
of the velocity in the EIH equations of motion, resulting in a
$\BigO{N^2}$ algorithm.
\item \emph{1PN GC Crossterms} for the integration of supermassive black
holes in galactic nuclei in which one massive object includes the
cross terms with the low-mass objects, but the low-mass objects are
not relativistic.
\end{itemize}

In \HermiteGRX,\, we implemented several numerical schemes, which
include
\begin{itemize}
\item \emph{Hermite}, for the standard Hermite predictor-corrector
integrator with variable but shared time step
\citep{1991ApJ...369..200M}.
\item \emph{SymmetrizedHermite}, which is a time-symmetrized version
of the Hermite integrator \citep{1995ApJ...443L..93H}.
\item \emph{RegularizedHermite}, which includes regularized close
  approaches for pairs of particles using KS regularization. This
  implementation still uses the standard Hermite scheme for time
  integrations.
\item \emph{SymmetrizedRegularizedHermite}, which adopts
RegularizedHermite, but with symmetrized time step
\citep{1996AJ....112.1697F}.
\end{itemize}

\subsubsection{Implementation in the Astrophysics Multipurpose
Software Environment}
\label{sec:AMUSE}

The Astrophysics Multipurpose Software Environment is a numerical
framework for multiscale and multiphysics simulations
\citep{2018araa.book.....P}. \AMUSE\, uses numerical implementations
for a wide variety of physical processes, including gas dynamics, star
formation, stellar evolution, gravitational dynamics, circumstellar
disk evolution, and radiative transport.  Other physical processes,
such as stellar and binary evolution, the Galactic tidal field or
hydrodynamical processes can be accommodated through the {\AMUSE}
framework.  One complication in multibody dynamics is the formation of
substructures, such as binaries and triples.

The treatment of such local condensations, but also of close
encounters in unsoftened systems, is a much lower-level issue and
requires special treatment.  {\AMUSE} does contain $N$-body modules
that include specialized treatment of close encounters, but most do
not, and because the guiding principle behind {\AMUSE} is to separate
functionality as much as possible, {\PhFour} and \Hermite\, like many
other modules, relies on the {\multiples} module to manage close
encounters and any long-lived binary and multiple systems that
arise. In {\multiples}, a binary or stable multiple, once identified,
is treated as an unperturbed object, possibly with the inclusion of
secular internal evolution terms until it has a close encounter with
another object in the system. At that point, the interaction is
treated as a few-body scattering, which eventually results in the
creation of new stable objects that are then reinserted into the
{\PhFour} or {\Hermite} integration. The {\multiples} module is
described in more detail in \citep[][Sect. 4.5]{2018araa.book.....P}.

{\HermiteGRX} can also be combined with {\multiples} through the
{\AMUSE} framework, but due to the local regularization strategies
using quaternions (see section\,\ref{sec:regularization}), this is not
always necessary.

In \AMUSE, at least two independently developed implementations for
each of the domain-specific solvers are available. This Noah's Ark
approach \citep{2009NewA...14..369P} allows the user to swap one
simulation code for another without any further changes to the runtime
environment (codes, scripts, or underlying hardware) and without the
need to recompile. It is nonintrusive in the sense that underlying
numerical implementations do not require any modifications or
recompilation.

The environment is tuned for running on high-performance
architectures. It also includes support for GPU and massive task-based
parallelism (using message-passing parallelism or open
multiprocessing parallelism).

\subsubsection{Stopping conditions}
\label{sec:stoppingconditions_parallelization}

In the \AMUSE\, framework, codes are interfaced to allow the
generation of a homogeneous and self-consistent simulation environment
for performing multiscale and multiphysics simulations. Many of the
codes in \AMUSE\, are not build for this purpose, but for operating
within a specific domain and parameter range. Due to the interaction
with other codes, the underlying simulation engines (called community
codes) may be forced to operate outside their usual domain range. If
this goes well, the particular community code crashes with the
appropriate memory core dump. In AMUSE, however, such an interrupt is
caught by the framework without resulting in a crash.

In a general simulation environment, this would be the moment another
code takes over to continue the calculation in a different part of
parameter space, another temporal or spatial domain, or including
different physics.  In \AMUSE,\, this problem is addressed by
introducing {\em \textup{stopping conditions}} to interrupt the
particular simulation domain that the underlying code is unsuited to
handle.

In \PhFour\, and \HermiteGRX,\, we implemented three different stopping
conditions. When any of these interrupts is initiated, the code
returns control to the \AMUSE\, framework, where the event can be
handeled appropriately. By default, the stopping conditions are
turned off. The three stopping conditions are described below.
\begin{itemize}
\item {\tt collision\_detection}\\
All particles have a property called {\tt radius}, which is used to
check for collisions between pairs of particles. At each
integration step, we check whether two stars approach each other
within the sum of their radii. For this, we assume that the particles
within that internal code-time step move in a straight line,
\begin{equation}
\vect{r}(s) = \vect{r}_b + (\vect{r}_e - \vect{r}_b).
\end{equation}
Here $\vect{r}_b$ and $\vect{r}_e$ are the positions at the beginning
and end of the current time step, respectively. If the relative
distance between two particles is smaller than the sum of their radii
for some value of $s$, the interrupt {\tt collision\_detection} is
initiated.

\item {\tt Wall\_clock\_time\_out\_detection} 

The interrupt is initiated
when the code takes too long in terms of the wall-clock time to
evolve to a specified model time.
\item {\tt maximum\_number\_of\_integration\_steps\_detection}. 

This is initiated
when the evolution to the required model time takes more than a
predetermined number of time steps to reach the end time of the
simulation.
\end{itemize}


\subsubsection{Parallelization by message-passing in \HermiteGRX}

Solving the EIH equations of motion scales $\propto N^3$, making it a
rather slow $N$-body code. In addition to regularization, we speed the
code up by parallelizing it. We do this by multithreading with
C++11. Communication between threads can be done in shared memory,
which is implemented through the standard library.

\subsubsection{Parallelization by message-passing in \PhFour}

The most important departure of {\PhFour} from {\kira} is the use of MPI parallelism and GPU acceleration in ph4 to increase
performance on parallel architectures and GPU-supported systems.
Although {\kira} in {\starlab} was designed to run on a single
processor, it can operate in parallel using MPI
\citep{2008NewA...13..285P}, but its normal mode of operation is with
special-purpose GRAPE-6
\citep{1996ComPh..10..352M,1998sssp.book.....M} or GPU through the
sapporo library.

A significant difference from the {\kira} formulation is that in
\PhFour,\, all global ($j$-data) calculations are implemented as MPI
parallel tasks with an arbitrary number of workers, and each worker is
optionally GPU accelerated using the {\sapporotwo} library
\citep{2009NewA...14..630G,2015ComAC...2....8B}. This allows an
arbitrary number of GPUs to be configured per MPI worker (subject to the availability of hardware), with all options settable
from the Python-level interface to {\PhFour}. The resulting boost in
speed makes the GPU-accelerated version of \PhFour\, one of the best-performing direct ($N^2$) $N$-body codes in the {\AMUSE} suite (see
Fig 10 in \cite{PortegiesZwart2013456}). We illustrate the higher
speed of the GPU-accelerated version of \PhFour\, in
Fig.\,\ref{fig:scaling_with_N}.

\subsubsection{Parallelization by message-passing in \Brutus}

In Brutus, the $i$-parallellization scheme is implemented, that is, in the
double-force loop, the outer for-loop is parallelized (
\cite{2002NewA....7..373M}). The numbers in arbitrary-precision data
type are converted into an array of characters that are subsequently
communicated through MPI. When it is received, the data type is converted
back into arbitrary-precision variables.


\subsubsection{Speed of light}\label{sect:speedoflight}

\HermiteGRX, \PhFour,\, and \Brutus\, operate internally in
dimensionless $N$-body units, for which $G=1$, the total mass $M=1,$
and the virial radius $R=1$. When we scale to physical units, like in a
$10^6$\,\MSun\, star cluster with a 1\,pc virial radius, we have to
introduce scaling between the physical units and
the $N$-body units. In \AMUSE,\, this is done with a {\tt converter}
\citep{2018araa.book.....P}.

In \HermiteGRX,\, we use the relative speed of light parameter $\zeta$ in
terms of the mean velocity in N-body units ($\frac{1}{2}\sqrt{2}$),
and size
\begin{equation}
v = \sqrt{G m \over r}.
\end{equation}

The speed of light in physical units also has to be converted into
$N$-body units. In \HermiteGRX,\, we do this by defining the relative
speed of light, or the relativisticality of the conditions, as
\begin{equation}
\zeta = \frac{1}{2}\sqrt{2}/v
.\end{equation}
For $m = 10^6$\,\MSun\  and $r = 1$\,au, we obtain
$\zeta = {\cal O}(2.4 \cdot 10^{-5})$, and for $r = 1$\,pc, we obtain
$\zeta = {\cal O}(0.01)$. Here the system becomes Newtonian for
$\zeta \rightarrow 1$, and lower values of $\zeta$ mean a more
relativistic system. The assumption of the Tailor-series expansion
tends to break down for $\zeta \apgt 0.3$ (see
sect.\,\ref{Sect:validity} or better, read
\citep{2011PNAS..108.5938W}).

In the main paper we express the speed of light in terms of the
velocity dispersion $v/c$, which is more natural. The conversion from
$\zeta$ to $v/c$ is $v_{\rm nbody} = 1/\sqrt{2}$:
\begin{equation}
v/c = v_{\rm nbody}/c_{\rm nbody} = v_{\rm nbody}/(\zeta c)
.\end{equation}

\subsubsection{Example}
\label{sec:Example}

In Listing~\ref{src:HGRX_main.py} and \ref{src:HGRX_loop.py}, we showed
a rudimentary AMUSE script for calculating the evolution of $N$ black
holes to 1-PN EIH equations of motion (including the cross terms).
Listing\,\ref{src:HGRX_main.py} showed how to start the code, and
Listing\,\ref{src:HGRX_loop.py} showed the main event loop. Here we
adopted the symmetrized regularized Hermite integrator with the full
EIH equations of motion to first order. We adopted a radius of the
particles of $10GM/c^2$.


{\small
\begin{lstlisting}[linerange={43-91},float,caption=source listing for
simulation gravitating system., label=src:HGRX_loop.py]
def gravity(bodies,
            converter,
            integrator,
            dt_param=0.1,
            t_end = 200| units.yr):

    gravity = HermitePN(converter)
    gp = gravity.parameters
    gp.dt_param = dt_param
    gp.perturbation = '1PN_EIH'
    gp.integrator = integrator
    gp.light_speed = constants.c
    gp.num_threads = 1

    gravity.particles.add_particles(bodies)
    sc = gravity.stopping_conditions.\
        collision_detection
    sc.enable()
    Etot_init = gravity.kinetic_energy +\
        gravity.potential_energy
    from_gravity = gravity.particles.\
        new_channel_to(bodies)

    dt = 0.001 * t_end
    while gravity.get_time()<t_end:

        time = gravity.model_time + dt
        gravity.evolve_model(time)
        from_gravity.copy()

        Ekin = gravity.kinetic_energy 
        Epot = gravity.potential_energy

        Etot = Ekin + Epot
        dE = (Etot_init-Etot)/Etot
        print("T =", gravity.get_time(),
              "M =", bodies.mass.sum(),
              "E = ", Etot,
              " Q = ", -Ekin/Epot)
        print("dE =", dE)

        if sc.is_set():
            print("Collision detected=",
                  len(sc.particles(0)))
            resolve_collision(sc,
                              gravity,
                              bodies)
    gravity.stop()
\end{lstlisting}
}

{\small
\begin{lstlisting}[linerange={197-210},float,caption=Source listing for
simulation gravitating system., label=src:HGRX_main.py]
mass = 1.e+6 |units.MSun
size = 0.001|units.pc
conv = nbody_system.nbody_to_si(mass,
                                size)
bodies = new_plummer_model(10, conv)
bodies.radius = Schartzschield(bodies.mass)
integrator = 'SymmetrizedRegularizedHermite'
    
gravity(bodies,
        integrator,
        converter,
        dt_param = o.dt_param,
        t_end = o.t_end)
\end{lstlisting}
}

\begin{figure}
\centering
\includegraphics[width=\columnwidth]{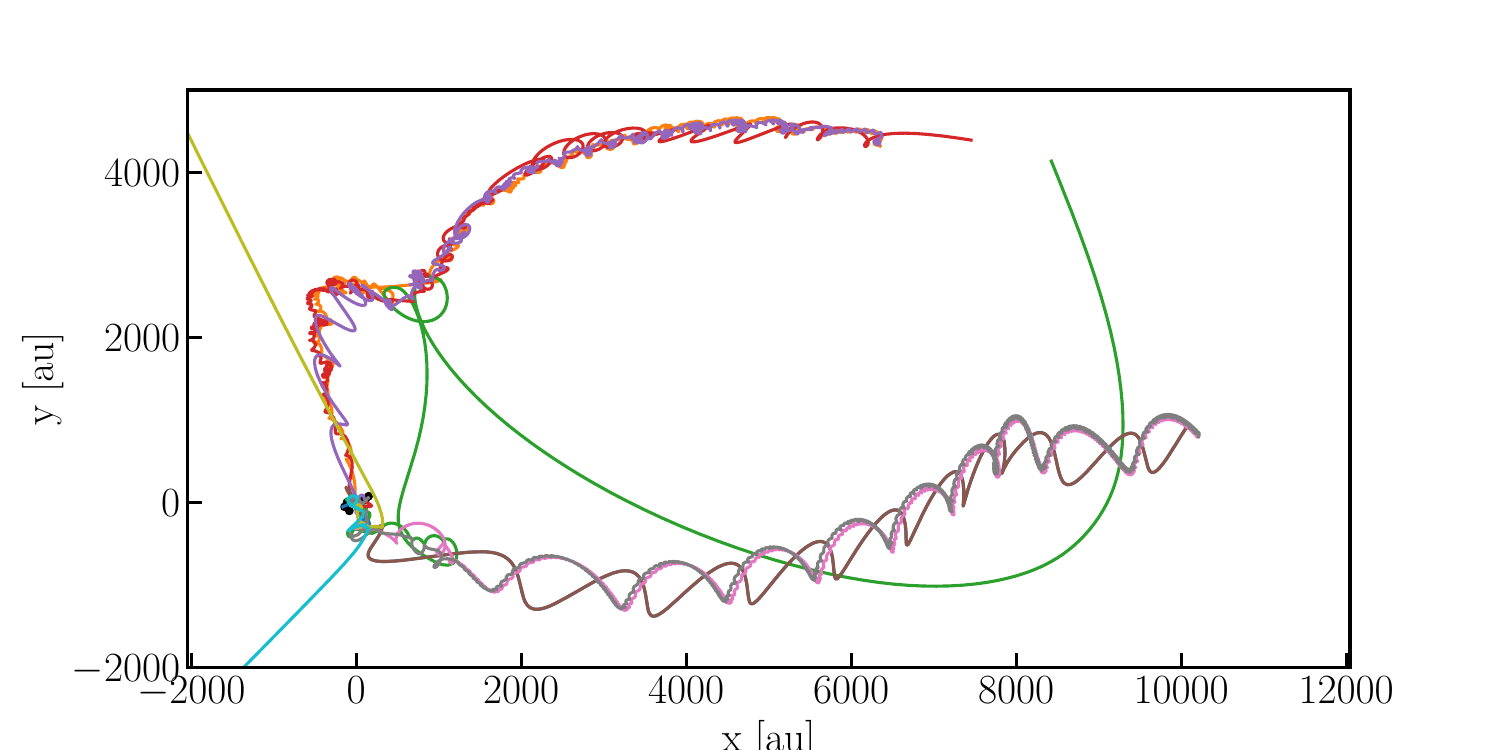}
\caption{Integration of an $N=10$ Plummer sphere with a virial radius
of 1\,mpc and a total mass of $10^7$ equal-mass black holes for 200
years. }
\label{fig:N10SMBH}
\end{figure}

\section{Validation of the code}
\label{sec:validation}

\subsection{Two-body systems}

\subsubsection{Integrator performance}

The Newtonian solution for the equations of motion for two particles
was described by \cite{Kepler:1609}.  In the absence of general
relativity, the solution is static, with the exception of the mean
anomaly.  As a first test, we check the conservation of these
theoretically conserved Keplerian elements for one orbit.

\begin{figure*}
  \centering
  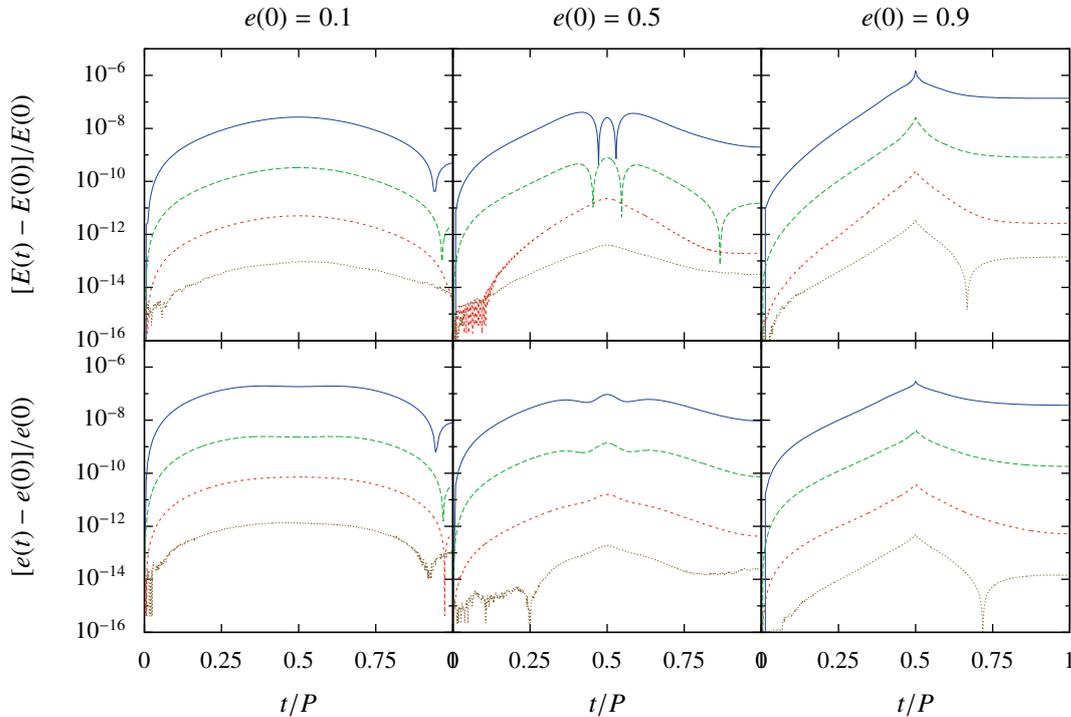
\caption{Relative error in the energy (top row of panels) and
  eccentricity (bottom row) for integrating a two-body system using
  the Hermite integrator without post-Newtonian terms for initial
  eccentricities $e_{0}\in \{0.1, 0.5, 0.9\}$ (from left to right), for
  time-step parameters $\eta\in\{0.03, 0.01, 0.003, 0.001\}$ (top to
  bottom in blue, green, red, and orange, respectively). }
\label{fig:newtonianintegrationerrorshermite}
\end{figure*}

\begin{figure*}
\centering
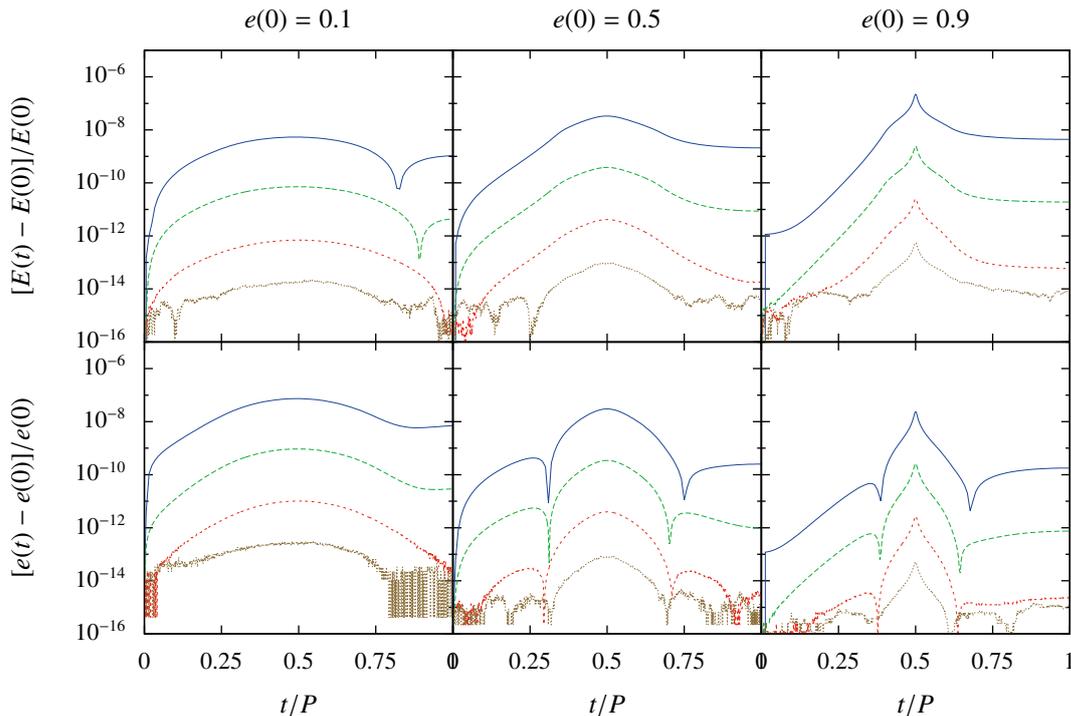
\caption{Relative integration errors in the energy (top row of panels)
  and eccentricity (bottom row) for one orbit using the regularized
  Hermite integrator for Newton's equations of motion for initial
  eccentricities $e_{0}\in \{0.1, 0.5, 0.9\}$ (left to right panels,
  respectively), for various time step parameters
  $\eta\in\{0.03, 0.01, 0.003, 0.001\}$ (in colour: blue, green, red
  and orange, respectively). }
\label{fig:newtonianintegrationerrorsregularizedhermite}
\end{figure*}

We varied the initial eccentricity $e_{0}$ and time-step parameter $\eta$
and adopted masses of $M = 10^6 \unit{M_\odot}$,
$m = 50 \unit{M_\odot}$, an initial semimajor axis $a_0 = 1\unit{mpc}$,
and an initial eccentrity $e_{0}\in \{0.1, 0.5, 0.9\}$.  For the time-step
parameter $\eta\in\{0.03, 0.01, 0.003, 0.001\}$ , and we integrated for one
orbital period \citep{Kepler:1609},
\begin{equation}
P=2\pi \sqrt{\frac{a_0^3}{G (M + m)}}.\end{equation}
In
figs.\,\ref{fig:newtonianintegrationerrorshermite}~and~\ref{fig:newtonianintegrationerrorsregularizedhermite}
we show the relative errors in energy and eccentricity in these
integrations for the unregularized and the regularized Hermite integrator,
respectively.  For each calculation, the error in the energy and
eccentricity reaches a maximum near pericenter.

The relative error in the regularized Hermite integration in
fig.\,\ref{fig:newtonianintegrationerrorsregularizedhermite} is one
orders of magnitude smaller than the unregularized integrator error
(see Figure~\ref{fig:newtonianintegrationerrorshermite}).
As intended in its design, the regularized Hermite integrator
performes equally in terms of conserving energy and angular momentum
compared to the nonregularized integrator for low-eccentricity
orbits, and considerably better for eccentric orbits.

\begin{figure*}[t]
\centering
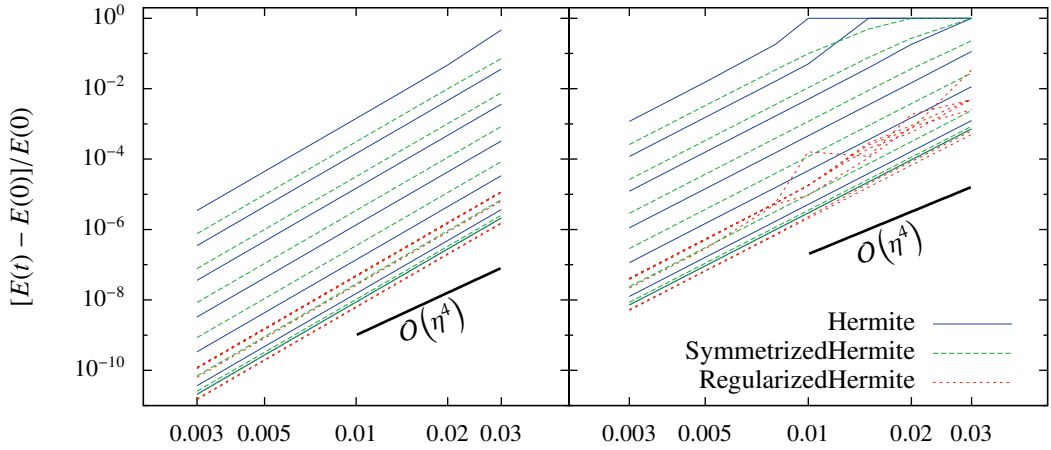
\caption{Relative energy errors for integration for
  $t_{\mathrm{end}}/P=10^3$ and $t_{\mathrm{end}}/P=3.4\times10^5$ for
  initial eccentricities
  $e_{0} \in \{0.01, 0.1, 0.5, 0.9, 0.99, 0.999, 0.9999\}$ as a function
  of the time-step parameter $\eta$. The various integrators are
  indicated with colors (see legend). }
\label{fig:newtonian_conserved_elements}
\end{figure*}

In fig.\,\ref{fig:newtonian_conserved_elements} we present the secular
drift in energy as a function of $\eta$ for initial eccentricities
$e_{0} \in \{0.01, 0.1, 0.5, 0.9, 0.99, 0.999, 0.9999\}$.  The same
set of initial conditions were adopted in \cite{2014MNRAS.443..355H}.
These integrations were performed for $t_{\mathrm{end}}/P=10^3$ and
$t_{\mathrm{end}}/P=3.4\times10^5$.  As expected for a fourth-order
integrator, the integration error scales with $\BigO{\eta^4}$.  The
regularized Hermite integrator outperforms the other integrators in
terms of energy conservation for high eccentricity.  The
nonregularized integrators have a secular growth of the energy error.
We therefore prefer the regularized integration for the long-term
evolution of highly eccentric orbits.

\subsubsection{Post-Newtonian corrections}
\label{sec:relativistictwobody}

General relativity changes the dynamics of astronomical systems. This
results in variations in the evolution of the orbital elements for
two-body systems.

The secular evolution in the argument of periastron forms one of the
major tests for general relativity.  The osculating elements,
instantaneous orbital parameters under the influence of a perturbing
acceleration, can be derived from the perturbing acceleration using
Lagrange's planetary equations
\citep{Lagrange1772,merritt2013dynamics}.

\begin{figure*}
  \centering
  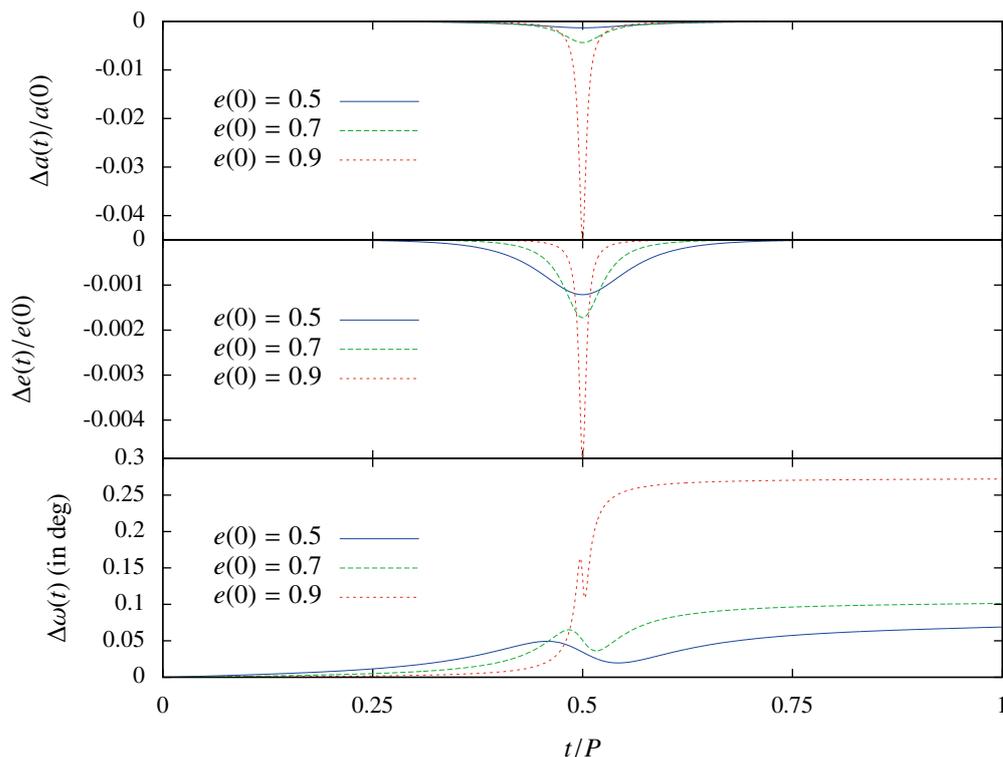
\caption{Osculating elements as a function of time for a binary with a
  stellar mass 50\,\MSun\, in orbit around a $10^6$\,\MSun\,
  supermassive black hole in a $a=1$\,mpc orbit with an initial
  eccentricity $e_{0}\in\{0.5, 0.7, 0.9\}$.  Only the argument of
  periastron, $\omega$, shows a secular variation (bottom panel).}
\label{fig:osculatingelements}
\end{figure*}

We numerically integrated Lagrange's planetary equations for the
first-order post-Newtonian perturbation using Euler's method
\citep{Euler1760}.  We decreased the time step until the solution
convergenced. For the earlier adopted binary, we used initial
eccentrities $e_{0}\in\{0.1,0.5,0.9\}$ . The resulting theoretical
osculating elements are presented in
fig.\,\ref{fig:osculatingelements}.  The post-Newtonian terms become
more important for larger eccentricities because near periastron, the
relative velocity of the particles becomes large while the distance
becomes smaller for higher eccentricities.

\begin{figure*}
\centering
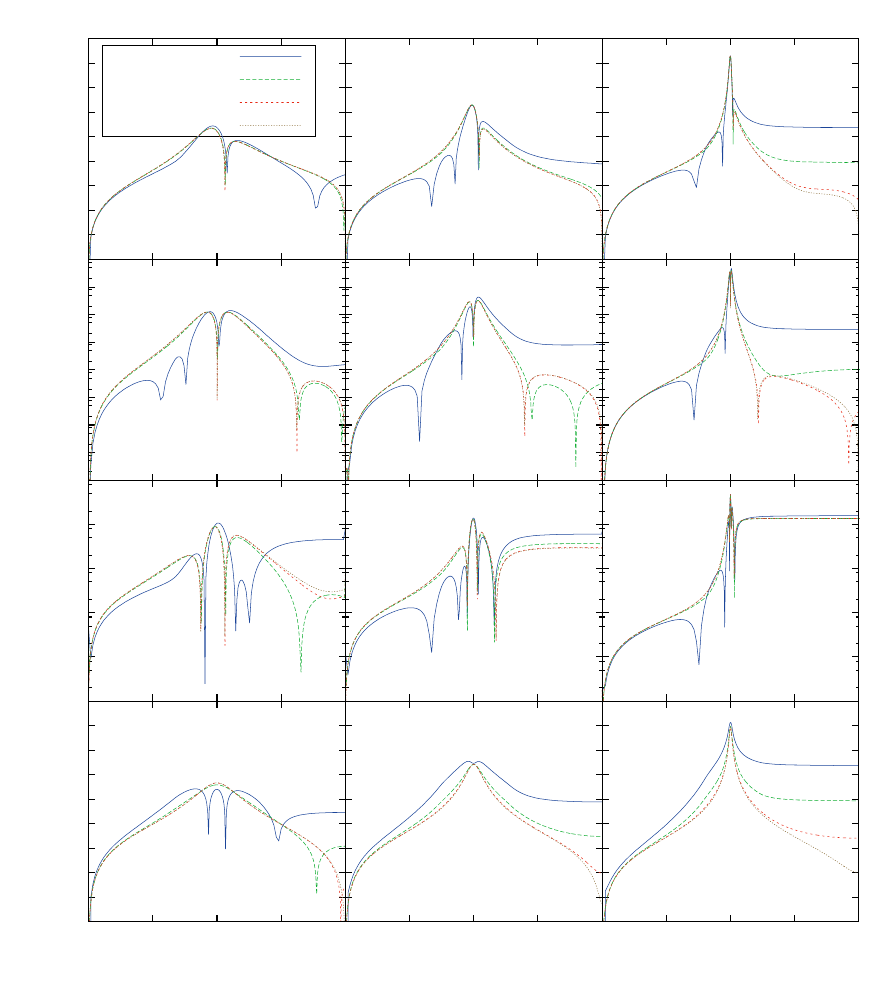
\caption{Relative errors in semimajor axis, eccentricity, argument of
  pericenter and post-Newtonian energy for a relativistic binary
  composed of a 50\,\MSun\, star in orbit around a $10^6$\,\MSun\,
  supermassive black hole in a $a=1$\,mpc orbit with an initial
  eccentricity $e_{0}\in\{0.5, 0.7, 0.9\}$.  The initial eccentricity
  was chosen to be $e_{0}\in\{0.5, 0.7, 0.9\}$. The integration was
  done using a regularized Hermite integrator, using a time step
  parameter of $\eta\in\{0.03, 0.01, 0.003, 0.001\}$ to show
  convergence (blue, green, red, and orange, respectively).}
\label{fig:postnewtonianconservedelements}
\end{figure*}

Direct numerical integration of the equations of motion should
reproduce these osculating elements, and in particular, the secular
change in the argument of periastron. We used the same binary as before
and a post-Newtonian perturbation to integrate one orbit.  For a two-body system, the EIH equations of motion reduce to the
pairwise approximation.  In
fig.\,\ref{fig:postnewtonianconservedelements} we present the
relative error of the osculating elements integrated using the
regularized Hermite and compare them to the theoretical prediction.
We also show the integration error in total energy, including the
post-Newtonian energy (eq.\,\ref{eq:energy}).

The error in the osculating element remains finite, even for very
low values of $\eta$.  The discrepancy is largest near pericenter
and smallest near apocenter.  The maximum relative error we
observe in fig.\,\ref{fig:postnewtonianconservedelements} in the first
post-Newtonian correction is several orders of magnitude smaller than
the theoretical predictions, making an implementation error
improbable. The discrepancy between the theoretical value and the
numerical results may well be caused by round-off, in particular since
the post-Newtonian corrections require quite a large number of
operations per step and the time step is small.  With a time-step
parameter $\eta = 10^{-3}$ and $\sim 10^3$ operations per
post-Newtonian evaluation, we expect the round-off error to grow by
some $\text{six}$ orders of magnitude over one orbital period. With the adopted
16 mantissa implementation, we then arrive at a mean error of about $\BigO{10^{6}}$, which is consistent with the observed energy
error (bottom row of panels in
fig.\,\ref{fig:postnewtonianconservedelements}).

For validation and verification, we recalculated the same initial
conditions using ARCHAIN \citep{mikkola2008implemnting}. The results
are indistinguishable from our implementation. The discrepancy
between the numerical $N$-body result and the converged semianalytic
solution then manifests itself in two independently developed codes.
We argue that the conserved energy corresponding to the equations of
motions truncated to first post-Newtonian order contains some second
post-Newtonian terms that are ignored. Because these terms have order
$\BigO{v^4/c^4}$ , they tend to be important for higher eccentricity and
near pericenter, which is precisely what we observe in our
simulations.

We caution about judging the accuracy of an
$N$-body simulation based on energy conservation alone, in particular
when considering the second-order post-Newtonian terms
\citep{2018CNSNS..61..160P}.  On the other hand, the secular evolution
of the energy does not seem to be affected. For the 1-PN terms
(including the cross terms) adopted in the main paper, the enery is
conserved.

\section{Relativistic von Zeipel-Lidov-Kozai effect}

There are only a few known solutions to the three-body problem. In addition to several semianalytic solutions to resonant cases, such as we
find in periodic braids \citep{Moore93braidsin,0951-7715-11-2-011},
there is also a theory about the general behavior of hierarchical
three-body systems. In this section, we focus on the latter, in
particular since there is a rich body of literature about the
associated phenomena observed in hierarchical triples. We refer to
this theory as von Zeipel-Lidov-Kozai cycles
\citep{1910AN....183..345V,1962PSS..9..719L,1962AJ.....67..591K}, and
there is copious literature about the theory
\citep{2020CosRe..58..249E,2021MNRAS.500.3481H}, its deeper
consequences \citep{2019A&A...627A..17D}, or the observational aspects
\citep{2020arXiv201010534S}. Here we adopt the von Zeipel-Lidov-Kozai
effect of testing the $\text{three}$-body methods for Newtonian and relativistic
dynamics.

\subsection{Newtonian von Zeipel-Lidov-Kozai problem}

Numerical and analytical dynamical stability arguments
\citep{georgakarakos2008stability} indicate that most triples with
comparable masses and mutual distances are dynamically unstable and
ultimately decay into a binary and a single star. Counterexamples
exist, however, in which the three stars form a stable and periodic
braid. Some of these solutions are even stable to second-order
post-Newtonian order \citep{lousto2008three}. Although dynamically
unstable triples are rare, they are of considerable theoretical
importance. From an observational perspective, they are also of
interest because they lead to relatively high-velocity stars
or stellar mergers.

\begin{figure*}[th]
  \centering
  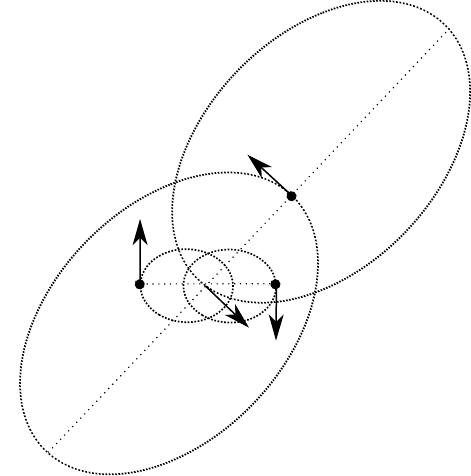
\caption{Schematic image of a hierarchical binary, consisting of
the inner binary with masses $m_1$ and $m_2$, orbiting each other
with a semimajor axis $a_1$ and an eccentricity $e_1$, orbited by
a third mass $m_3$, with semimajor axis $a_2$ and eccentricity
$e_2$. This figure is not to scale, as typically $a_2\gg a_1$.}
\label{fig:hierarchicalbinary}
\end{figure*}

Except for braids, stable triples are always hierarchical in the sense
that they can be described as an inner binary and a third body that
orbits the center of mass of the inner binary at a distance much
larger than the separation of the inner binary, as shown schematically
in Figure~\ref{fig:hierarchicalbinary}. Such hierarchical triples are
rather common, and \cite{tokovinin2014from} estimated their fraction
among solar-type stars in the solar neighborhood $\sim 13$\,\%.

An important aspect in the dynamics of hierarchical triples is the
periodic exchange of orbital angular momentum between the inner and
outer binary. Such coupling is only effective when the relative
inclination between the two orbital planes of the inner and outer
orbits exceed some critical value
\begin{equation}
i_{\mathrm{rel, crit}} = \arccos{\left( \sqrt{\tfrac35} \right)} \approx 39.2^\circ.
\end{equation}
In this case, the inner binary and the relative inclination evolve
periodically. To first nonzero order, this periodicity conserves
\begin{equation}
L_z \propto \sqrt{1-e_1^2} \cos{(i_{\mathrm{rel}})} = {\textrm constant} 
\label{eq:kozailinearmomentum}
.\end{equation}
During such von Zeipel-Lidov-Kozai cycles, the orbital energies remain
constant, which leads to constant semimajor axes. During such a
cycle, the eccentricity of the inner binary can reach values as high
as $e_1\sim1-10^{-6}$, as we show in
Figures~\ref{fig:kozaiflipunregularized} and
\ref{fig:kozaiflipregularized}. Such highly eccentric orbits are
easily subject to tidal effects or the emission of gravitational
waves, and could lead to stellar collisions
\citep{2016MNRAS.456.4219A}. Fortunately, such high-eccentricity
encounters are not expected to naturally occur in large $N$-body
systems, except in the presence of hierarchical multiple
subsystems. If such high eccentricities are relevant, the entire
integration scheme, the Taylor expansion adopted for the
post-Newtonian terms, and the possibility of tidal effect should all be
reconsidered.

One great advantage of von Zeipel-Lidov-Kozai cycles is the possibility
of deriving the secular evolution analytically by averaging over the
inner and outer orbits. Here we assume that the orbital parameters
vary slowly compared to the outer orbit: the timescale on which the
orbital angular momentum of the inner binary varies is small compared
to the inner orbital period \citep{antonini2014black}. Direct
numerical integration of the equations of motion of hierarchical
triples remains important for validating the underlying assumption on the
system's hierarchy.

\begin{figure*}[th]
  \centering
  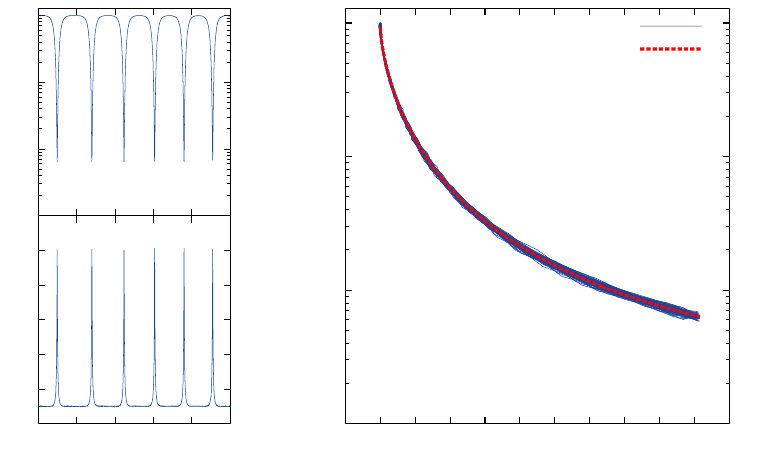
\caption{A few Kozai-Lidov cycles with their distinctive signature of
high-eccentricity (low $1-e_1$) spikes. On the right,
the eccentricity and relative inclination are plotted against each
other from $t=0$ to $t=10000\unit{yr}$, including the analytical
prediction (based on the initial conditions and conservation of
linear momentum in Equation~\ref{eq:kozailinearmomentum}). The
thickness of this line is due to numerical integration
errors. Initial condition is a binary of masses
$m_1=1\unit{M_{\mathrm{Jup}}}$ and $m_2=1\unit{M_\odot}$,
semimajor axis $a_1=0.005\unit{AU,}$ and eccentricity $e_1=0.001$,
orbited by a third body of mass $m_3=10^6 \unit{M_\odot}$ at
semimajor axis $a_2=51.4\unit{AU}$, eccentricity $e_2=0.7,$ and
relative inclination $i_{\mathrm{rel}}=95^\circ$. Integration was
done using a regularized Hermite integrator with a time-step
parameter of $\eta=0.01$.}
\label{fig:kozaicycle}
\end{figure*}

In Figure~\ref{fig:kozaicycle} we present the result of several such
numerical integration of some von Zeipel-Lidov-Lidov cycles together
with the relation between $e_1$ and $i_{\mathrm{rel}}$.  The
lowest-order approximation for von Zeipel-Lidov-Kozai cycles is the
result of the quadrupole term in the multipole expansion that is used
in the derivation. The corresponding timescale is approximately
\citep{naoz2013resonant}
\begin{equation}
t_{\mathrm{quad}}^{\mathrm{Newton}} \sim \frac{2\pi a_2^3 (1-e_2^2)^{\tfrac32} \sqrt{m_1+m_2}}{a_1^{\tfrac32} m_3 \sqrt{G}},
\end{equation}
in which the eccentricity reaches a maximum value of
\begin{equation}
e_{1, \max} = \sqrt{1-\tfrac53 \cos^2(i_{\mathrm{tot}}}).
\label{eq:maxecc}
\end{equation}
The latter expression is only valid in the limit in which
the octopole terms vanish, the test particle quadrupole order limit
\citep{naoz2013secular}, in which $a_2 \gg a_1$. When the
octupole-level terms become important, they can be seen as a
modulation of von Zeipel-Lidov-Kozai cycles. The importance of the
octupole-level variations can be quantified by considering the ratio
of the octupole to quadrupole-level coefficients,
\begin{equation}
\frac{C_3}{C_2} = \frac{15}{4} \frac{\epsilon_M}{e_2}.
\end{equation}
Here $C_2$ and $C_3$ are the quadrupole- and octupole-level
coefficients given by~\cite{naoz2013secular}, and $\epsilon_M$ is the
relative importance of the octupole-level term in the secularized
Hamiltonian,
\begin{equation}
\epsilon_M = \left(\frac{m_1-m_2}{m_1+m_2}\right) \left(\frac{a_1}{a_2}\right) \left(\frac{e_2}{1-e_2^2}\right).
\label{eq:octupoleimportance}
\end{equation}
This suggests that octupole-level variations are important for
eccentric inner-binaries with high-mass components. We note here that
$\epsilon_M$ is independent of the mass of the third (outer) body
$m_3$, but depends on its orbital parameters $a_2$ and $e_2$. The
timescale of the octupole variation can be defined in a similar fashion,
\begin{equation}
t_{\mathrm{oct}}^{\mathrm{Newton}} \sim \frac{4}{15} \epsilon_M^{-1} t_{\mathrm{quad}}^{\mathrm{Newton}}.
\end{equation}
\cite{naoz2013secular} demonstrated that this octupole variation can
have consequences on the maximum eccentricity reached during the
evolution of the system. The induction of variations in the relative
inclination $i_{\mathrm{tot}}$ over time can lead to a flip in the
inner binary's orbit. The maximum eccentricity is reached at the
moment the flip occurs (see eq.~\ref{eq:maxecc}).

\begin{figure*}[th]
\centering
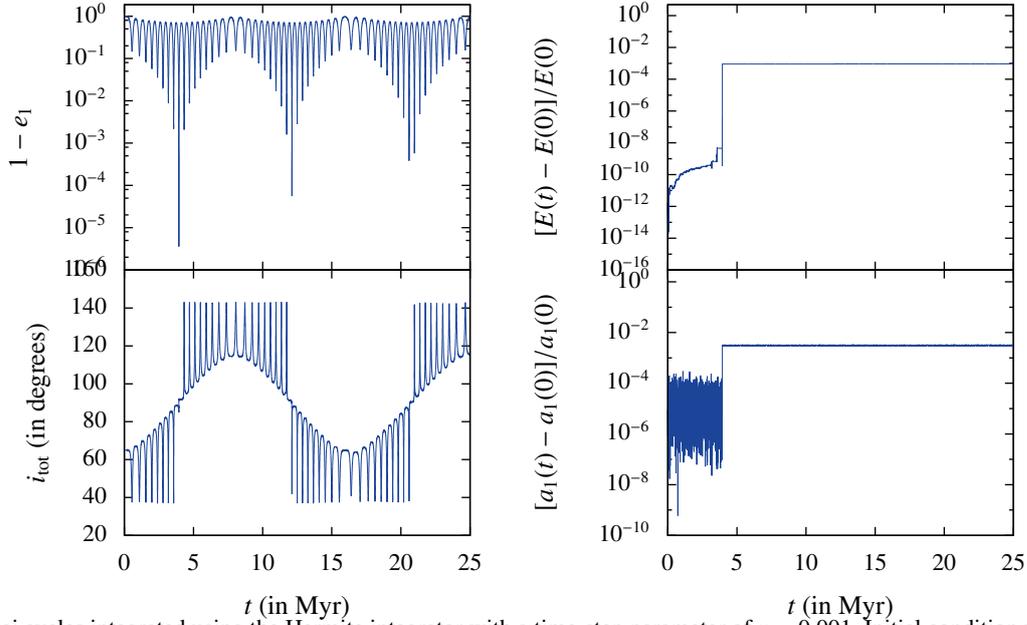
\caption{Kozai cycles integrated using the Hermite integrator with a
time-step parameter of $\eta=0.001$. Initial condition was a binary
with masses $m_1=1\unit{M_{\mathrm{Jup}}}$ and $m_2=1\unit{M_\odot}$
with semimajor axis $a_1=6\unit{AU}$ and eccentricity $e_1=0.001$
orbited by a third body of mass $m_3=40\unit{M_{\mathrm{Jup}}}$ at a
distance $a_2=100\unit{AU}$ and eccentricity $e_2=0.6$. The initial
relative inclination was $i_{\mathrm{tot}}=65^\circ$. These initial
conditions are identical to Figure~3 of~\cite{naoz2013secular}. The integration errors in both the energy and the semimajor
axis of the inner binary rapidly rise at $t\sim4\unit{Myr}$, where
the inner binary reaches maximum eccentricity.}
\label{fig:kozaiflipunregularized}
\end{figure*}

We demonstrated in sect.\,\ref{sec:validation} that the unregularized
Hermite scheme is prone to introducing integration errors for highly
eccentric orbits. This may pose a problem while integrating triples
that are subject to von Zeipel-Lidov-Kozai cycles, for which the inner
binary can become highly eccentric. We illustrated this in
fig.\,\ref{fig:kozaiflipunregularized}, wherethe inner binary reaches an eccentricity in excess of $1-10^{-6}$  at $t\sim4\unit{Myr}$
,
leading to a relative integration error of about eight orders of magnitude
larger than when integrating a circular orbit. The result of
regularized Hermite scheme for the same triple is presented in
fig.~\ref{fig:kozaiflipregularized}, showing a considerably better
conservation of energy, and it is also faster.

\begin{figure*}[th]
\centering
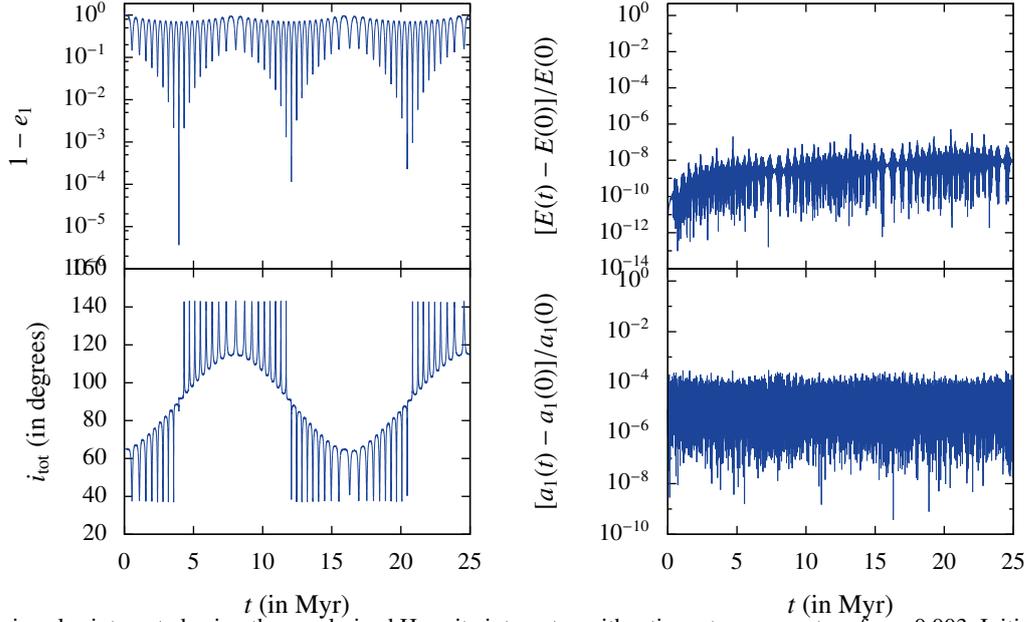
\caption{Kozai cycles integrated using the regularized Hermite
integrator with a time-step parameter of $\eta=0.003$. Initial
conditions are identical to those in
Figure~\ref{fig:kozaiflipunregularized}, using a similar wall-clock
run time. The large integration errors at high eccentricities are now
absent, as regularization is applied to the inner binary.}
\label{fig:kozaiflipregularized}
\end{figure*}

\subsection{Relativistic von Zeipel-Lidov-Kozai problem}

Triple star systems that are subject to von Zeipel-Lidov-Kozai cycles
may be affected by general relativity, in particular, when this leads
to highly eccentric orbits. Since these cycles are the result of
secular resonances between an inner and an outer orbit, variations in
the parameters of the inner binary on a timescale similar to the
secular resonance tend to quench the effect and reduce the extremes
in the von Zeipel-Lidov-Kozai cycles.

The timescale on which the inner binary orbit is affected by general
relativity is \citep{naoz2013resonant}
\begin{equation}
t_{\mathrm{i}}^{\mathrm{1PN}} \sim 2\pi \frac{a_1^{\tfrac{5}{2}} c^2 (1-e_1^2)}{3G(m_1+m_2)^{\tfrac{3}{2}}}.
\end{equation}
We define the relative (dimensionless) parameter ${\cal R}$ as the ratio
between the 1-PN terms and Newtonian quadrupole timescales for a
circular inner orbit,
\begin{equation}
{\cal R} = \left.\frac{t_{\mathrm{i}}^{\mathrm{1PN}}}{t_{\mathrm{quad}}^{\mathrm{Newton}}}\right|_{e_1=0} = \frac13 \frac{(a_1/{\cal R}_1)^4}{(a_2/{\cal R}_3)^3} \frac{1}{(m_3/m_1)^2 (1-e_2)^{\tfrac{3}{2}}}.
\end{equation}
Here ${\cal R}_1$ and ${\cal R}_2$ are the gravitational radii of the inner binary
and the outer orbiting tertiary body. They are given by
\begin{subequations}
\begin{equation}
{\cal R}_1 = \frac{G (m_1+m_2)}{c^2},\\
\end{equation}
\end{subequations}
and
\begin{subequations}
\begin{equation}
{\cal R}_2 = \frac{G m_3}{c^2},
\end{equation}
\end{subequations}
respectively.

A maximum in the eccentricity of the inner orbit is induced when the
timescale for which the inner orbit evolves due to general relativity
is on the same order as one von Zeipel-Lidov-Kozai cycle due to
classical Newtonian resonance. The criteria for this to happen are
${\cal R} \sim 1$ and $m_3 \gg m_1 > m_2$ \citep{naoz2013resonant}.

\begin{figure*}[th]
  \centering
     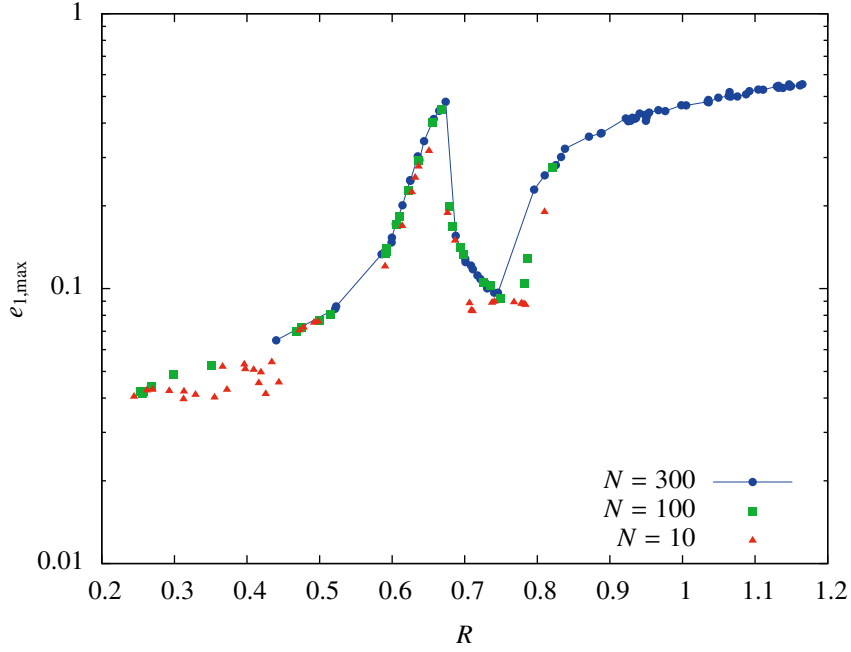
\caption{Maximum eccentricity reached after
$t_{\mathrm{end}}=N t_{\mathrm{quad}}^{\mathrm{Newton}}$, where
$N\in\{10,100,300\}$ are denoted by the colors. The initial
conditions are $m_1=1\unit{M_\odot}$, $m_2=0.001\unit{M_\odot}$,
$m_3=10^6 \unit{M_\odot}$, $a_1=10^5 {\cal R}_1$,
$i_{\mathrm{tot}}=65^\circ$, $e_1=0.001$, $e_2=0.7$, and $a_2$ was
varied to simulate different ${\cal R}$. An
eccentricity excitation around ${\cal R}=0.65$ is evident. Simulations were
performed using a regularized Hermite integrator with a time-step
parameter of $\eta=0.003$ for $N=10$ and $N=100$ and $\eta=0.002$
for $N=300$. Each data point is an independent simulation that
took a wall-clock time of $\sim20\unit{min}$ for $N=10$,
$\sim3\unit{hours}$ for $N=100,$ and $\sim12\unit{hours}$ for
$N=300$.}
\label{fig:maxeccresonant}
\end{figure*}

In fig.~\ref{fig:maxeccresonant} we compare \HermiteGRX\, with the
results presented in Naoz et al. (2013b, see their
fig.5)\nocite{naoz2013resonant}. For the initial conditions, we used a
star with planet that orbit a supermassive black hole. The inner
binary, the star and planet, have masses $m_1=1\unit{M_\odot}$ and
$m_2=0.001\unit{M_\odot}$, with semimajor axis $a_1=10^5 {\cal R}_1$,
eccentricity $e_1=0.001$, and inclination $i_{\mathrm{rel}}=65^\circ$
with respect to the outer orbit. The third body is a supermassive
black hole with mass $m_3=10^6 \unit{M_\odot}$ and orbits the inner
binary with a semimajor axis $a_2$ and eccentricity $e_2=0.7$ for
various values of ${\cal R}$. The system is evolved to
$t_{\mathrm{end}}= s t_{\mathrm{quad}}^{\mathrm{Newton}}$ for three
value of $s$, where we adopted a time-step parameter $\eta=0.003$ for
$s=10$ and $s=100$, and $\eta=0.002$ for $s=300$. The maximum
eccentricity reached during this time interval for various values of
${\cal R}$ is plotted in fig.\,\ref{fig:maxeccresonant}. Some minor
deviation from the theoretical curve, in particular for $s=10$ is
caused by our incomplete sampling because we only determined the
eccentricity of the inner orbit near apocenter. In addition, the
distance between the third body to the inner binary also introduces
variations in the eccentricity of the inner orbit.

In fig.\,\ref{fig:maxeccresonant} we show the resonant-like
eccentricity excitation discussed in \cite{naoz2013resonant}. Our
calculations are integrated directly, whereas \cite{naoz2013resonant}
conducted an orbit-averaged integration, using test particles with
1-PN including terms only up to \BigO{a_1^{-2}} and \BigO{a_2^{-2}}
and the term, whereas we perform a direct $N$-body integration of the
full EIH equations of motion to 1-PN with finite masses.

The first peak in the resonant structure of the eccentricity in
fig.~\ref{fig:maxeccresonant} is shifted to ${\cal R}\sim0.65$ with
respect to to ${\cal R}\sim0.55$ in fig.\,5 of
\cite{naoz2013resonant}.  The calculations performed by
\cite{naoz2013resonant} adopted orbit averaging, which gives a
considerable speed-up compared to direct integration. On the other
hand, however, this may lead to missing the collision because the
maximum eccentricity is reached in a time interval that is shorter
than the orbital period of the outer binary. Orbit averaging over the
outer orbit then lacks the resolution to resolve the maximum
eccentricity in the inner orbit, whereas in the direct $N$-body
integration presented in fig.\,\ref{fig:maxeccresonant}, we do resolve
the evolution of the eccentricity of the inner orbit.

\begin{figure*}[th]
\centering
  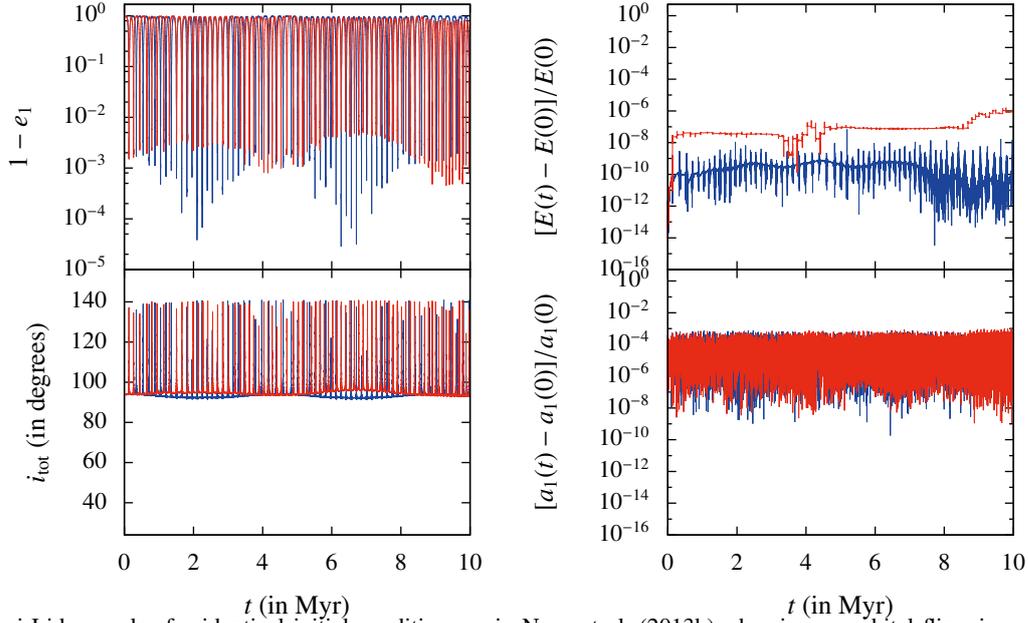
\caption{Kozai-Lidov cycles for identical initial conditions as in
\cite{naoz2013resonant}, showing no orbital flips, in contrast to
the orbital flips shown in that paper. The initial conditions of
this direct numerical integration consist of a similar-mass inner
binary ($m_1=10\unit{M_\odot}$, $m_2=8 \unit{M_\odot}$,
$a_1=10\unit{AU}$, $e_1=0.001$) orbited by a third body
($m_3=30\unit{M_\odot}$, $a_2=502\unit{AU}$, $e_2=0.7$), inclined
by $i_{\mathrm{rel}}=94^\circ$ relative to the inner
binary. Integration was done using Newtonian (blue line) and the EIH
equations of motion (green striped line) using a regularized Hermite
integrator with a time-step parameter $\eta=0.0003$. }
\label{fig:kozai1pnorbitalflips}
\end{figure*}

\cite{naoz2013resonant} further discussed the possibility of orbital
flips when including relativistic effects, even in the absence of
considerable variations at the Newtonian octupole moments, which form
the usual cause of orbital flips. In their fig.\,7, they presented a
specific case for an inner binary with $m_1=10\unit{M_\odot}$,
$m_2=8 \unit{M_\odot}$, $a_1=10\unit{AU}$, and $e_1=0.001$ that is
orbited by a $m_3=30\unit{M_\odot}$ tertiary body with semimajor axis
$a_2=502\unit{AU}$ and eccentricity $e_2=0.7$ and inclined by
$i_{\mathrm{rel}}=94^\circ$ to the plane of the inner
binary. According to eq.~\ref{eq:octupoleimportance}, the high mass
ratio of the inner binary suppresses the octupole-level (Newtonian)
effects. By integrating these initial conditions, including the EIH
equations of motion using the regularized Hermite integrator with a
time-step parameter $\eta = 0.0003$, we do not observe such a
orbital flip, as we show in fig.~\ref{fig:kozai1pnorbitalflips}.

The origin of this discrepancy is not so clear. The Newtonian case
does not show orbital flips, and we see no direct argument for the
presence of orbital flips when adopting the EIH equations of motion.
On the other hand, when integrating the EIH equations of motion, we
acquire a considerable error in the total energy when the binary
reaches its highest eccentricity. This is caused by the large
perturbation of the post-Newtonian terms when the inner binary reaches
pericenter. As a result, this system is hard to integrate
numerically. Our integration with a time-step parameter $\eta=0.0003$
integrated for $1\unit{Myr}$ took $\sim5 \unit{days}$ on a regular
workstation and reached a minimum relative inclination of
$i_{\mathrm{rel, min}}=93^\circ$.

Overall, our regularized Hermite integrator performs
well and gives results that are consistent with previous
calculations. Discrepancies with secular evolution calculations can be
understood from the lack of inner-orbit resolution in the latter.
We therefore see no reason to doubt our implementation of the
regularized and post-Newtonian terms.

\end{appendix}

\end{document}